\documentclass[useAMS,usenatbib]{mn2e}
\usepackage{txfonts}
\usepackage{graphicx}
\usepackage{natbib}
\usepackage{amsbsy}

\def\del#1{{}}

\newcommand{\dd}{\mathrm{d}}
\newcommand{\bra}{\langle}
\newcommand{\ket}{\rangle}
\newcommand{\ltsima}{$\; \buildrel < \over \sim \;$}
\newcommand{\lsim}{\lower.5ex\hbox{\ltsima}}
\newcommand{\gtsima}{$\; \buildrel > \over \sim \;$}
\newcommand{\gsim}{\lower.5ex\hbox{\gtsima}}
\newcommand{\eps}{\varepsilon}
\newcommand {\apgt} {\ {\raise-.5ex\hbox{$\buildrel>\over\sim$}}\ }
\newcommand {\aplt} {\ {\raise-.5ex\hbox{$\buildrel<\over\sim$}}\ } 

\newcommand{\ev}{\mathrm{eV}}
\newcommand{\gev}{\mathrm{GeV}}
\newcommand{\mev}{\mathrm{MeV}}
\newcommand{\tev}{\mathrm{TeV}}
\newcommand{\mpc}{m_{\mathrm{p}}c}
\newcommand{\mpcc}{m_{\mathrm{p}}c^2}
\newcommand{\p}{\mathrm{p}}

\newcommand{\e}{\mathrm{e}}

\newcommand{\inj}{\mathrm{inj}}

\newcommand{\CR}{\mathrm{CR}}
\newcommand{\KN}{\mathrm{KN}}

\newcommand{\mr}{\mathrm}
\newcommand{\DM}{\mathrm{DM}}
\newcommand{\thh}{\mathrm{th}}
\newcommand{\CMB}{\mathrm{CMB}}

\newcommand{\IC}{\mathrm{IC}}
\newcommand{\mecc}{m_\e c^2}
\newcommand{\mug}{\mathrm{\mu G}}
\newcommand{\LL}{\mathcal{L}}
\newcommand{\rvir}{R_\mathrm{vir}}
\newcommand{\mvir}{M_\mathrm{vir}}

\newcommand{\vel}{\upsilon}

\newcommand{\M}{{\mathcal M}}

\newcommand{\cc}{\rmn{c}}
\newcommand{\vv}{\rmn{v}}
\newcommand{\pic}{\rmn{pIC}}
\newcommand{\vr}{\boldsymbol{R}}
\newcommand{\vrp}{\boldsymbol{R}_\perp}
\newcommand{\pig}{\pi^0-\gamma}
\newcommand{\aei}{\alpha_{\rmn{e},i}}

\title[Simulating the gamma-ray emission from galaxy clusters]
{Simulating the gamma-ray emission from galaxy clusters: a universal cosmic ray spectrum and spatial distribution}

\author[A. Pinzke and C. Pfrommer] 
  {Anders Pinzke$^{1,2}$\thanks{e-mail:apinzke@fysik.su.se (AP); pfrommer@cita.utoronto.ca (CP)}
    and Christoph Pfrommer$^3$\footnotemark[1]\\
    $^1$Department of Physics, Stockholm University, AlbaNova University Center, SE - 106 91 Stockholm, Sweden\\
    $^2$The Oskar Klein Centre for Cosmoparticle Physics, Department of Physics, Stockholm University, AlbaNova University Center, SE - 106 91 Stockholm, Sweden\\
    $^3$Canadian Institute for Theoretical Astrophysics, University of Toronto, 60 St. George Street, Toronto, Ontario, M5S 3H8, Canada}
    
\begin{document}
\pagerange{\pageref{firstpage}--\pageref{lastpage}} \pubyear{2008}
\maketitle
\label{firstpage}

\begin{abstract}
  Entering a new era of high-energy $\gamma$-ray experiments, there is an
  exciting quest for the first detection of $\gamma$-ray emission from clusters
  of galaxies.  To complement these observational efforts, we use
  high-resolution simulations of a broad sample of galaxy clusters, and follow
  self-consistent cosmic ray (CR) physics using an improved spectral
  description. We study CR proton spectra as well as the different contributions
  of the pion decay and inverse Compton emission to the total flux and present
  spectral index maps. We find a universal spectrum of the CR component in
  clusters with surprisingly little scatter across our cluster sample. When CR
  diffusion is neglected, the spatial CR distribution also shows approximate
  universality; it depends however on the cluster mass. This enables us to
  derive a semi-analytic model for both, the distribution of CRs as
  well as the pion-decay $\gamma$-ray emission and the secondary radio emission
  that results from hadronic CR interactions with ambient gas protons. In
  addition, we provide an analytic framework for the inverse Compton emission
  that is produced by shock-accelerated CR electrons and valid in the full
  $\gamma$-ray energy range. Combining the complete sample of the brightest
  X-ray clusters observed by ROSAT with our $\gamma$-ray scaling relations, we
  identify the brightest clusters for the $\gamma$-ray space telescope Fermi and
  current imaging air \v{C}erenkov telescopes (MAGIC, HESS, VERITAS). We
  reproduce the result in \citet{2008MNRAS.385.1242P}, but provide somewhat more
  conservative predictions for the fluxes in the energy regimes of Fermi and
  imaging air \v{C}erenkov telescopes when accounting for the bias of
  `artificial galaxies' in cosmological simulations. We find that it will be
  challenging to detect cluster $\gamma$-ray emission with Fermi after the
  second year but this mission has the potential of constraining interesting
  values of the shock acceleration efficiency after several years of
  surveying. Comparing the predicted emission from our semi-analytic model to
  that obtained by means of our scaling relations, we find that the $\gamma$-ray
  scaling relations underpredict, by up to an order of magnitude, the flux from
  cool core clusters.
\end{abstract}


\section{Introduction}

\subsection{General background}

In the cold dark matter (CDM) universe, large scale structure grows
hierarchically through merging and accretion of smaller systems into larger
ones, and clusters are the latest and most massive objects that had time to
virialise.  This process leads to collisionless shocks propagating through the
intra-cluster medium (ICM), accelerating both protons and electrons to highly
relativistic energies \citep{1983RPPh...46..973D, 1987PhR...154....1B,
  1999ApJ...520..529S}.  High resolution X-ray observations by the {\it Chandra}
and {\it XMM-Newton} satellites confirmed this picture, with most clusters
displaying evidence for significant substructures, shocks, and contact
discontinuities (e.g.,~\citealp{2002ARA&A..40..539R,2005RvMP...77..207V,
  2007PhR...443....1M}).  In addition, observations of radio halos and radio
relics demonstrate the presence of synchrotron emitting electrons with energies
reaching $\sim$ 10 GeV in more than 50 clusters \citep{2003ASPC..301..143F,
  2008SSRv..134...93F}, although their precise origin in radio halos is still
unclear.  Similar populations of electrons may radiate $\gamma$-rays efficiently
via inverse Compton (IC) upscattering of the cosmic microwave background photons
giving rise to a fraction of the diffuse $\gamma$-ray background observed by
EGRET \citep{2000Natur.405..156L, 2000ApJ...545..572T, 2002MNRAS.337..199M,
  2003MNRAS.342.1009M, 2005AIPC..745..567I}.  Although there is no clear
observational evidence yet for a relativistic proton population in clusters of
galaxies, these objects are expected to contain significant populations of
relativistic protons originating from different sources, such as structure
formation shocks, radio galaxies, and supernovae driven galactic winds. The ICM
gas should provide ample target matter for inelastic collisions of relativistic
protons leading to $\gamma$-rays \citep{1996SSRv...75..279V,
  1997ApJ...477..560E, 2003MNRAS.342.1009M, 2003A&A...407L..73P,
  2004A&A...413...17P, 2008MNRAS.385.1211P, 2008MNRAS.385.1242P, Kushnir:2009vm}
as well as secondary electron injection \citep{1980ApJ...239L..93D,
  1982AJ.....87.1266V, 1999APh....12..169B, 2000A&A...362..151D,
  2001ApJ...562..233M, 2004A&A...413...17P, 2008MNRAS.385.1211P, Kushnir:2009vm,
  2009JCAP...09..024K}. These hadronic collision processes should illuminate the
presence of these elusive particles through pion production and successive decay
into the following channels:
\begin{eqnarray}
  \pi^\pm &\rightarrow& \mu^\pm + \nu_{\mu}/\bar{\nu}_{\mu} \rightarrow
  e^\pm + \nu_{e}/\bar{\nu}_{e} + \nu_{\mu} + \bar{\nu}_{\mu}\nonumber\\
  \pi^0 &\rightarrow& 2 \gamma \,\nonumber
\end{eqnarray}
This reaction can only unveil those cosmic ray protons (CRs) which have a total
energy that exceeds the kinematic threshold of the reaction of $E_\rmn{thr} =
1.22$~GeV.  The magnetic fields play another crucial role by confining
non-thermal protons within the cluster volume for longer than a Hubble time,
i.e. any protons injected into the ICM accumulates throughout the cluster's
history \citep{1996SSRv...75..279V, 1997ApJ...477..560E, 1997ApJ...487..529B}.
Hence, CRs can diffuse away from the production site, establishing a smooth
distribution throughout the entire ICM which serves as efficient energy
reservoir for these non-gravitational processes \citep{1999APh....12..169B,
  2000A&A...362..151D, 2001ApJ...559...59M}.

There is only little known theoretically about the spectral shape of the CR
population in the ICM. It is an interesting question whether it correlates with
injection processes or is significantly modified by transport and
re-acceleration processes of CRs through interactions with magneto-hydrodynamic
(MHD) waves.  The most important processes shaping the CR spectrum as a function
of cluster radius are (1) acceleration by structure formation shock waves
\citep{1998ApJ...502..518Q, 2000ApJ...542..608M, 2006MNRAS.367..113P}, MHD
turbulence, supernova driven galactic winds \citep{2009Natur.462..770V}, or
active galactic nuclei (AGN), (2) adiabatic and non-adiabatic transport
processes, in particular anisotropic diffusion, and (3) loss processes such as
CR thermalization by Coulomb interactions with ambient electrons and
catastrophic losses by hadronic interactions.  The spectral distribution of CRs
that are accelerated at structure formation shocks should be largely described
by a power-law with a spectral index of the one-dimensional distribution given
by
\begin{eqnarray}
\label{eq:alpha_rc}
  \alpha_\inj = \frac{r_\cc+2}{r_\cc-1},
\end{eqnarray}
where $r_\cc$ is the shock compression factor. Strong (high Mach number) shocks
that inject a hard CR population occur either at high redshift during the
formation of the proto-clusters or today at the boundary where matter collapses
from voids onto filaments or super-cluster regions. In contrast, merger shocks
show weak to intermediate strength with typical Mach numbers in the range of
$\M\simeq 2\ldots 4$ \citep{2003ApJ...593..599R, 2006MNRAS.367..113P,
  2008ApJ...689.1063S}.  AGNs or supernova remnants are expected to inject CRs
with rather flat spectra, $\alpha_\inj \approx 2.2 - 2.4$
\citep{1996SSRv...75..279V, 2002cra..book.....S, 2003A&A...399..409E}, but it is
not clear whether they are able to build up a homogeneous population of
significant strength.

The CRs offer a unique window to probe the process of structure formation due to
its long cooling times. While the thermal plasma quickly dissipates and erases
the information about its past history, the CR distribution keeps the fossil
record of violent structure formation which manifests itself through the
spectrum that is shaped by acceleration and transport processes.  The cluster
$\gamma$-ray emission is crucial in this respect as it potentially provides the
unique and unambiguous evidence of a CR population in clusters through observing
the pion bump in the $\gamma$-ray spectrum. This knowledge enables determining
the CR pressure and whether secondary electrons could contribute to the radio
halo emission. In the $\gamma$-ray regime, there are two main observables, the
morphological appearance of the emission and the spectrum as a function of
position relative to the cluster center.  The morphology of the pion induced
$\gamma$-ray emission should follow that seen in thermal X-rays albeit with a
slightly larger extent \citep{2008MNRAS.385.1211P}. The primary electrons that
are accelerated directly at the structure formation shocks should be visible as
an irregular shaped IC morphology, most pronounced in the cluster periphery
\citep{2003MNRAS.342.1009M, 2008MNRAS.385.1211P}.

\subsection{The $\gamma$-ray spectrum of a galaxy cluster}

\begin{figure}
\begin{minipage}{1.0\columnwidth}
  \includegraphics[width=1.0\columnwidth]{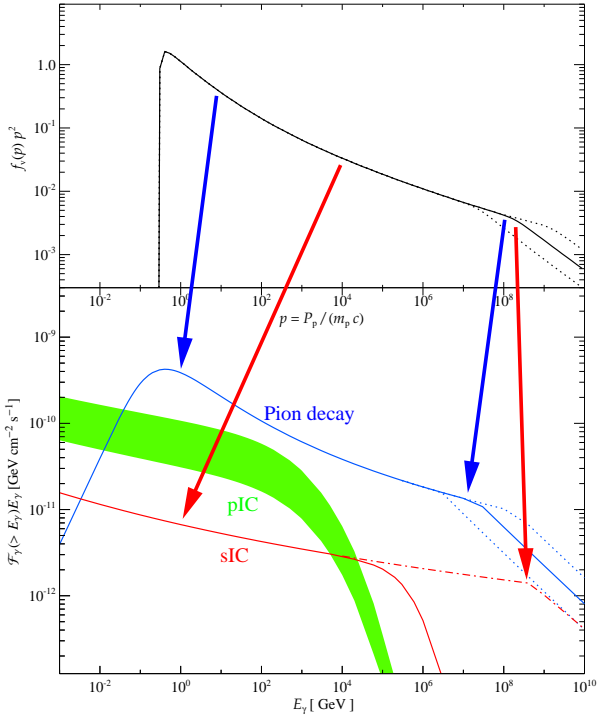}
  \caption{\emph{Upper Panel:} CR spectral distribution within the virial radius
    of a Coma-like cluster. It shows three distinct features: a cutoff around
    GeV energies, a concave shape, and a steepening at high energies due to
    diffusive losses.  \emph{Lower Panel:} we show the intrinsic $\gamma$-ray
    number flux weighted by the photon energy that does not take into account
    photon propagation effects. The arrows indicate the spectral mapping from
    the CR spectrum to the photon spectrum. The pion decay flux is denoted in
    blue color, the secondary inverse Compton (sIC) in red, and the primary
    inverse Compton (pIC) emission in green. Due to the large uncertainty in the
    diffusion coefficient $\kappa$, we demonstrate how varying $\kappa$ by a
    factor of two changes the corresponding $\gamma$-ray spectrum at high
    energies (dotted lines). The green band shows how the pIC emission changes
    if we vary the maximum electron injection efficiency from 0.05 (top) to 0.01
    (bottom).}
      \label{fig:sketch_CR_gamma}
  \end{minipage}
\end{figure}

How do the spectral electron and proton distributions map onto the $\gamma$-ray
spectrum?  We show the CR spectrum within the virial radius of a simulated
Coma-like galaxy cluster in the upper part of Fig.~\ref{fig:sketch_CR_gamma}. It
is shaped by diffusive shock acceleration at structure formation shocks,
adiabatic transport and the relevant CR loss processes\footnote{The physics will
  be thoroughly developed in this work but we will review the main
  characteristics here for introduction.}. Three distinct features are visible
in the spectrum: a cutoff close to the proton rest mass at $\mpcc\simeq 1$~GeV,
a concave shape for proton energies above $\mpcc$ and a steepening due to
diffusive losses at energies $E_\p\gtrsim 10^{16}\, \ev \,\times [\kappa_0/(
  10^{29} \mbox{ cm}^2\mbox{ s}^{-1})]^{-3}$, where $\kappa_0$ is the value of
the diffusion coefficient at 1~GeV. The dotted lines represents different values
of the diffusion coefficient which is varied by a factor two from its fiducial
value.  The low energy cutoff is due to a balance of Coulomb and hadronic losses
at energies around a GeV \citep{2008A&A...481...33J}. As shown in the present
paper and in an upcoming work by Pinzke \& Pfrommer (in prep.), the concave
curvature is a unique shape\footnote{A CR distribution with uniform spectral
  index was also hinted at by \citet{2001ApJ...559...59M} and
  \citet{2002MNRAS.337..199M}, where the dominating strong shocks caused a
    constant spectral index of about $\alpha\simeq2$. Their underlying Mach
    number distribution, however, is in conflict with that obtained by
  independent other works \citep{2003ApJ...593..599R, 2007MNRAS.378..385P,
    2008ApJ...689.1063S} and the spectral shape that we find (see
  Section~\ref{sect:comparison} for more details).} that is caused by the cosmic
Mach number distribution in combination with adiabatic transport processes.
These features are mapped onto the pion decay $\gamma$-ray emission spectra as a
consequence of hadronic CR interactions.

This can be seen in the lower part of Fig.~\ref{fig:sketch_CR_gamma}, where the
arrows indicate the spectral mapping from the CR spectrum to the photon
spectrum. In a hadronic interaction, CRs produce pions that decay into photons
with an energy that is on average smaller by a factor eight compare to the
original CR energy (see Section~\ref{sec:CR-modeling}). At CR energies that are
larger than the hadronic reaction threshold, the CR power-law behavior is
linearly mapped onto the {\em pion decay induced $\gamma$-ray} spectrum (solid
blue). This emission component clearly dominates the total photon spectrum and
therefore shapes the total emission characteristics in the central parts of the
cluster, where the densities are high.  Note that this spectrum is an intrinsic
spectrum emitted at the cluster position and converted to a flux while assuming
a distance of 100~Mpc without taking into account photon propagation
effects. Depending on the cluster redshift, the finite mean free path of
high-energy $\gamma$-rays to $e^+e^-$-pair production on infra red (IR) and
optical photons limits the observable part of the spectrum to energies
$E_\gamma\lesssim 10$~TeV for clusters with redshifts $z\sim0.03$, and smaller
energies for higher redshift objects \citep{2008A&A...487..837F}.

Secondary CR electrons and positrons up-scatter cosmic microwave background
(CMB) photons through the IC process into the $\gamma$-ray regime, the so-called
{\em secondary inverse Compton emission (sIC)}. This emission component
originates from the flat high-energy part of the CR spectrum and produces a
rather flat sIC spectrum up to the Klein-Nishina (KN) regime. At large
electron energies, we enter the KN regime of IC scattering where the electron
recoil effect has to be taken into account. It implies less efficient energy
transfer in such an elastic scattering event compared to the Thomson regime and
leads to a dramatic steepening of the sIC spectrum at $\gamma$-ray energies
around 100~TeV (solid red line). The dash-dotted red line shows the hypothetical
sIC spectrum in the absence of the KN effect (which is never realized in
Nature). However, it clearly shows that the diffusive CR break is not observable
in the sIC component for large clusters (while it can move to energies below the
KN break for small enough clusters, causing a faster steepening there).  The
spectrum shown in green color represents the energy weighted photon spectrum
resulting from the IC process due to electrons accelerated at structure
formation and merger shocks, the {\em primary inverse Compton emission
  (pIC)}. The exponential cutoff is due to synchrotron and IC losses which lead
to a maximum energy of the shock-accelerated electrons. The green pIC band shows
the effect of the maximum electron injection efficiency, where we use an
optimistic value of $\zeta_\rmn{e,max}=0.05$ (see
e.g. \citealt{2003ApJ...585..128K}) in the top and a value of
$\zeta_\rmn{e,max}=0.01$ at the bottom. This more realistic value is suggested
to be the theoretically allowed upper limit for the injection efficiency that is
consistent with the non-thermal radiation of young supernova remnants
\citep{2010ApJ...708..965Z}.

This work studies the spectral and morphological emission characteristics of the
different CR populations in the $\gamma$-ray regime. We concentrate on
observationally motivated high-energy $\gamma$-ray bands. (1) The energy regime
accessible to the {\em Fermi $\gamma$-ray space telescope} with a particular
focus on $E_\gamma = 100 \, \mev$ and (2) the energy regime accessible to imaging
air \v{C}erenkov telescopes (IACTs) assuming a lower energy limit of $100 \,
\gev$. In Section~\ref{sect:setup} we describe the setup of our simulations,
explain our methodology and relevant radiative processes considered in this
work.  In Section~\ref{sect:morphology}, we study emission profiles and maps, as
well as spectral index maps. We then present the CR spectrum and spatial
distribution and show its universality across our simulated cluster sample in
Section~\ref{sect:CRp_spec}. This allows us to derive a semi-analytic framework
for the cluster $\gamma$-ray emission in Section~\ref{sect:analytic_model} which
we demonstrate on the Perseus and Coma galaxy clusters. Furthermore, we study
the mass-to-luminosity scaling relations (Section~\ref{sect:scaling}) and predict
the $\gamma$-ray flux from a large sample of galaxy clusters for the GeV and TeV
energy regimes in Section~\ref{sect:prediction}. We compare our work to previous
papers in this field and point out limitations of our approach in
Section~\ref{sect:discussion_comparison}.  We conclude our findings in
Section~\ref{sect:conclusions}. Throughout this work we use a Hubble constant of
$H_0 =70\,\rmn{km}\,\rmn{s}^{-1}\,\rmn{Mpc}^{-1}$, which is a compromise between
the value found by the Hubble key project \citep[$H_0
=72$,][]{2001ApJ...553...47F} and from that one inferred from baryonic acoustic
oscillation measurements \citep[$H_0 =68$,][]{2009arXiv0907.1660P}.

\section{Setup and formalism}
\label{sect:setup}
We follow the CR proton pressure dynamically in our simulations while taking
into account all relevant CR injection and loss terms in the ICM, except for a
possible proton production from AGN and supernova remnants. In contrast, we
model the CR electron population in a post-processing step because it does not
modify the hydrodynamics owing to its negligible pressure contribution. We use a
novel CR formalism that allows us to study the spectral properties of the CR
population more accurately.

\subsection{Adopted cosmology and cluster sample}
\label{sec:cosmology}

The simulations were performed in a $\Lambda$CDM universe using the cosmological
parameters: $\Omega_{\mr m}=\Omega_\DM + \Omega_{\mr b}=0.3, \, \Omega_{\mr
  b}=0.039, \, \Omega_\Lambda=0.7, \, h=0.7, \, n_{\mr s}=1$, and
$\sigma_8=0.9$. Here, $\Omega_\rmn{m}$ denotes the total matter density in units
of the critical density for geometrical closure today, $\rho_\rmn{crit} = 3
H_0^2 / (8 \upi G)$. $\Omega_\rmn{b}$, $\Omega_\DM$ and $\Omega_\Lambda$ denote
the densities of baryons, dark matter, and the cosmological constant at the
present day. The Hubble constant at the present day is parametrized as $H_0 =
100\,h \mbox{ km s}^{-1} \mbox{Mpc}^{-1}$, while $n_{\mr s}$ denotes the
spectral index of the primordial power-spectrum, and $\sigma_8$ is the {\em rms}
linear mass fluctuation within a sphere of radius $8\,h^{-1}$Mpc extrapolated to
$z=0$.

Our simulations were carried out with an updated and extended version of the
distributed-memory parallel TreeSPH code GADGET-2 \citep{2005MNRAS.364.1105S,
  2001NewA....6...79S}. Gravitational forces were computed using a combination
of particle-mesh and tree algorithms.  Hydrodynamic forces are computed with a
variant of the smoothed particle hydrodynamics (SPH) algorithm that conserves
energy and entropy where appropriate, i.e. outside of shocked regions
\citep{2002MNRAS.333..649S}.  Our simulations follow the radiative cooling of
the gas, star formation, supernova feedback, and a photo-ionizing background
\citep[details can be found in][]{2007MNRAS.378..385P}.  

The clusters have originally been selected from a low-resolution
dark-matter-only simulation \citep{2001MNRAS.328..669Y}. Using the `zoomed
initial conditions' technique \citep{1993ApJ...412..455K}, the clusters have
been re-simulated with higher mass and force resolution by adding
short-wavelength modes within the Lagrangian regions in the initial conditions
that will evolve later-on into the structures of interest.  We analyzed the
clusters with a halo-finder based on spherical overdensity followed by a merger
tree analysis in order to get the mass accretion history of the main progenitor.
The spherical overdensity definition of the virial mass of the cluster is given
by the material lying within a sphere centered on a local density maximum, whose
radial extend $R_\Delta$ is defined by the enclosed threshold density condition
$M (< R_\Delta) / (4 \pi R_\Delta^3 / 3) = \rho_\rmn{thres}$. We chose the
threshold density $\rho_\rmn{thres}(z) = \Delta\, \rho_\rmn{crit} (z)$ to be a
constant multiple $\Delta=200$ of the critical density of the universe
$\rho_\rmn{crit} (z) = 3 H (z)^2/ (8\pi G)$. In the remaining of the paper, we
use the terminology $\rvir$ instead of $R_{200}$. Our sample of simulated galaxy
clusters consists of 14 clusters that span a mass range from $8\times 10^{13}
\mathrm{M_\odot}$ to $3\times 10^{15} \mathrm{M_\odot}$ where the dynamical
stages range from relaxed cool core clusters to violent merging clusters
(cf. Table~\ref{tab:cluster_sample}). Each individual cluster is resolved by
$8\times 10^4$ to $4\times 10^6$ particles, depending on its final mass. The SPH
densities were computed from the closest 48 neighbors, with a minimum smoothing
length ($h_\rmn{sml}$) set to half the softening length. The Plummer equivalent
softening length is $7 \,\rmn{kpc}$ in physical units after $z=5$, implying a
minimum gas resolution of approximately $1.1\times 10^{10} \mathrm{M_\odot}$
\citep[see also][]{2007MNRAS.378..385P}.

\begin{table}
\caption{Cluster sample.}
\begin{tabular}{l l l l r r}
\hline
\hline
Cluster & sim.'s & dyn. state$^{(1)}$ & $M_{\rm vir}^{(2)}$ & $\rvir^{(2)}$ & $kT_{\rm vir}^{(3)}$ \\
& & & [$\rmn{M}_\odot$] & [Mpc] & [keV] \\
\hline
1  & g8a  & CC    & $2.6\times 10^{15}$ &   2.9~~ & 13.1 \\
2  & g1a  & CC    & $1.9\times 10^{15}$ &   2.5~~ & 10.6 \\
3  & g72a & PostM & $1.6\times 10^{15}$ &   2.4~~ & 9.4  \\
4  & g51  & CC    & $1.5\times 10^{15}$ &   2.4~~ & 9.4  \\
                                                     
5  & g1b  & M     & $5.2\times 10^{14}$ &   1.7~~ & 4.7  \\
6  & g72b & M     & $2.2\times 10^{14}$ &   1.2~~ & 2.4  \\
7  & g1c  & M     & $2.0\times 10^{14}$ &   1.2~~ & 2.3  \\
8  & g8b  & M     & $1.5\times 10^{14}$ &   1.1~~ & 1.9  \\
9  & g1d  & M     & $1.3\times 10^{14}$ &   1.0~~ & 1.7  \\
                                                     
10 & g676 & CC    & $1.3\times 10^{14}$ &   1.0~~ & 1.7  \\
11 & g914 & CC    & $1.2\times 10^{14}$ &   1.0~~ & 1.6  \\
12 & g1e  & M     & $9.1\times 10^{13}$ &  0.93   & 1.3  \\
13 & g8c  & M     & $8.5\times 10^{13}$ &  0.91   & 1.3  \\
14 & g8d  & PreM  & $7.8\times 10^{13}$ &  0.88   & 1.2  \\
\hline
\end{tabular}  \begin{quote}
 Notes:\\ (1) The dynamical state has been classified through a combined
 criterion invoking a merger tree study and the visual inspection of the X-ray
 brightness maps. The labels for the clusters are M--merger, PostM--post
 merger (slightly elongated X-ray contours, weak cool core (CC) region developing),
 PreM--pre-merger (sub-cluster already within the virial radius), CC--cool
 core cluster with extended cooling region (smooth X-ray profile).  (2) The
 virial mass and radius are related by $M_\Delta(z) = \frac{4}{3} \pi\,
 \Delta\, \rho_\rmn{crit}(z) R_\Delta^3 $, where $\Delta=200$ denotes a
 multiple of the critical overdensity $\rho_\rmn{crit}(z) = 3 H (z)^2/ (8\pi
 G)$.  (3) The virial temperature is defined by $kT_{\rm vir} = G M_{\rm vir} \,
 \mu\, m_\p / (2 R_{\rm vir})$, where $\mu$ denotes the mean molecular weight.
\end{quote}
\label{tab:cluster_sample}
\end{table}

\subsection{Modeling of CR protons and induced radiative processes}
\label{sec:CR-modeling}

Our simulations follow cosmic ray physics in a self-consistent way
\citep{2006MNRAS.367..113P, 2007A&A...473...41E, 2008A&A...481...33J}.  We model
the adiabatic CR transport process such as compression and rarefaction, and a
number of physical source and sink terms which modify the cosmic ray pressure of
each CR population separately. The most important source considered\footnote{For
  simplicity, in this paper we do not take into account CRs injected into the
  inter-stellar medium from supernova remnants (see \citet{2009arXiv0909.3267T}
  for a discussion of this topic).} for acceleration is diffusive shock
acceleration at cosmological structure formation shocks, while the primary sinks
are thermalization by Coulomb interactions, and catastrophic losses by
hadronization.  Collisionless structure formation shocks are able to accelerate
ions and electrons in the high-energy tail of their Maxwellian distribution
functions through diffusive shock acceleration \citep[for reviews
  see][]{1983RPPh...46..973D, 1987PhR...154....1B, 2001RPPh...64..429M}.  In the
test particle picture, this process injects a CR distribution with a power-law
in momentum and a slope that depends on the instantaneous sonic Mach number of
the shock. The overall normalization of the injected CR distribution depends on
the adopted sub-resolution model of diffusive shock acceleration
\citep[e.g.,][]{2007A&A...473...41E}; in particular it depends on the maximum
acceleration efficiency $\zeta_\rmn{max,p} = \eps_{\CR,\rmn{max}}/
\eps_\rmn{diss}$ which is the maximum ratio of CR energy density that can be
injected relative to the total dissipated energy density at the shock.  We
assume that in the saturated regime of shock acceleration, 50 percent of the
dissipated energy at strong shocks is injected into cosmic ray protons.  While
there are indications from supernova remnant observations of one rim region
\citep{2009Sci...325..719H} as well as theoretical studies
\citep{2005ApJ...620...44K} that support such high efficiencies, to date it is
not clear whether these efficiencies apply in an average sense to strong
collisionless shocks or whether they are realized for structure formation shocks
at higher redshifts. This high efficiency rapidly decreases for weaker shocks
(decreasing Mach number) and eventually smoothly approaches zero for sonic waves
\citep{2007A&A...473...41E}.  Our paper aims at providing a quantitative
prediction of the $\gamma$-ray flux and hence the associated CR flux that we
expect in a cluster depending on our adopted acceleration model. Non-detection
of our predicted emission will limit the CR acceleration efficiency and help in
answering these profound plasma astrophysics questions about particle
acceleration efficiencies.

We significantly revised the CR methodology and allow for multiple non-thermal
cosmic ray populations of every fluid element (Pinzke \& Pfrommer, in
prep.). Each CR population $f_i(p,\vr)$ is a power-law in particle
momentum\footnote{The true CR particle momentum is denoted by $P_\p$, but we
  loosely refer to $p=P_\p/(m_\p c)$ as the particle momentum.},
\begin{eqnarray}
\label{eq:single_CRp_spectrum}
f_i(p,\vr) = C_i(\vr)\,p^{-\alpha_i} \,\theta(p-q_i)\,,
\end{eqnarray}
characterized by a fixed slope $\alpha_i$, a low-momentum cutoff $q_i$, and an
amplitude $C_i(\vr)$ that is a function of the position of each SPH particle
through the variable $\vr$. For this paper we have chosen five CR populations
with the spectral index distribution $\boldsymbol{\alpha} =(2.1, 2.3, 2.5, 2.7,
2.9)$ for each fluid element (a convergence study on the number of CR
populations is presented in the appendix~\ref{app:convergence}).  This approach
allows a more accurate spectral description\footnote{The total CR proton
  spectrum is a sum of the spectra of the individual SPH particles within a
  certain volume, and since our sample contains a large number of SPH particles
  with varying normalization, the CRp spectral index is a statistically well
    defined continuous quantity.} as the superposition of power-law
spectra enables a concave curvature of the composite spectrum in logarithmic
representation.  Physically, more complicated spectral features such as bumps
can arise from the finite lifetime and length scale of the process of diffusive
shock acceleration or incomplete confinement of CRs to the acceleration
region. These effects imprint an upper cutoff to the CR population locally that
might vary spatially and which translates into a convex curvature in
projection. Additionally, interactions of pre-existing CRs with MHD waves can
yield to more complex spectral features. Future work will be dedicated to study
these topics.

In addition to the spectral features mentioned above, we model in the
post-processing the effect of high-energy CR protons that are no longer confined
to a galaxy cluster as these are able to diffuse into the ambient warm-hot
intergalactic medium (WHIM). In this paper we define WHIM to be the region
within $\rvir < R < 3\,\rvir$, which is a subset of the entire WHIM
\citep{2001ApJ...552..473D}. Assuming particle scattering off magnetic
irregularities with the Kolmogorov spectrum, we obtain the characteristic
scaling of the diffusion coefficient $\kappa \simeq \kappa_0\, (E/E_0)^{1/3}$,
where we normalize $\kappa$ at $E_0 = 1\,\rmn{GeV}$. One can estimate the
characteristic proton energy $E_\rmn{p,~break}$ at which the spectrum steepens
\citep{1996SSRv...75..279V, 1997ApJ...487..529B},
\begin{eqnarray}
\label{eq:break}
E_\rmn{p,~break} &\approx& \frac{E_0 R^6}{\left(6\, \kappa_0\, \tau\right)^3} \\
&\approx& 
10^8 \,\gev \,\left(\frac{R}{3 \mbox{ Mpc}}\right)^6
\left(\frac{\kappa_0}{ 10^{29} \mbox{ cm}^2\mbox{ s}^{-1}}\right)^{-3}
\left(\frac{\tau}{\tau_\rmn{Hubble}}\right)^{-3}.\nonumber
\end{eqnarray}
For the reminder, we adopt a value of the diffusivity that is scaled to
$R=2\,\rvir$ for each cluster, as this volume is expected to fall within the
virialised part of the cluster past the accretion shock region
\citep{2007MNRAS.378..385P, Molnar:2009rk} and traps CRs in a cluster for time
scales longer than a Hubble time. This choice also has the property that the
diffusion break is at energies $E_\p>100$~GeV; hence it does not interfere with
the pion decay as well as secondary IC emission in the energy regime accessible
to IACTs as we will show in the following.  The momentum of a photon that
results from pion decay is given by
\begin{eqnarray}
\label{eq:pion_energy}
P_\gamma \approx \frac{K_\p}{2}\frac{P_\p}{\xi} \approx \frac{P_\p}{8}.
\end{eqnarray}
This approximate relation is derived using the inelasticity $K_\p \approx 1/2$
and multiplicity $\xi \approx 2$ for the $p+p\rightarrow \pi^0$ channel together
with the two photons in the final state.  Secondary electrons that are injected
in hadronic CR interactions Compton up-scatter CMB photons. A break in the
parent CR spectrum would imprint itself in the sIC spectrum if there are no
other effects that modify the spectrum at lower energies.  Compared to the pion
decay emission, this break manifests at slightly higher energies (for parameters
adopted in Fig.~\ref{fig:sketch_CR_gamma}). The momentum of the electrons $P_\e$
depends on the proton momentum $P_\p$ through the relation given by hadronic
physics
\begin{eqnarray}
\label{eq:sCRe_energy}
P_\rmn{e} \approx \frac{K_\p}{4}\frac{P_\p}{\xi} \approx \frac{P_\p}{16}.
\end{eqnarray}
Here we used the $p+p\rightarrow \pi^\pm$ channel together with the four
particles in the final state of the charged pion decay ($e^\pm$,
$\nu_{e}/\bar{\nu}_{e}$, $\nu_{\mu}$, $\bar{\nu}_{\mu}$).  Combining the
classical inverse Compton formulae from CR electrons with energies $E_\e >
1\,\gev$
\begin{eqnarray}
\label{eq:ICenergy}
E_\rmn{IC} = \frac{4}{3}E_\CMB \left(\frac{E_\e}{\mecc}\right)^2\,,
\end{eqnarray}
with the energy relation in equation~(\ref{eq:sCRe_energy}) we obtain a break in
the sIC spectrum. This steepening of the CR spectrum take place at high photon
energies $E_\rmn{sIC,~break}\simeq 10^{17} \, \ev$ where we choose CMB photons
with the energy $E_\CMB = h\nu_\rmn{CMB}\simeq 0.66$~meV as source for the
inverse Compton emission using Wien's displacement law. It turns out that these
energies are deeply in the Klein-Nishina regime. This means that in the rest
frame of the energetic electron, the Lorentz boosted photon energy is comparable
to or larger than the electron rest mass, $E_\rmn{IC} = \gamma_\e h
\nu_\rmn{init} \sim m_\e c^2$, so that the scattering event becomes
elastic. This implies a less efficient energy transfer to the photon and
manifests itself in a break in the resulting IC spectrum. While the number flux
scales as $\mathcal{F}\sim E_\rmn{IC}^{-(\alpha_\e-1)/2}$ in the Thomson-limit
for $E_\rmn{IC}\ll 30$~TeV, it steepens significantly to $\mathcal{F}\sim
E_\rmn{IC}^{-\alpha_\e}\rmn{log}(E_\rmn{IC}^{})$ in the extreme KN-limit for
$E_\rmn{IC} \gg 30$~TeV, where $\alpha_\e$ is the spectral index of the (cooled)
CR electron distribution \citep{1970RvMP...42..237B}.

\subsection{Magnetic fields}

High energy CR electrons with $\gamma_\e > 200$ loose their energy by means of
IC scattering off CMB photons as well as through interactions with cluster
magnetic fields which results in synchrotron emission. The relative importance
of these two emission mechanisms depends on the {\em rms} magnetic field
strength, $B$, relative to the equivalent field strength of the CMB, $B_{\rm
  CMB} = 3.24\, (1+z)^2\mug$, where $z$ denotes the redshift. In the peripheral
cluster regions, where $B\ll B_{\rm CMB}$, the CR electrons loose virtually all
their energy by means of IC emission. In the central cluster regions, in
particular in the dense centers of cool cores, the magnetic energy density is
probably comparable or even larger than the energy density of the CMB
\citep{2005A&A...434...67V}, $\eps_\rmn{ph}=B_{\rm CMB}^2/(8\pi)$. Hence in
these regions, the radio synchrotron emission carries away a fraction of the CR
electrons' energy losses; an effect that reduces the level of IC emission.  We
model the strength and morphology of the magnetic fields in the post-processing
\citep{2008MNRAS.385.1211P} and scale the magnetic energy density field
$\epsilon_B$ by the thermal energy density $\epsilon_\thh$ through the relation
\begin{eqnarray}
\label{eq:magnetic_energyd}
\epsilon_B =
\epsilon_{B_0}\left(\frac{\epsilon_\thh}{\epsilon_{\thh_0}}\right)^{2\alpha_B},
\end{eqnarray}
where $\epsilon_{B_0}= B_0^2/(8\pi)$ and $\epsilon_{\thh_0}$ denote the core
values. If not mentioned otherwise, we use the magnetic decline $\alpha_B=0.5$
and the central magnetic field $B_0=10 \, \mug$ throughout this paper. The
central thermal energy density $\epsilon_{\thh_0}=3P_{\thh_0}/2$, is calculated
by fitting the modified $\beta$-model
\begin{eqnarray}
\label{eq:radial_pressure}
P(R) =  P_{\thh_0} \left[1+\left(\frac{R}{R_\rmn{core}}\right)\right]^{-\beta}
\end{eqnarray}
to the radial pressure $P(R)$. The parametrization in
equation~(\ref{eq:magnetic_energyd}) is motivated by both cosmological MHD SPH
simulations \citep{1999A&A...348..351D} and radiative adaptive mesh refinement
MHD simulations \citep{Dubois:2008mz}. Rather than applying a densities scaling
as those simulations suggest, we use a scaling with thermal gas energy density
which is not affected by the over-cooled centers in radiative simulations that
do not take into account AGN feedback.

\subsection{CR electron acceleration and inverse Compton emission}
\label{section:CRe_IC}

\subsubsection{Modeling diffusive shock acceleration}
\label{section:diffusive_shock_acceleration}
Collisionless cluster shocks are able to accelerate ions and electrons through
diffusive shock acceleration \citep[for reviews see][]{1983RPPh...46..973D,
  1987PhR...154....1B, 2001RPPh...64..429M}.  Neglecting non-linear shock
acceleration and cosmic ray modified shock structure, the process of diffusive
shock acceleration uniquely determines the spectrum of the freshly injected
relativistic electron population in the post-shock region that cools and finally
diminishes as a result of loss processes. The $\gamma$-ray inverse Compton
emitting electron population cools on such a short time scale $\tau_\rmn{sync} <
10^8$~yrs (compared to the long dynamical time scale $\tau_\rmn{dyn} \sim
2$~Gyr) that we can describe this by instantaneous cooling.\footnote{Assuming a
  magnetic field of a few $\mug$ and an electron density $n_\e =
  10^{-3}\,\rmn{cm}^{-3}$, for further discussion see e.g. appendix in
  \citet{2008MNRAS.385.1211P}} In this approximation, there is no steady-state
electron population and we would have to convert the energy from the electrons
to inverse Compton and synchrotron radiation. Instead, we introduce a virtual
electron population that lives in the SPH-broadened shock volume only; this is
defined to be the volume where energy dissipation takes place. Within this
volume, which is co-moving with the shock, we can use the steady-state solution
for the distribution function of relativistic electrons and we assume no
relativistic electrons in the post-shock volume, where no energy dissipation
occurs. Thus, the cooled CR electron equilibrium spectrum can be derived from
balancing the shock injection with the IC/synchrotron cooling: above a GeV it is
given by
\begin{equation}
\label{eq:f_eq}
f_\e(p_\e) = C_\e\, p_\e^{-\alpha_\e}, \quad
C_\e \propto \frac{\rho}
{\eps_B + \eps_\rmn{ph}}
\end{equation}
Here, we introduced the dimensionless electron momentum $p_\e=P_\e/(m_\e c)$,
where $P_\e$ is the electron momentum,  $\alpha_\e = \alpha_\inj + 1$ is the
spectral index of the equilibrium electron spectrum, $\rho$ is the gas density,
$\eps_B$ is the magnetic energy density, and $\eps_\rmn{ph}$ denotes the photon
energy density, taken to be that of CMB photons. The primary CRe distribution in
equation~(\ref{eq:f_eq}) is calculated in the post-processing with a spectrum
reflecting the current Mach number of the shock (without the assumption of
spectral bins). Superposing the individual spectra of a large number of SPH
particles, each with a spectrum reflecting the accelerating shock, produces a
well defined total spectrum with a running index in general. A more detailed
discussion of this simplified approach can be found in
\citet{2008MNRAS.385.1211P}.

Once the radiative cooling time due IC and synchrotron emission becomes
comparable to the diffusive acceleration time scale, the injection spectrum
experiences a high-energy cutoff \citep{webb84}. The electrons start to pile-up
at this critical energy; the super-exponential term describing the maximum
energy of electrons reached in this process, however, effectively cancels this
pile-up feature which results in a prolonged power-law up to the electron cutoff
momentum $p_\e\sim p_\rmn{max}$ \citep{2007A&A...465..695Z}.  We account for
this effect by using the following parametrization of the shock injected
electron spectrum,
\begin{eqnarray}
\label{eq:fe_exp_sect2}
f_\e(x,p_\e) &=& C_\e p_\e^{-\alpha_\rmn{inj}} \left[1+j(x,
  p_\e)\right]^{\delta_\e}\exp\left[\frac{-p_\e^2}{p_\rmn{max}^2(x)}\right]\,,
\end{eqnarray}
where $x$ is the distance from the shock surface, $j(x, p_\e)$ and $\delta_\e$
describe the characteristic momentum and shape of the pile-up region.  The
continuous losses cause the cutoff to move to lower energies as the electrons
are transported advectively with the flow downstream. Integration over the
post-shock volume causes the cutoffs to add up to a new power-law that is
steeper by unity compared to the injection power-law (equation~\ref{eq:f_eq}).
Hence, the shock-integrated distribution function -- as defined in
equation~(\ref{eq:fe_cool}) and displayed in Fig.~\ref{fig:elcc_comp} -- shows
three regimes: (1) at low energies (but large enough in order not to be affected
by Coulomb losses) the original injected power-law spectrum, (2) followed by the
cooled power-law that is steeper by unity, $\alpha_\e = \alpha_\inj + 1$, and
(3) an ultimate cutoff that is determined from the magnetic field strength and
the properties of the diffusion of the electron in the shock (we refer the
reader to Section~\ref{sec:fe_inj} for a detailed discussion). Note that only
the last two regimes are important for $\gamma$-ray IC emission.

The fact that we observe X-ray synchrotron emission at shocks of young supernova
remnants \citep{2006ApJ...648L..33V, 1999ApJ...525..357S} necessarily requires
the existence of high-energy CR electrons with $E_\e \simeq 25$~TeV. The
non-thermal synchrotron emission generated by CR electrons with energies $E_\e >
\mbox{GeV}$ is given by
\begin{equation}
  \label{eq:synchrotron}
  h\,\nu_\rmn{synch} = \frac{3 e B h}{2\pi\, m_\rmn{e} c}\,\gamma^2\,,
\end{equation}
where $\gamma \sim 5\times10^7$ ($E_\e \sim 25\,$TeV) and magnetic fields of
order $100\,\umu$G are required to reach X-ray energies of order 10~keV.  To
keep the highly relativistic electrons from being advected downstream requires
efficient diffusion so that they can diffuse back upstream crossing the shock
front again. We therefore use the most effective diffusion, refereed to as Bohm
diffusion limit, as the electron propagation model at the shock. Balancing Bohm
diffusion with synchrotron/IC cooling of electrons enables us to derive a
maximum energy of the accelerated CR electrons at the position of the shock
surface \citep[derived in appendix, see also][]{webb84, 1998A&A...332..395E}
\begin{eqnarray}
\label{eq:Ecut}
E_\rmn{max} = \left[B\,e\,u^2\,m_\e c \,\tau_\rmn{loss}
\frac{r_\cc-1}{r_\cc\left(r_\cc+1\right)}\right]^{0.5}\,,
\end{eqnarray}
where diffusion parallel to the magnetic field has been assumed. Here $u$
denotes the flow velocity in the inertial frame of the shock ($u =
\upsilon_\rmn{s}$), and the electron loss time scale due to synchrotron and
inverse-Compton losses reads
\begin{eqnarray}
\label{eq:tauloss}
\tau_\rmn{loss}^{-1} = \frac{\dot{E}}{E}=
\frac{4 \sigma_\rmn{T}\,\gamma}{3m_\e c} ({\eps_B + \eps_\rmn{ph}})\,,
\end{eqnarray}
where $\sigma_\rmn{T}$ is the Thompson cross-section. $r_\cc$ denotes the shock
compression and is given by
\begin{eqnarray}
\label{eq:Rcompress}
r_\cc = \frac{\left(\gamma_\mathrm{th}+1\right)\mathcal{M}^2}{\left(\gamma_\mathrm{th}-1\right)\mathcal{M}^2 + 2}\,,
\end{eqnarray}
where $\mathcal{M}$ denotes the sonic Mach number. Inserting typical numbers for
different cluster regions show that the electron cutoff energy, $E_\rmn{max}$,
only varies within a factor two inside $\rvir$ (Table~\ref{tab:diffusion}). The
equivalent cutoff energy in the IC spectrum can easily be derived from
equation~(\ref{eq:ICenergy}), yielding $E_{\IC, \rmn{cut}} \simeq 10 \,\rmn{TeV}
\times (E_\rmn{max} / 50\,\rmn{TeV})^2$.

\begin{table}
  \caption{Electron energy cutoff in different regions of a typical cluster.}
\begin{tabular}{c c  c c c}
\hline
\hline
 Cluster variables$^{(1)}$ & $< 0.03\,\rvir$  & $0.3\,\rvir$  &  $\rvir$  &   WHIM  \\
\hline
$h_\rmn{sml}\,[\rmn{kpc}]$                         & $< 30$ & 60 & 120 & 240 \\
$n\,[10^{-3}\times\,\rmn{particles}\,\rmn{cm}^{-3}]$   & 4 & 0.4 & 0.04  & 0.004 \\
$r_\cc$                                             & 1.3 & 2.3 & 3.0  & 3.7 \\
$\mathcal{M}$                                       & 1.2 & 2 & 3 & 6 \\
$T\,[\rmn{keV}]$                                     & 15 & 9  & 5 & 0.2 \\
$c_\rmn{sound}\,[\rmn{km}\,\rmn{s}^{-1}]$            & 2400 & 3100  & 3500 & 1400 \\
$B\,[\mug]$                                         & 10 & 2.5  & 0.6 & 0.04 \\
\hline
\hline
$E_\rmn{max}\,[\rmn{TeV}]$                           & 50 & 100 & 65 & 6.5 \\
\hline
\end{tabular}
\begin{quote}
 Notes: \\ 
 (1) Other constants used:  mean molecular weight $\mu = 0.588\,m_\p$.\\
 \label{tab:diffusion}
  \end{quote}
\end{table}

\subsubsection{IC emission}

Following \citet{1979rpa..book.....R}, we calculate the inverse Compton emission
from electrons that up-scatter CMB photons. It should be noted that we neglect
the inverse Compton emission induced by starlight and dust, which might
contribute significantly in the inner 10 kpc of the cluster. The inverse Compton
source density $\lambda_\rmn{IC}$ in units of produced photons per unit time
interval and volume for a simple power-law spectrum of CRes
(equation~\ref{eq:f_eq}) scales as
\begin{equation}
  \lambda_\rmn{IC} \propto C_{\e} \eps_\rmn{ph} E_\rmn{IC}^{-\alpha_{\nu}},
  \label{eq:jnu}
\end{equation}
where $\alpha_{\nu} = (\alpha_{\mathrm{e}} - 1)/2$.  When we account for the
competition between radiative cooling and diffusive acceleration of electrons in
the shock region, the shape of $\lambda_\rmn{IC}$ in the high-energy regime
changes. Using the cooled electron distribution of
equation~(\ref{eq:fe_cool}), we construct an effective integrated source
function for primary inverse Compton emission (see Section~\ref{sec:pIC} for
a self-consistent and extensive description),
\begin{eqnarray}
\label{eq:pICfit}
\lambda_\pic &=& \tilde{\lambda}_0(\zeta_\rmn{e,max},C_\e)\,f_\mathrm{IC}(\alpha_\e)\,
f_\mathrm{KN}(E_\IC,\alpha_\e)\,
\left(\frac{E_\IC}{k T_\mathrm{CMB}}\right)^{-\frac{\alpha_\e-1}{2}} \nonumber\\
&\times&\left(1+0.84\sqrt{\frac{\,E_\IC}{E_{\IC, \rmn{cut}}}}\,\right)^
{\delta_{\rmn{IC}}(E_\IC, \alpha_\e)}
\rmn{exp}\left(-\sqrt{\frac{4.07\,E_\IC}{E_{\IC, \rmn{cut}}}}\right), 
\end{eqnarray}
where Bohm diffusion has been assumed. The normalization constants
$\tilde{\lambda}_0(\zeta_\rmn{e,max},C_\e)$ and $f_\mathrm{IC}(\alpha_\e)$ are
derived in equations~(\ref{eq:IC_l_norm}) and (\ref{eq:IC_f_norm}),
respectively.  The KN suppression of the IC spectrum is captured by
$f_\mathrm{KN}(E_\IC,\alpha_\e)$ (equation~\ref{eq:KN_fit}), and the shape of
the transition region from the Thompson to the KN regime is given by
$\delta_{\rmn{IC}}(E_\IC, \alpha_\e)$
(equation~\ref{eq:fit_trans_KN}). Following recent work that carefully models
the non-thermal radiation of young supernova remnants
\citep{2010ApJ...708..965Z}, we typically adopt a maximum electron injection
efficiency of $\zeta_\rmn{e,max}=0.01$. We note that this value seems to be at
the upper envelope of theoretically allowed values that match the supernova
data.

\subsection{Multiphase structure of the ICM}
\label{sect:galaxies}
The ICM is a multiphase medium consisting of a hot phase which attained its
entropy through structure formation shock waves dissipating gravitational energy
associated with hierarchical clustering into thermal energy. The dense, cold
phase consists of the true interstellar medium (ISM) within galaxies and at the
cluster center as well as the ram-pressure stripped ISM that has not yet
dissociated into the ICM \citep{2009MNRAS.399..497D}.  All of these phases
contribute to the $\gamma$-ray emission from a cluster.  Physically, the
stripped ISM should dissociate after a time scale that depends on many unknowns
such as details of magnetic draping of ICM fields \citep{2008ApJ...677..993D,
  2009arXiv0911.2476P} on galaxies or the viscosity of the ICM. In SPH
simulations, this dissociation process is suppressed or happens only
incompletely in our simulations leaving compact galactic-sized point sources
that potentially biases the total $\gamma$-ray luminosity high.  On the other
hand, once these stripped compact point sources dissociate, the CRs diffuse out
in the bulk of the ICM, and produce $\gamma$-rays by interacting with protons of
the hot dilute phase. This flux, however, is negligible since
\begin{eqnarray}
  \label{eq:diff-CRs}
 L_\rmn{diff.-CRs} 
&\propto& \int\,\mathrm{dV} \,  n_\rmn{diff.-CRs}\, n_\rmn{icm}
\propto \int\, \mathrm{dV} \, n_\rmn{ism-CRs}\,   n_\rmn{ism} \, 
\left(\frac{n_\rmn{icm}}{n_\rmn{ism}}\right)^2\nonumber\\
&\sim& L_\rmn{icm-CRs}\, \left(\frac{n_\rmn{icm}}{n_\rmn{ism}}\right)^2,
\quad\mbox{where}~\frac{n_\rmn{icm}}{n_\rmn{ism}} \sim 10^{-3}.
\end{eqnarray}
Here $n_\rmn{icm}$ ($n_\rmn{ism}$) denotes the gas number density in the ICM
(ISM), $n_\rmn{ism-CRS}$ the CR number density in the ISM, and
$n_\rmn{diff.-CRS}$ the CR number density of the CRs that diffused out of their
dense ISM environment into the ambient ICM that is in pressure equilibrium with
the ISM. In the second step of equation~(\ref{eq:diff-CRs}) we accounted for the
adiabatic expansion that these CRs would experience as they diffused out. In the
last step we assumed $L_\rmn{ism-CRs} \sim L_\rmn{icm-CRs}$, i.e.{\ }that the CR
luminosity of all compact galactic sources in a cluster is of the same order as
the CR luminosity in the diffuse ICM; a property that is at least approximately
true in our simulations as we will show later on.  Leaving all gaseous point
sources would definitively be too optimistic, removing all of them would be too
conservative since cluster spiral galaxies should contribute to the total
$\gamma$-ray emission (which defines our so-called {\em optimistic} and {\em
  conservative} models). Hence, we perform our analysis with both limiting
cases, bracketing the realistic case. The effect from the gaseous point sources
is largest in low mass clusters, where they constitute a few percent of the
total ICM mass. For high-mass clusters this fraction is lower, and constitutes
only about one percent. In practice, we cut multiphase particles with an
electron fraction $x_\e < 1.153$ and a gas density above the star forming
threshold $2.8\times10^{-25}\,\rmn{g}\,\rmn{cm}^{-3}$. If nothing else is
stated, we use our conservative model without galaxies throughout the paper.

\section{Characteristics of $\gamma$-ray emission}
\label{sect:morphology}
From surface brightness $S_\gamma$ maps that are obtained by line-of-sight
integration of the source functions, we study the $\gamma$-ray emission to
characterize the morphology of clusters. Additionally, we use emission profiles
to compare pIC, sIC, and pion decay induced emission for different clusters.

\begin{figure*}
\begin{minipage}{2.0\columnwidth}
  \includegraphics[width=0.5\columnwidth]{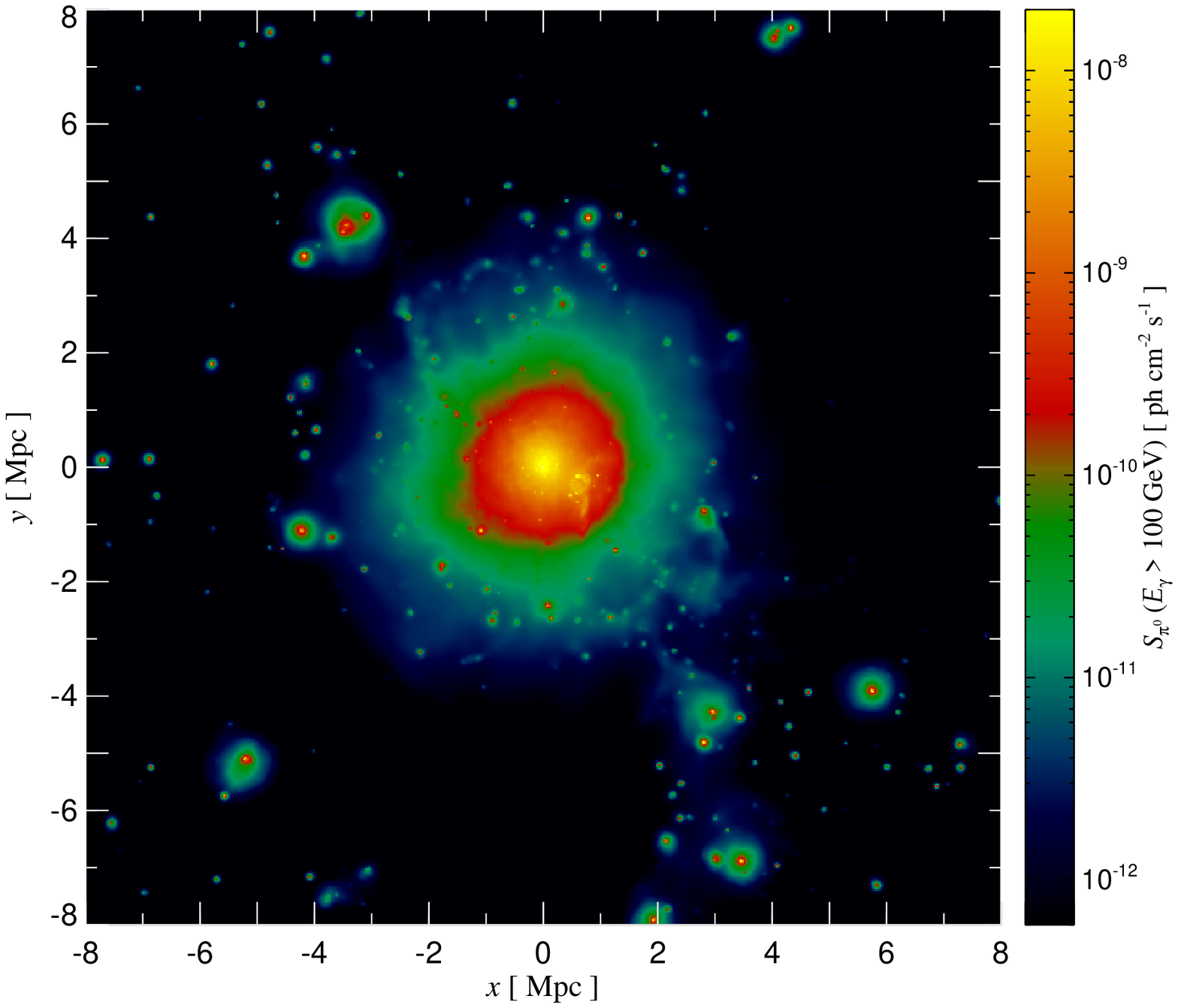}
  \includegraphics[width=0.5\columnwidth]{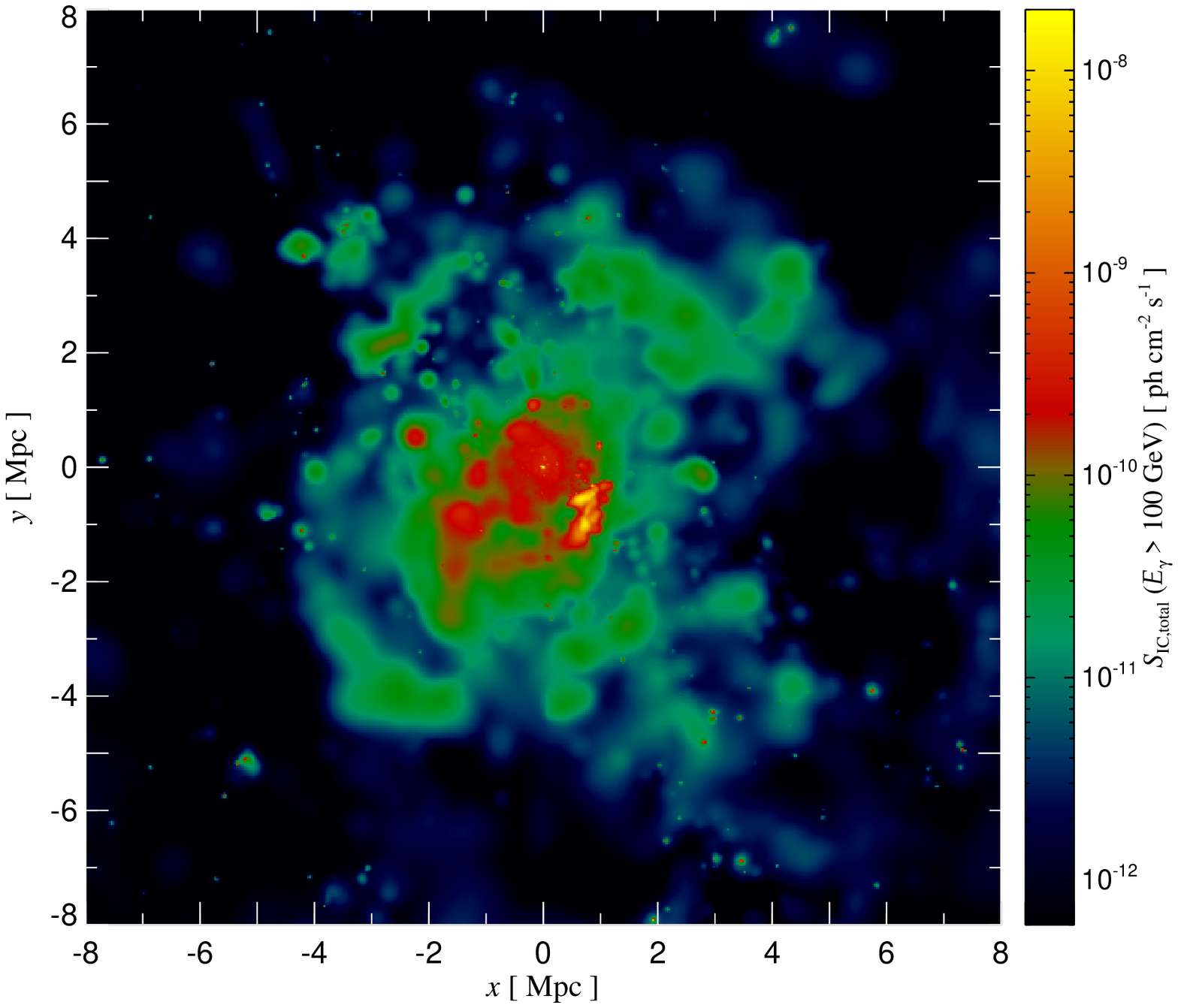}
  \caption{The $\gamma$-ray emission above $100 \, \gev$ of our Coma-like
    cluster g72a is shown.  We show the pion decay $\gamma$-ray emission that
    originates from hadronic CR interactions with ambient gas protons on the
    left.  On the right, we plot the inverse Compton emission from both, primary
    and secondary CR electrons.  Comparing the different $\gamma$-ray emission
    components, we note that the pion decay has a very regular morphology and
    clearly dominates the cluster region. In contrast, the emission from primary
    electrons shows an irregular morphology that traces the structure formation
    shock waves and dominates in the virial periphery and the warm-hot
    intergalactic medium.}
  \label{fig:emission_components_maps}
\end{minipage}
\end{figure*}
\begin{figure*}
\begin{minipage}{2.0\columnwidth}
  \includegraphics[width=0.5\columnwidth]{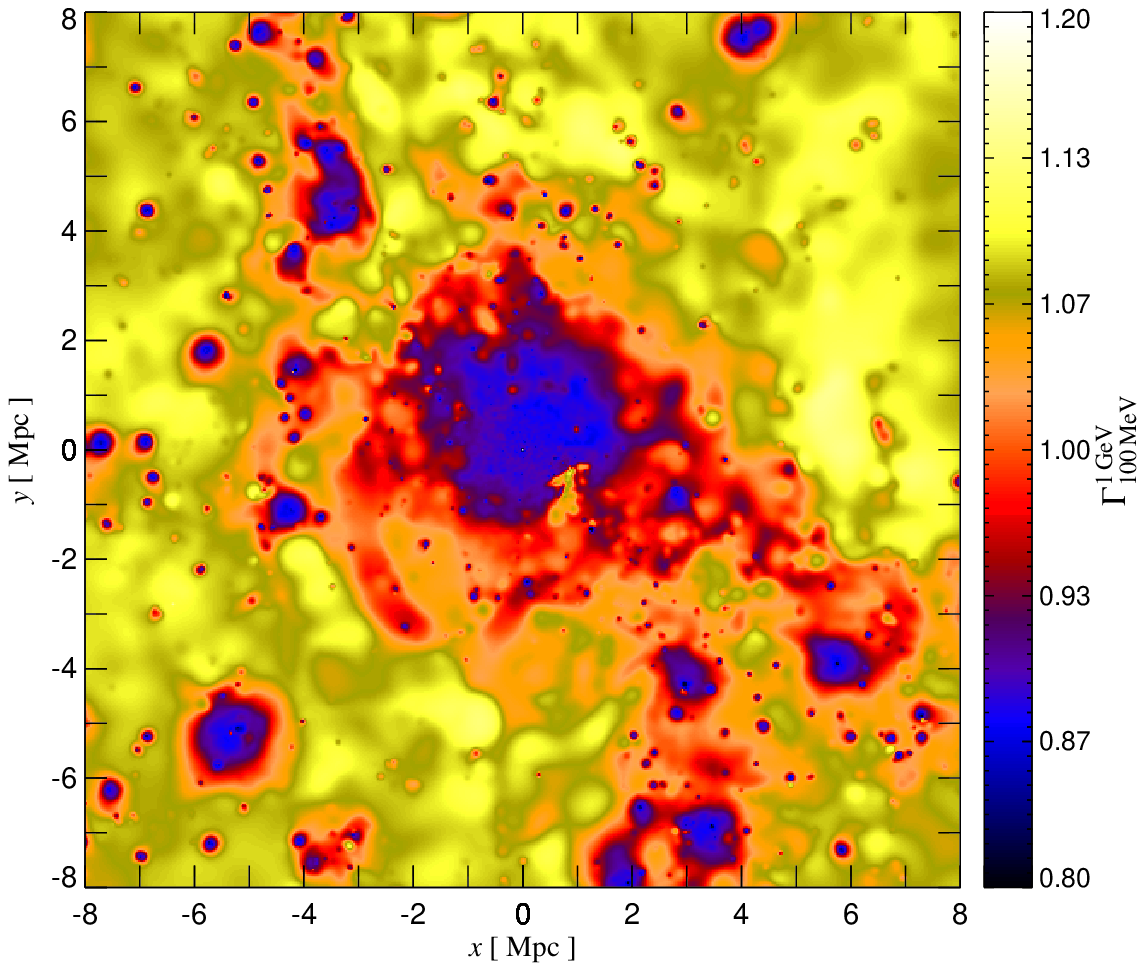}
  \includegraphics[width=0.5\columnwidth]{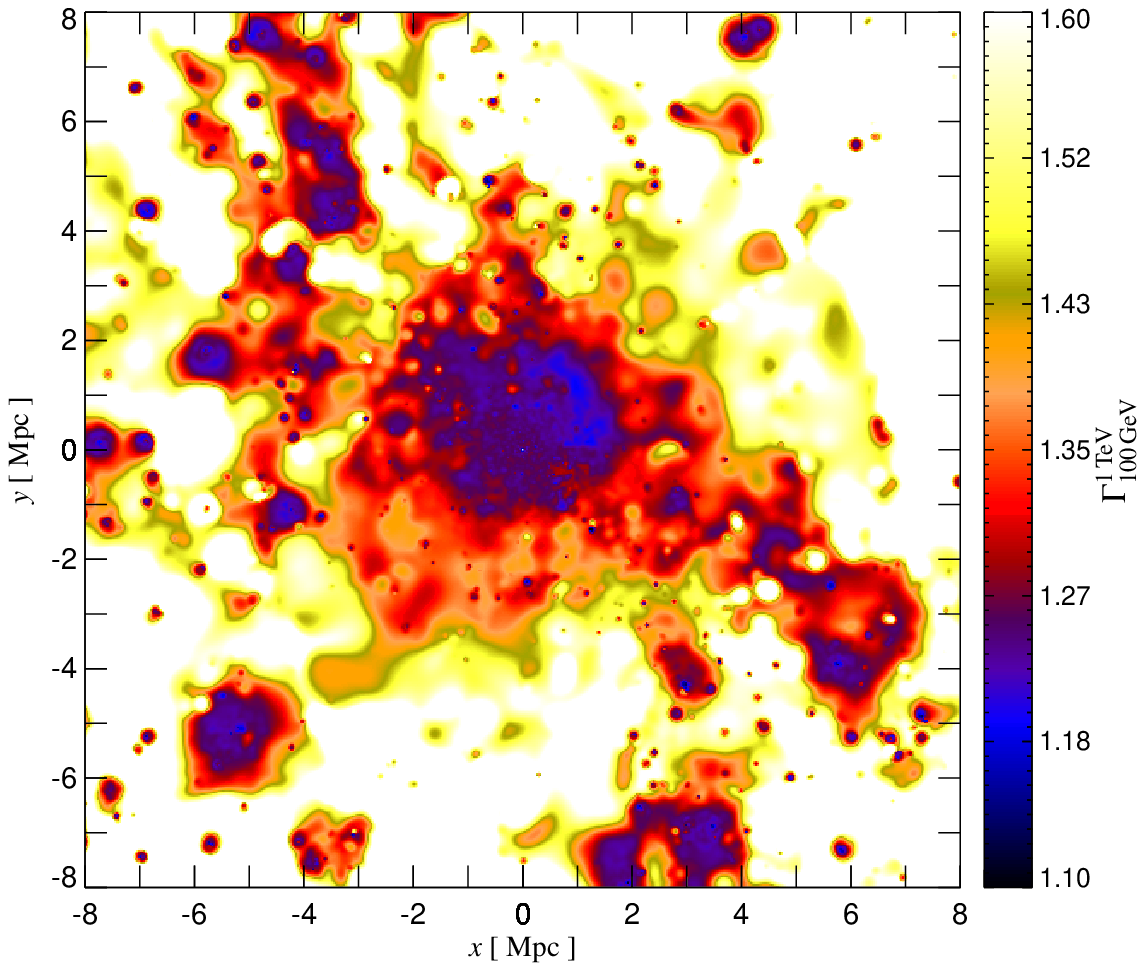}
  \caption{Projected energy dependent photon index for the galaxy cluster
    g72a. On the left, we show the photon index between $100 \, \mev - 1 \,
    \gev$ and on the right, the photon index between $100 \, \gev - 1 \,
    \tev$. Note that the central part (where pion decay dominates) shows little
    variations which reflects the spectral regularity of the CR distribution. At
    the periphery and beyond, there are larger fluctuations due to the
    inhomogeneity of the pIC emission.}
\label{fig:alpha_maps}
\end{minipage}
\end{figure*}

\subsection{Morphology of $\gamma$-ray emission}
The left side of Fig.~\ref{fig:emission_components_maps} shows the morphology of
the $\gamma$-ray emission above $100 \, \gev$ that results from hadronic CR
interactions with ambient gas protons. The right side shows the primary and
secondary IC emission for the post-merger cluster g72a. The comparison of the
two panels shows that the central parts are dominated by the pion induced
$\gamma$-ray emission. It has a very regular morphology that traces the gas
distribution. There is a transition to the pIC emission as the dominant emission
mechanism outside the cluster in the WHIM at a level depending on the dynamical
state of the cluster.  The pIC emission is very inhomogeneous which can be
easily understood since it derives from primary CR electrons that are directly
accelerated at structure formation shocks. Structure formation is not a steady
process, it rather occurs intermittently.  The morphology of the $\gamma$-ray
emission above $100 \, \mev$ for g72a was investigated in
\citet{2008MNRAS.385.1211P}. It shows a very similar morphology, indicating a
similar power-law CR spectrum.

\label{sec:emission_profiles}
\begin{figure*}
\begin{minipage}{2.0\columnwidth}
  \includegraphics[width=0.5\columnwidth]{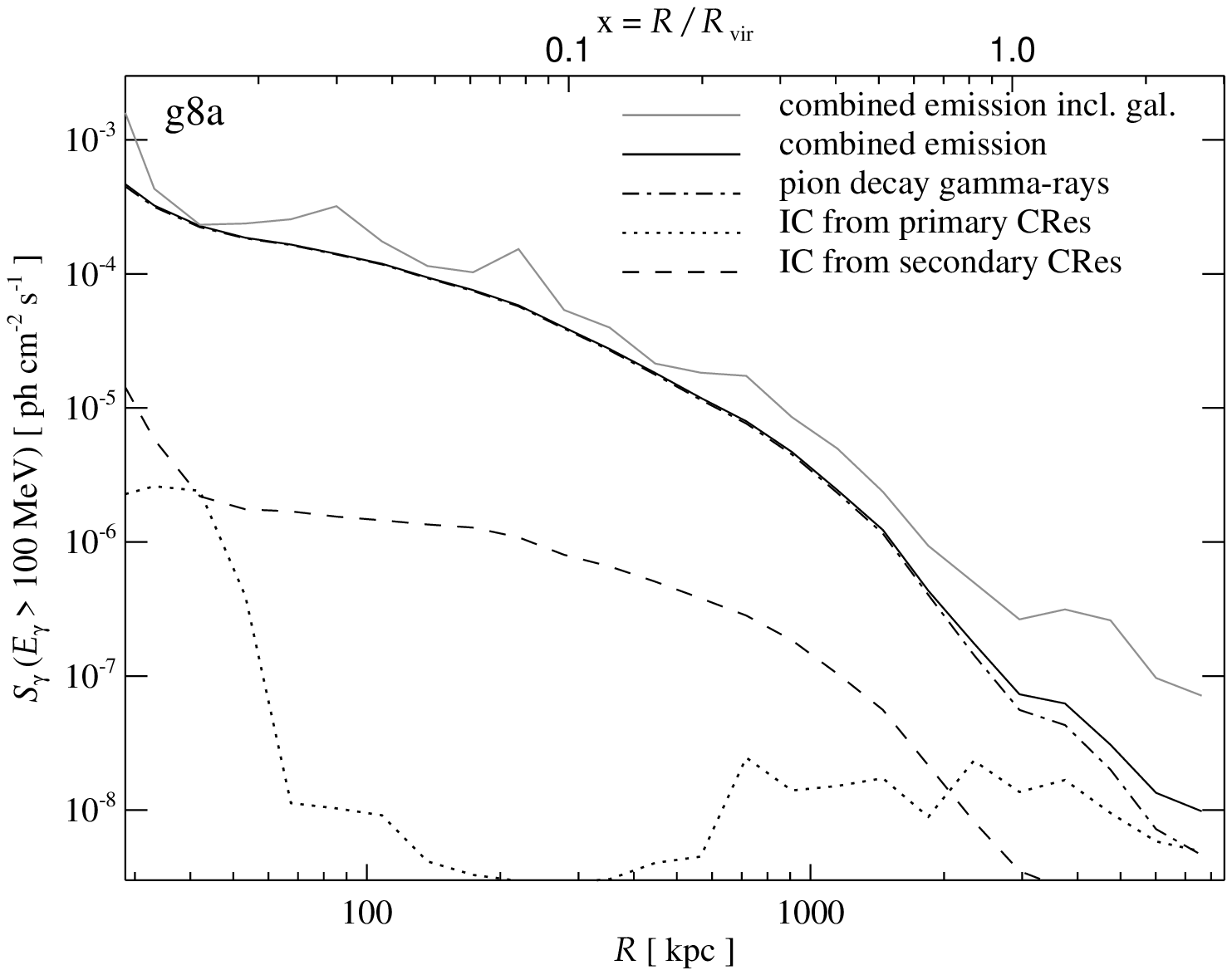}  
  \includegraphics[width=0.5\columnwidth]{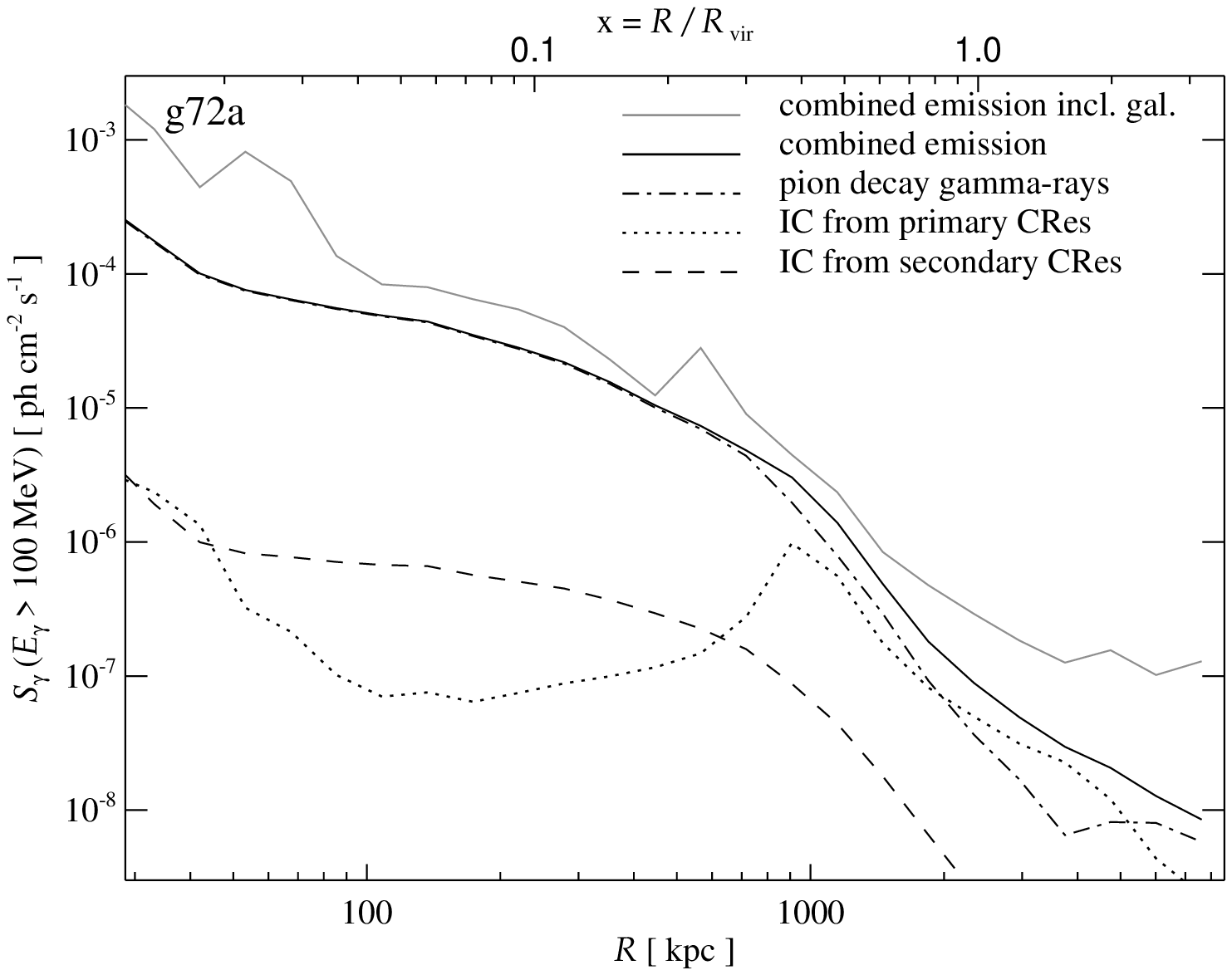}
  \caption{Azimuthally averaged profiles of the $\gamma$-ray emission components
    above $100 \, \mev$ for two different clusters: large cool core (CC) g8a
    (left panel) and large merger g72a (right panel). Shown are the dominating
    pion decay induced emission (dash-dotted lines), the IC emission of primary
    CR electrons accelerated directly at structure formation shocks (dotted
    lines), the secondary IC electrons resulting from CRp-p interactions (dashed
    lines) and the sum of all emission components (thick line). We cut the
    galaxy emission from our individual emission components in both panels and
    additionally show the total emission including galaxies (grey solid
    line).  \label{fig:emission_components_profiles}}
\end{minipage}
\end{figure*}

The projected energy dependent photon index is a well defined continuous
quantity as it is defined through
\begin{eqnarray}
\label{eq:photon_spectral_index}
\Gamma \equiv -\frac{\dd \mathrm{log} S_\gamma}{\dd \mathrm{log} E_\gamma}\qquad
\longrightarrow\qquad \Gamma_{E_1}^{E_2}(\vrp) =
-\frac{\mathrm{log}\left[\frac{S_\gamma(E_1,\vrp)}{S_\gamma(E_2,\vrp)}\right]}
{\mathrm{log}\left(\frac{E_1}{E_2}\right)},
\end{eqnarray}
where $S_\gamma(E_\gamma,\vrp)$ is the surface brightness of $\gamma$-rays with
energies $>E_\gamma$ at the projected radius $\vrp$. Using $S_\gamma$ maps, such
as the ones in Fig.~\ref{fig:emission_components_maps}, we can extract the
photon index $\Gamma_{E_1}^{E_2}$ between the energies $E_1$ to $E_2$. Close to
the pion bump (see Fig.~\ref{fig:sketch_CR_gamma}) at $m_{\pi^0}/2 \simeq 68.5
\, \mev$ (energy of a photon originating from a decaying $\pi^0$ at the
threshold of the hadronic p-p reaction) the photon spectrum has a convex
shape. This is characterized by a flatter photon index compared to the
asymptotic limit, $E_\gamma\gg m_{\pi^0} c^2$.  Calculating $\Gamma_{E_1}^{E_2}$
at two discrete energies results in a slightly steeper value for
$\Gamma_{E_1}^{E_2}$ than its continuous counterpart $\Gamma$ at $E_1$.

Since we are interested in comparing the morphology of clusters to spectra, we
calculate the projected photon index for g72a (see Fig.~\ref{fig:alpha_maps}).
We compare $\Gamma_{E_1}^{E_2}$ for an energy regime accessible to the {\em
  Fermi space telescope} of $100 \, \mev - 1 \, \gev$ (left panel) to $100 \,
\gev - 1 \, \tev$ accessible for IACTs (right panel). In the {\em Fermi regime},
we find a median value of $\Gamma_{100\,\mev}^{1\,\gev}\simeq 0.9$ in the
central regions of the cluster and a value $\Gamma_{100\,\mev}^{1\,\gev}\simeq
1.1$ in the periphery.  The reason is that the total emission in the central
regions of the cluster is dominated by the pion decay emission at 100~MeV with a
lower spectral index than pIC due to the pion bump.  In the periphery and the
WHIM, where the pIC contributes substantially to the total emission,
intermediate shocks with $\mathcal{M} \sim 4$ are typical
\citep{2006MNRAS.367..113P}. Using the spectral index of the electron
equilibrium spectrum, $\alpha_\e = \alpha_\inj + 1$, where $\alpha_\inj \simeq
2.2$ for $\mathcal{M}\simeq 4$ and $S_\IC \sim E_\IC^{-(\alpha_\e-1)/2} \simeq
E_\IC^{-1.1}$, results in the observed steepening of
$\Gamma_{100\,\mev}^{1\,\gev}$ in the periphery.

We now turn to the energy region important for IACTs with the photon index
$\Gamma_{100\,\gev}^{1\,\tev}$. In the central regions the photon index traces
the proton spectral index $\Gamma_{100\,\gev}^{1\,\tev}=\alpha-1 \sim 1.25$
since this spatial region is dominated by the asymptotic regime of the pion
emission. Moving towards the periphery, the photon index steepens to
$\Gamma_{100\,\gev}^{1\,\tev}>1.4$, despite the presence of strong external
shocks as well as accretion shocks that efficiently accelerate electrons on
these large scales. The reason for this steepening is the exponential cutoff in
the pIC emission. Increasing the energy or the distance from the cluster results
in an even steeper photon index.

\subsection{Emission profiles}

The profiles of different non-thermal $\gamma$-ray emission processes without
galaxies are shown in Fig.~\ref{fig:emission_components_profiles} for, a large
CC cluster (g8a, left) and large post-merging cluster (g72a, right). The {\em
  secondary IC emission} traces the dominating {\em pion decay emission} because
to zeroth order, both components depend on $n_\mathrm{gas} n_\CR$, where
$n_\mathrm{gas}$ is the gas number density and $n_\CR$ the CR number
density. This would be exactly true if the magnetic field was smaller than
$B_\rmn{CMB} \simeq 3 \umu$G. In this case, the CRe population would exclusively
cool by means of IC emission. Since we assume the central magnetic field to be
larger than $B_\rmn{CMB}$, a fraction of the CRe energy is radiated through
synchrotron emission into the radio, causing a larger discrepancy of the sIC
emission compared to the pion emission in the center.

In contrast to the centrally peaked secondary emission components, the average
{\em primary IC emission} shows a rather flat surface brightness profile which
can be nicely seen in our cool core cluster g8a. This is because we see the
strong accretion shocks that efficiently accelerate CRes (in terms of the
injected energy density) in projection.  There are noticeable exceptions in the
centers of both clusters: accreting small sub-clumps dissipate their
gravitational energy through weak shocks in the larger core regions of
clusters. However, once these weak shock waves encounter the (over-cooled)
centers of our simulated clusters they transform into strong (high Mach number)
shock waves. These inject a hard population of primary CR electrons which causes
the centrally peaked and bright pIC emission. We also observe an excess pIC
emission at a radius of $\sim 1$~Mpc in g72a. This can be traced back to a
prominent merger shock wave with a Mach number up to $\M\simeq 3.5$ that
accelerates primary CRes (see also Fig.~\ref{fig:emission_components_maps}).

The {\em total emission} flattens out in the cluster exterior close to $\rvir$ because
of two reasons: (1) there the pIC emission contributes significantly to the total
emission because of strong merger shocks as well as accretion shocks, and (2)
subhalos in the periphery that have not yet merged with the larger halo
contribute to the pion decay induced emission. In this regime, the halo-halo
correlation term starts to dominate the average density profile of a cluster
with its characteristic flattening \citep{2008MNRAS.388....2H}. This naturally
translates to the pion emission profile that tightly correlates with the gas
density distribution.

\begin{figure*}
\begin{minipage}{2.0\columnwidth}
 \includegraphics[width=0.5\columnwidth]{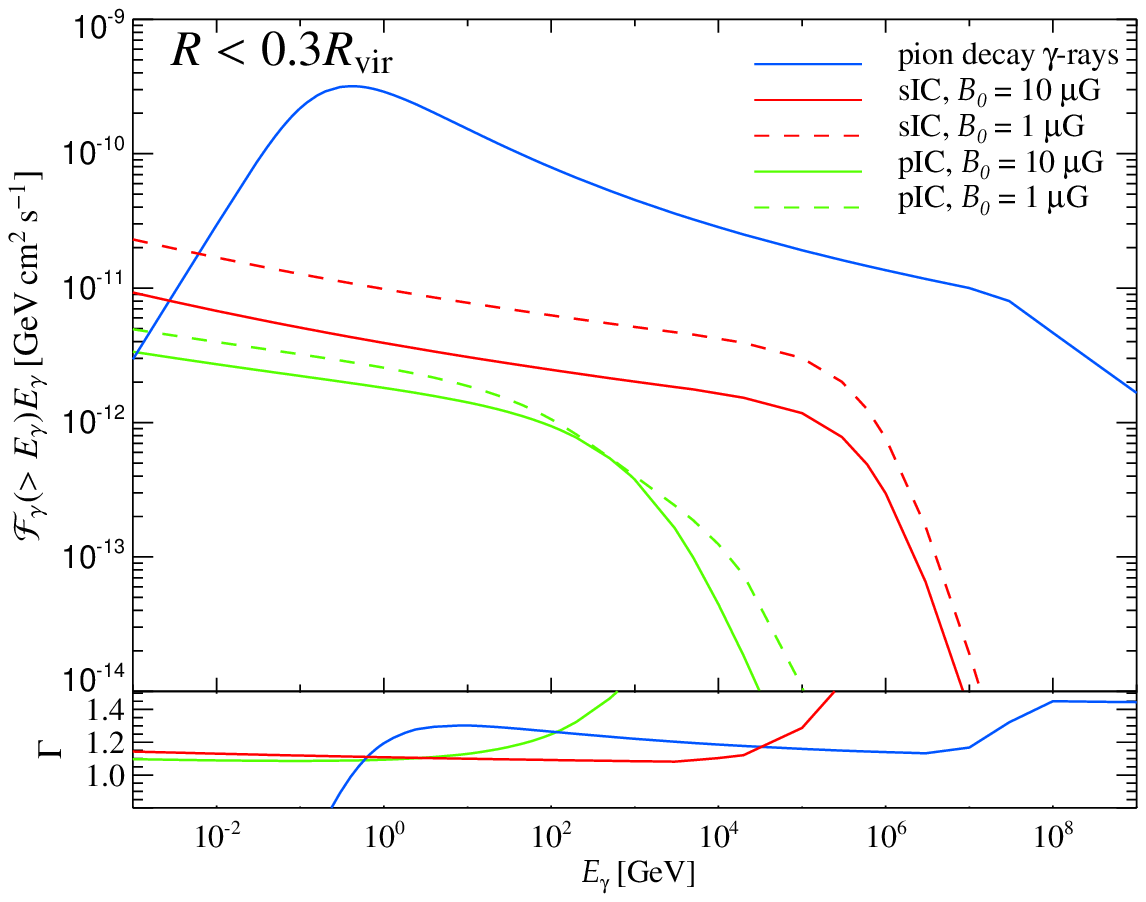}
 \includegraphics[width=0.5\columnwidth]{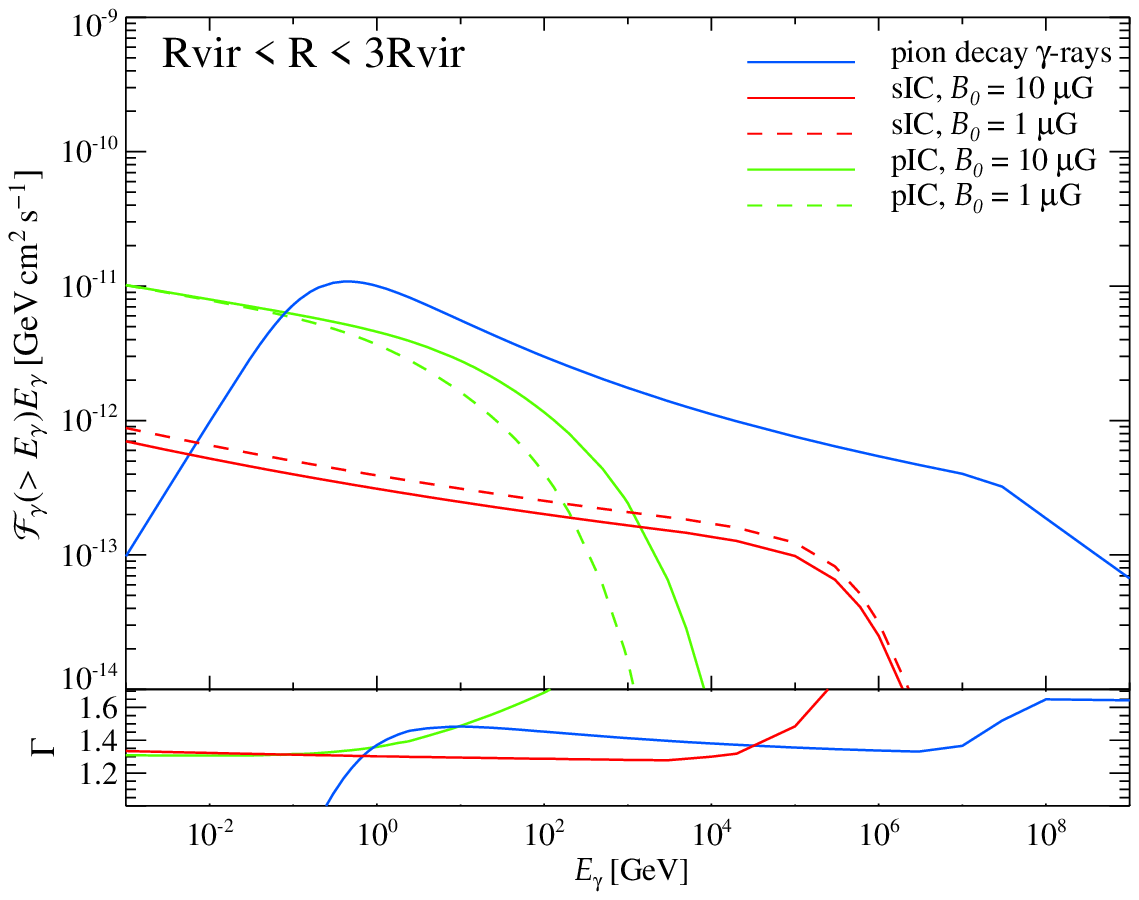}
 \caption{The $\gamma$-ray number flux weighted by photon energy of our
   Coma-like cluster g72a: core region (left panel) and WHIM region (right
   panel). The pion decay flux is shown in blue color, the secondary IC in red,
   and the primary IC emission in green. The solid IC lines assume a central
   magnetic field strength of $B_0=10\,\mug$ while the dashed lines assume
   $B_0=1\,\mug$. The lower panels show the photon spectral index $\Gamma$ for
   each emission component, where the colors are the same as in the upper
   panel. Note that in the core region ($R < 0.3\, R_\rmn{vir}$) the pion decay
   induced flux dominates the emission. In the WHIM region ($R_\rmn{vir} < R <
   3\,\rvir$) the pion decay and sIC emission components are much smaller
   compared to the central part. Contrary, the pIC component is boosted in the
   WHIM due to accretion shocks onto our simulated cluster. Nevertheless, its
   emission falls slightly short of the pion decay.}
\label{fig:flux_core_whim}
\end{minipage}
\end{figure*}

The {\em ratio between pion decay and sIC emission} can be estimated
analytically by calorimetric considerations.  To this end we compare the CR
energy spectrum, $E^2\,f(E)$, at the respective CR energies which give rise to
$\gamma$-ray emission at some specific photon energy, say
$E_\gamma=1$~GeV. Additionally we have to include a factor, $f_B$, that accounts
for the possibility that the CRes do not only emit IC $\gamma$-rays but also
radio synchrotron radiation.  Using the hadronic branching ratios for the
production of pions, $R_{\pi^0}\simeq 1/3$ and $R_{\pi^\pm}\simeq2/3$, as well
as the multiplicities for the decay products in the respective decay channels,
$\xi_{\pi^0 \to \gamma}\simeq 2$ and $\xi_{\pi^\pm \to e^\pm}\simeq1/4$, we
obtain
\begin{eqnarray}
\label{eq:pion_sIC_ratio}
\left.\frac{E_\rmn{p,\pi^0}^2\,s_{\pi^0}^{}}
{E_\rmn{p,sIC}^2\,s_\rmn{sIC}^{}}\right|_{1\,\gev} 
\simeq 
 \frac{R_{\pi^0}\,\xi_{\pi^0 \to \gamma}}{R_{\pi^\pm}\,\xi_{\pi^\pm \to e^\pm}}
\left(\frac{E_\rmn{p,\pi^0}}{E_\rmn{p,sIC}}\right)^{-\alpha_{\pi^0}^\rmn{sIC}+2}f_\rmn{B} 
 \simeq 30\,f_\rmn{B}.
\end{eqnarray}
Here $\alpha_{\pi^0}^\rmn{sIC} \simeq 2.3$ is the CR proton spectral index
between the CR energy $E_\rmn{p,\pi^0}=8.0\,\gev$ that give rise to pion decay
flux at 1 GeV and the CR energy $E_\rmn{p,sIC}=8.0\,\tev$ that gives rise to sIC
flux at 1 GeV. The $\gamma$-ray source function for pion decay and sIC 
(\citep{2004A&A...413...17P}) is denoted by $s_{\pi^0}$ and $s_\IC$, respectively. 
Finally, the factor $f_B=(B/B_\rmn{CMB})^2+1\simeq 1$, for
magnetic fields much smaller than the CMB equivalent magnetic field, otherwise
$f_\rmn{B}>1$. In the region close to $0.3\,\rvir$, the magnetic field is about
2.4~$\mug$ in our model (Table~\ref{tab:diffusion}), which implies an emission
ratio of about 50. For smaller photon energies than 1~GeV, the pion decay
$\gamma$-ray emission falls below the asymptotic power-law due to the
characteristic pion bump. This effect implies a lower ratio of about 40 (instead
of the expected ratio of about 100) for the emission above 100~MeV in
Fig.~\ref{fig:emission_components_profiles}.

We also show the {\em total surface brightness profile with galaxies} in
Fig.~\ref{fig:emission_components_profiles} (grey line).  The resulting profile
shows a boosted emission by about a factor two compared to the one where we
exclude galaxies from the surface brightness. The entire population of these
galaxies takes up only a negligible volume so that the volume weighted CR
pressure is almost the same in either case, when taking these galaxies into
account or not. We note that this bias needs to be addressed when deriving
average CR pressure contributions from the cluster's $\gamma$-ray emission
\citep{2009arXiv0909.3267T}.  Especially in the inner parts, the profile is very
inhomogeneous. Since galaxies follow an approximate Poisson distribution and
since the inner radial bins of the profile sample only few galaxies, we
naturally obtain a larger Poissonian scatter across the inner radial bins.

\subsection{Emission spectra from the cluster core and WHIM}

The central parts of clusters are characterized by high gas and CR densities,
and magnetic fields -- at least compared to average values of the ICM. Even
though the cluster core region only makes up a fraction of the total volume of
the ICM, the high densities result in a significant $\gamma$-ray flux
contribution to the total flux from the cluster. In contrast to the cluster
center, the WHIM is characterized by on average low gas and CR densities, and
magnetic fields. The low densities cause a smaller total $\gamma$-ray flux from
this region compared to the cluster core regions. However, the WHIM of the
super-cluster region contains a large number of galaxies and groups that are
accreted onto the cluster. This generates more shocks compared to the cluster
core region. The cluster characteristics in the two regions give rise to
different normalizations of the individual $\gamma$-ray emission components, but
with a similar shape. The shape of the emission components from the different
regions agrees with that of the entire cluster as shown in
Fig.~\ref{fig:sketch_CR_gamma}. The emission can be summarized as follows. The
$\pi^0$-decay is characterized by the so-called pion bump followed by a concave
curvature and a diffusive break. The pIC emission component shows a power-law
followed by an exponential cutoff while the sIC component has a power-law with
similar index that is however followed by the Klein-Nishina break.

In Fig.~\ref{fig:flux_core_whim} we show the $\gamma$-ray number flux weighted
by photon energy from different regions of our Coma-like cluster g72a that we
place at the distance of 100~Mpc. The left figure shows the $\gamma$-ray number
flux within the core region ($R < 0.3\,\rvir$), where the {\em $\pi^0$-decay
  dominates over the sIC component that itself is sub-dominant to the pIC
  component}. The surface brightness profile of the $\pi^0$-decay is
sufficiently flat in the core region so that the $\gamma$-ray flux is dominated
by the outer scale around $R \sim 0.3\,\rvir$ where most of the volume is. Hence
the $\pi^0$-decay flux is largely insensitive to numerical inaccuracies of our
modeling of the physics at the very center of the cluster. Both the {\em pIC and
  sIC emission components} have a larger $\gamma$-ray flux in our models with
weak central magnetic fields ($B_0=1\,\mug$, $f_\rmn{B}\simeq1$) compared to our
models with strong central magnetic fields ($B_0=10\,\mug$, $f_\rmn{B}\simeq2$).
The sIC emission with $B_0=10\,\mug$ is characterized by $B \simeq B_\rmn{CMB}$
on scales around the core radius which is the region that dominate the
flux. Relative to the pion decay emission, the sIC is suppressed by a factor of
$\sim 60$ which can be understood by considering hadronic decay physics and the
fact that the CR energy spectrum, $E^2 f(E)$, is decreasing as a function of
proton energy (see equation~\ref{eq:pion_sIC_ratio}). Even though the pIC
emission component is sub-dominant, it shows a rather flat spectral index. This
implies only a few strong shocks that are responsible for the electron
acceleration. These merging shock waves are traversing the cooling core region
in the cluster center. We caution the reader that the over-cooling of the
cluster centers in our simulations possibly overestimates the true shock
strengths and numbers which also results in an artificially enhanced pIC
emission. At high energies, the electron cooling time is smaller than the time
scale for diffusive shock acceleration which causes an exponential cutoff in the
electron spectrum which is passed on to the pIC spectrum. The energy scale of
the cutoff $E_\rmn{IC, cut}$ (combining equations~\ref{eq:ICenergy} and
\ref{eq:Ecut}) scales with the magnetic field which causes the low magnetic
field model to turn down faster than the large magnetic field model. Note that
the second cutoff in the figure for the small central magnetic fields is caused
by a small fraction of the particles that have unusually high electron energy
cutoff. The lower panel shows the photon spectral index defined in
equation~(\ref{eq:photon_spectral_index}), where $\Gamma \simeq 1.3$ for the
$\pi^0$-decay emission after the pion bump that slowly flattens out with energy.
For both the pIC and sIC emission the photon spectral index $\Gamma \simeq 1.1$
above the MeV regime up to at about 100 GeV. The reason for the flat spectra
is that the pIC emission is generated by a few strong shocks in the center,
while the sIC emission is caused by protons in the flat high energy part of CR
spectrum.

Now we turn to the {\em $\gamma$-ray spectra in the WHIM} which are shown in the
right panel of Fig.~\ref{fig:flux_core_whim} for our g72a cluster. Here we
define the WHIM by the emission in the region $\rvir < R < 3\,\rvir$ as seen in
2D projection of the cluster.  We see that the pion decay and sIC are suppressed
by a factor that is larger than 10 compared to the flux within the core region
since the emission correlates with the CR- and gas densities. The suppression of
$\gamma$-ray flux emitted by the intergalactic medium is expected to be much
greater due to the large density decrease. The presence of small groups in the
super-cluster region with densities that are much larger than the average
density in the WHIM partially counteracts the flux suppression.  Contrary to the
pion decay and sIC component, the pIC emission is boosted by a factor of a few
compared to the center because of the larger spatial region in the WHIM that
contains a greater number of shocks. This leads to {\em comparable flux from the
  pIC and pion decay emission above the energy of the pion bump in the
  super-cluster region}. Note however, that this flux is still sub-dominant
compared to the pion decay flux emitted by the cluster core region.  The
different central magnetic fields do not play any significant role in the
power-law regime in the WHIM since $B\ll B_\rmn{CMB}$, which implies that the
CRes mainly cool through IC emission. However, note that the pIC cutoff is
shifted towards lower energies for these smaller magnetic fields since
$E_\rmn{IC,max}\propto B/(B^2+B_\rmn{CMB}^2)\propto B/B_\rmn{CMB}^2$ (as derived
by combining equations~\ref{eq:ICenergy} and \ref{eq:Emax}). In the lower panel
we show that the photon spectral index is steeper in the WHIM for all three
emission components compared to the core region. The photon index is about 1.3
for both pIC and sIC below 10~GeV and about 1.4 for the pion decay above the
energy of the pion bump. The reason for the steeper pIC is because most of the
energy that is injected into primary electrons comes from multiple
intermediate-strength shocks (accretion shocks), while in the cluster center the
pIC emission is build up from a few strong shocks in the over-cooled center
(merger shocks). The steeper sIC and pion decay photon indices are caused by the
slightly steeper CR spectrum present in the WHIM of the g72a cluster
(cf. Fig.~\ref{fig:proton_spectrum_radial}).

\section{The CR proton distribution}
\label{sect:CRp_spec}
In this section we investigate the CR proton spectrum that we obtain from our
simulations and discuss the relevant physics that gives rise to it.  We explore
the variance of the spectrum across our cluster sample and within individual
clusters and show that it obeys a universal spectral shape.  In addition, we
study the spatial profile of the CRs within a cluster as well as across our
cluster sample and find it to be approximately universal.  This universal
behavior enables us to construct a semi-analytic CR spectrum and to compute the
$\gamma$-ray spectrum as well as other secondary decay spectra (electrons,
neutrinos) semi-analytically.

\begin{figure}
\begin{minipage}{1.0\columnwidth}
 \includegraphics[width=1.0\columnwidth]{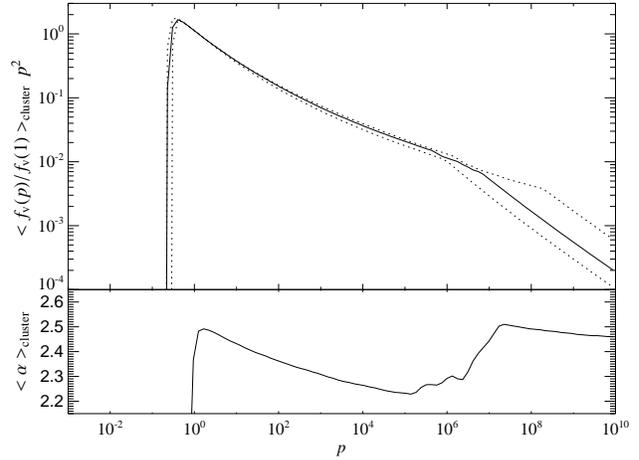}
 \caption{Spectral universality across our cluster sample: we show the median CR
   proton spectrum (solid line) as a function of dimensionless CR momentum
   $p=P_\p/\mpc$ for our sample of 14 galaxy clusters in
   Table~\ref{tab:cluster_sample} together with the 68 percentiles (dotted
   lines).  The CR spectrum of every cluster, $f_\vv(p)$, has been obtained by
   volume weighting the individual spectra of our SPH particles and normalizing
   them at $p = 1$.  Note the small variance between different clusters below
   the diffusive break, which indicates a universal CR spectrum for galaxy
   clusters. The large scatter in the $p=10^6-10^8$ momentum regime is a
   consequence of the strong radial dependence of the diffusive break. The lower
   panel shows the CR spectral index $\alpha = -\dd \log f / (\dd \log p)$ of the
   cluster sample.}
 \label{fig:proton_spectrum_all}
\end{minipage}
\end{figure}

\subsection{A universal CR spectrum}
In Fig.~\ref{fig:proton_spectrum_all} we show the median CR spectrum of all 14
galaxy clusters as well as the associated spectral index alpha. The CR spectrum
of every cluster, $f_\vv(p)$, has been obtained by volume weighting the
individual spectra of our SPH particles which have been normalized at the
dimensionless proton momentum $p =P_\p/\mpc = 1$.  Before discussing the
spectral shape we note that it shows a remarkably small variance across our
cluster sample which indicates a universal CR spectrum for galaxy clusters.
There are three important features in the spectra.

\begin{enumerate}
\item The spectrum shows a {\em low-momentum cutoff} due to efficient Coulomb
  cooling at these low momenta with $p\lesssim 1$: the CR energy is transferred
  into the thermal energy reservoir through individual electron scatterings in
  the Coulomb field of the CR particle \citep{1972Phy....58..379G}.  The Coulomb
  time scale of a mono-energetic CR population is very short, $\tau_\rmn{Coul} =
  E_\CR/\dot{E}_\rmn{CR,Coul}\simeq 0.03\,\rmn{Gyr} \times (p/0.1)^3\,(n_\e/
  10^{-3}\rmn{cm}^{-3})^{-1}$, where we show a momentum scaling that is valid
  only for the relevant non-relativistic regime. The Coulomb time scale for a
  power-law population of CRs can be significantly longer, $\tau_\rmn{Coul} =
  \eps_\CR(C,q,\alpha)/ \dot{\eps}_\CR(C,q,\alpha)_\rmn{Coul}\simeq 1\,\rmn{Gyr}
  \times (n_\e/ 10^{-3}\rmn{cm}^{-3})^{-1}$, where we assumed a low-momentum
  cutoff $q=0.1$ and $\alpha=2.5$ \citep{2007A&A...473...41E}. This, however, is
  still short compared to the dynamical time scale $\tau_\rmn{dyn}\sim
  2\,\rmn{Gyr} \times(n_\e/ 10^{-3}\rmn{cm}^{-3})^{-1/2}$. Hence, we expect the
  formation of an equilibrium cutoff of our CR spectrum around $q\simeq 0.1
  \ldots 1$ for the typical number densities $n_\e\sim (10^{-3} \ldots
  10^{-2})\,\rmn{cm}^{-3}$ that we encounter at the cluster cores. Note that in
  the presence of cooling processes only, the equilibrium cutoff $q$ is
  determined from the competition between Coulomb cooling and hadronic losses
  and converges around q = 1.685 if the cooling time is sufficiently short (see
  \citet{2008A&A...481...33J} for a detailed discussion). The reason for that is
  that Coulomb cooling shifts the cutoff to higher momenta, as the CRs with
  low momenta are transferred to the thermal reservoir. At high momenta, the
  cooling time due to hadronic interactions is shorter than the Coulomb cooling
  time.  Hadronic cooling effectively removes the CRs with high energy and moves
  the cutoff towards lower momenta.
\item In the momentum range between $p \simeq 1 - 10^6$, the spectrum has a
  {\em concave shape} in double-logarithmic representation, i.e. it is a
  decreasing function with energy in the GeV/TeV energy regime. This is
  quantified by the momentum dependent spectral index (shown in the lower panel
  of Fig.~\ref{fig:proton_spectrum_all}) which ranges from $\alpha\sim 2.5$ at
  energies above a GeV to $\alpha\sim 2.2$ around 100 TeV. This spectral shape
  is a consequence of the cosmological Mach number distribution that is mapped
  onto the CR spectrum \citep{2006MNRAS.367..113P}. This mapping depends on the
  shock acceleration efficiency as a function of shock strength as well as on
  the property of the transport of CRs into galaxy clusters. Nevertheless, we
  can easily understand the qualitative features: the typical shocks responsible
  for CR acceleration are stronger at higher redshift since they encounter the
  cold unshocked inter-galactic medium.\footnote{Note that after re-ionization,
    the UV background sets a temperature floor by photo-ionizing the
    intergalactic medium (IGM) to temperatures of about $T\sim 10^4$~K except
    for void regions where the densities are too small so that the
    photo-ionization heating rate is smaller than the expansion rate
    \citep{1996ApJS..105...19K}. As time progresses, formation shocks dissipate
    energy and heat the IGM to temperatures of $T\sim (10^5-10^7)$~K which is
    even heated further as it is accreted onto clusters. Since the sound-speed
    is proportional to $T^{0.5}$, we expect a higher sound-speed in
    (pre-)collapsed objects at low redshifts which results in weaker shocks at
    later times. This is quantified in the redshift evolution of the Mach number
    distribution \citep{2003ApJ...593..599R, 2006MNRAS.367..113P} that also
    confirms the argument given above.}  This implies the build-up of a hard CR
  population. Since the forming objects have been smaller in a hierarchically
  growing Universe, their gravity sources smaller accretion velocities which
  results in smaller shock velocities, $\vel$. Hence the energy flux through the
  shock surfaces, $\dot{E}_\rmn{diss}/R^2\propto \rho \vel^3$, that will be
  dissipated is much smaller than for shocks today. This causes a lower
  normalization of this hard CR population. With increasing cosmic time, more
  energy is dissipated in weaker shocks which results in a softer injection
  spectrum. Despite the lower acceleration efficiency, the normalization of the
  injected CR population is larger that that of the older flat CR population
  which yields an overall concave spectral curvature. We will study the details
  of the CR acceleration and transport that leads to that particular spectrum in
  a forthcoming paper (Pinzke \& Pfrommer, in prep.).
\item There is a {\em diffusive break} in the spectrum at high momenta where the
  spectral index steepens by 0.3. The CRs above these energies can diffusively
  escape from the cluster within a Hubble time. The particular value of the
  steepening assumes that the CRs scatter off Kolmogorov turbulence. Using twice
  the virial radius of each cluster, we find that the diffusive break varies
  between the momenta $p=10^6-10^8$, dependent on the characteristic size of a
  cluster (equation~\ref{eq:break}).
\end{enumerate}

\begin{figure}
\begin{minipage}{1.0\columnwidth}
  \includegraphics[width=1.0\columnwidth]{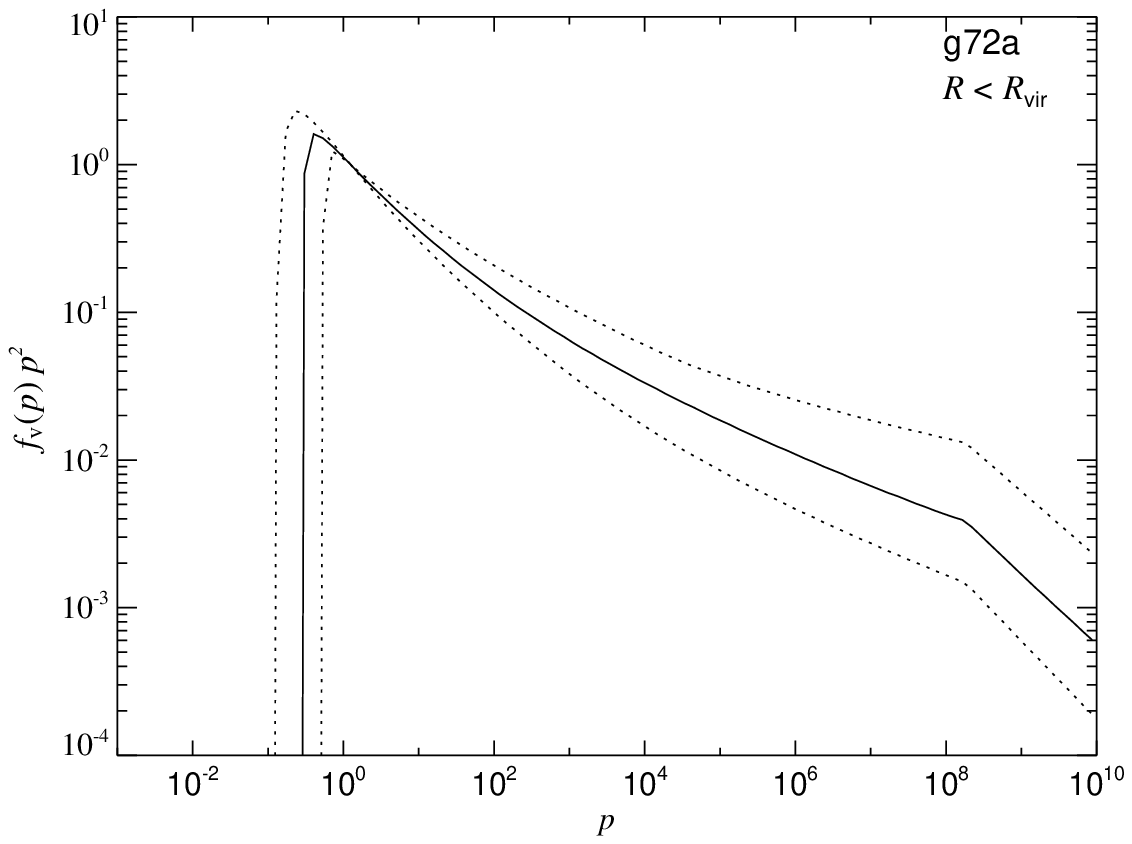}
  \includegraphics[width=1.0\columnwidth]{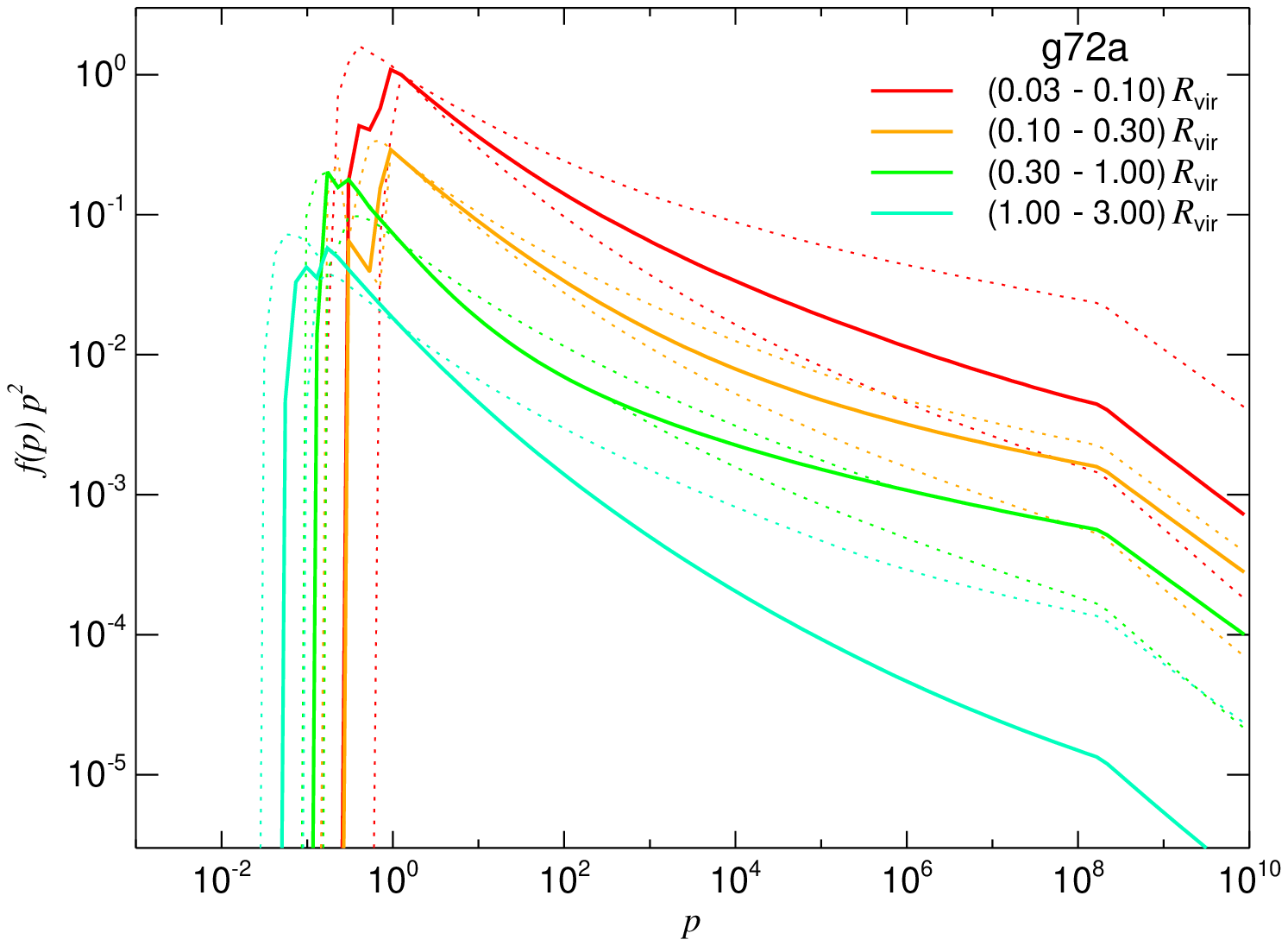}
  \caption{Spectral universality within a cluster -- variance and radial
    dependence: we show the CR proton spectrum of a large merging cluster, g72a, as a
    function of dimensionless CR momentum $p=P_\p/\mpc$. {\em Top:} median of the
    volume weighted and normalized CR spectrum $f_\vv(p)p^2$ (solid) and
    particle-by-particle variance as indicated by the 68 percentiles (dotted
    line).  {\em Bottom:} volume weighted CR proton spectrum (median and 68
    percentiles) for different radial bins. Those are represented by different
    colors: core region of the intra-cluster medium in red, the intermediate
    cluster scales in orange, the periphery of the ICM in green, and the WHIM in
    blue. The spectrum for the different radial bins has been offset vertically
    by a constant factor of four, with decreasing amplitude for increasing
    radius. The shift of the cutoff $q$ to a higher momentum for smaller radius
    is due to the smaller CR cooling time scales in denser regions. Note that
    the cosmic ray spectrum inside the virial radius is almost independent of
    radius. This behavior is not only observed for this cluster, but persistent
    across our cluster sample. \label{fig:proton_spectrum_radial}}
   \end{minipage}
\end{figure}

We now turn to the question on how universal is our CR spectrum within a
cluster. In Fig.~\ref{fig:proton_spectrum_radial} we find that the intrinsic
scatter of our CR spectra within a cluster is larger than the scatter among
clusters.  The reason being that formation shocks are intermittent as mass is
accreted in clumps an not continuously. Hence there is a high intrinsic variance
of the CR spectrum for similar fluid elements that end up in the same galaxy
cluster. However, averaging over all fluid elements that accrete onto a galaxy
cluster results in a very similar spectrum since different galaxy cluster
experience on average the same formation history.\footnote{The reason for our
  increased scatter at higher energies is that we normalize the spectrum of each
  SPH particle at $p=1$ where the softest CR populations dominate. Thus, at high
  energies the variance is mainly driven by the scatter in the hard CR
  populations, which have been accelerated in strong shocks at higher
  redshifts.} We note that the spectral variance of g72a is representative for
all the CR spectra in our sample.

We study the radial dependence of the CR spectrum for g72a at the bottom of
Fig.~\ref{fig:proton_spectrum_radial}.  Inside the cluster, the spectral shape
does not strongly depend on the radius. This is a crucial finding as it enables
us to {\em separate the spectral and the spatial part} of the CR
distribution. The level of particle-by-particle variance is similar to that of
the total cluster spectrum.  Also noticeable is the increasing low-momentum
cutoff $q$ as we approach the denser cluster center. This is due to enhanced
Coulomb losses and to a lesser extent increased adiabatic compression that the
CR distribution experienced when it was transported there. Outside the cluster
g72a, for radii $R>R_\rmn{vir}$, we observe a considerably steeper CR spectrum
at CR energies of $E\apgt 1$~TeV compared to the cluster region. We note that
this behavior is not universal and we observe a large scatter of the CR spectral
indices among our cluster sample at these large radii. This behavior might be
caused by the recent dynamical activity of the cluster under consideration but a
detailed characterization goes beyond the scope of this work and will be
postponed (Pinzke \& Pfrommer, in prep.).

\subsection{The spatial distribution of CRs}
\label{subsect:details_CRp_spec}

We have shown that the CR spectrum is separable in a spectral and spatial part.
To this end, we quantify the spatial part of the CR spectrum by taking the
volume weighted CR spectrum in each radial bin $i$ (see
Section~\ref{sect:formalism_analyic_model} in the appendix for derivation),
\begin{equation}
  \label{eq:f_vw}
  f_\vv(R_i) \equiv \bra f \ket_\vv(p=1,R_i)
  = C_\rmn{v}(R_i) = \tilde{C}_\rmn{M}(R_i)\,\frac{\rho_\vv(R_i)}{m_\p}.
\end{equation}
Here we assume $q < 1$, the subscripts M and V denote mass- and volume
weighted quantities, respectively, and we introduced the dimensionless
normalization of the CR spectrum,
\begin{equation}
\label{eq:tildeC}
\tilde{C}(\vr) =
m_\p\frac{C(\vr)}{\rho(\vr)}=
m_\p\left(\alpha-1\right)\,q^{\alpha-1}\frac{n_\CR(\vr)}{\rho(\vr)}\,,
\end{equation}
where $n_\CR$ is the CR number density. We now have to take into account
physical effects that shape the profile of $\tilde{C}$. Those include
acceleration of CRs at structure formation shocks, the subsequent adiabatic
transport of CRs during the formation of the halos as well as non-adiabatic CR
cooling processes; primarily hadronic interactions. At the same time we have to
consider the assembly of the thermal plasma and CRs within the framework of
structure formation that is dominated by CDM.  Hierarchical structure formation
predicts a difference in the halo formation time depending on the halo mass,
i.e.~smaller halos form earlier when the average mass density of the Universe
was higher. This leads to more concentrated density profiles for smaller halo
masses; an effect that breaks the scale invariance of the dark matter halo
profile \citep{2009ApJ...707..354Z}. The central core part is assembled in a
regime of fast accretion \citep{2006MNRAS.368.1931L}. This violent formation
epoch should have caused the CRs to be adiabatically compressed. In the further
evolution, some cluster have been able to develop a cool core which additionally
could have caused the CRs to be adiabatically contracted. On larger scales, the
gas distribution follows that of dark matter (at least in the absence of violent
merging events that could separate both components for time scales that are of
order of the dynamical time). On these scales, the CR number density roughly
traces the gas distribution \citep{2008MNRAS.385.1211P}. Overall, we expect the
spatial CR density profile relative to that of the gas density to scale with the
cluster mass. If non-adiabatic CR transport processes have a sufficiently weak
impact, these considerations would predict flatter $\tilde{C}$-profiles in
larger halos as these halos are less concentrated.

Our goal is to characterize the trend of $\tilde{C}$-profiles with cluster mass
and to test whether we can dissect a universal spatial CR profile. To this end,
we adopt a phenomenological profile that allows for enough freedom to capture
these features as accurately as possible. Hence, we parametrize $\tilde{C}(\vr)$
with shape parameters that include a flat central region given by $C_{\rm
  center}$, a transition region where the location is denoted by $R_{\rm trans}$
and the steepness by $\beta$, and finally a flat outer region denoted by
$C_\rmn{vir}$, through the equation
\begin{equation}
  \label{eq:tildeCM}
\tilde{C}_\rmn{M}(R) = \left(C_{\rm vir}- C_{\rm center}\right)\left(1 +
\left(\frac{R}{R_{\rm trans}}\right)^{-\beta}\right)^{-1} + C_{\rm
 center}\,.
\end{equation}
The core regions in our radiative simulations show too much cooling, and we
possibly lack of important CR physics such as anisotropic CR diffusion that
could smooth out any strong inhomogeneity at the center. Hence we adopt a
conservative limit of the central profile value of $C_{\rm center}=5\times
10^{-7}$.

\begin{figure}
\begin{minipage}{1.0\columnwidth}
   \includegraphics[width=1.0\columnwidth]{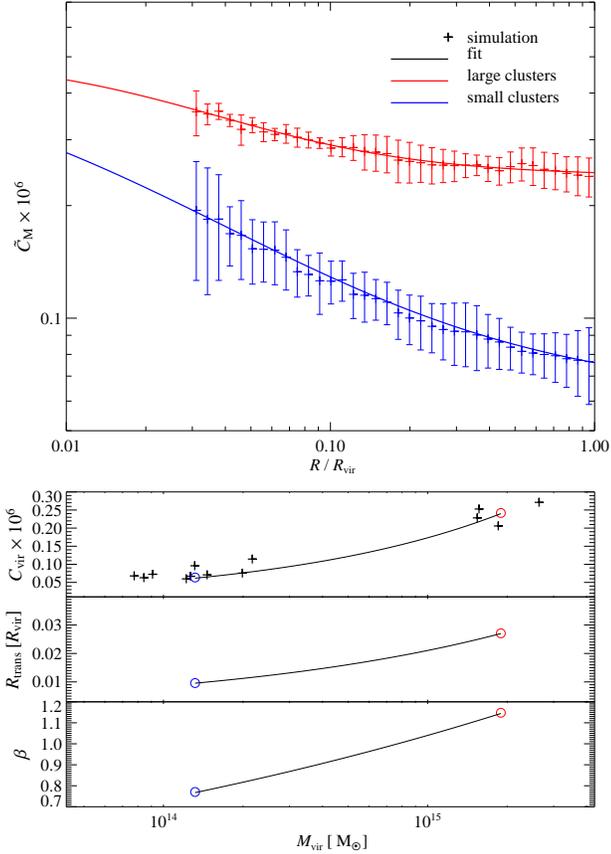}
   \caption{The top panel shows the profile of the dimensionless normalization
     of the CR spectrum, $\tilde{C}_\rmn{M}$. We show the mean
     $\tilde{C}_\rmn{M}$ and the standard deviation across our cluster sample
     which has been subdivided into two different mass intervals: large- (red),
     and low-mass clusters (blue) representing the mass range $1\times 10^{15} <
     \mvir/\mathrm{M_\odot} < 3\times 10^{15}$, and $7\times 10^{13} <
     \mvir/\mathrm{M_\odot} < 4\times 10^{14}$.  The solid lines show the best
     fit to equation~(\ref{eq:tildeCM}). The lower three panels show the mass
     dependence of the quantities which parametrize $\tilde{C}_\rmn{M}$ for low mass
     clusters (blue circles) and large mass clusters (red circles). The top
     small panel shows the asymptotic $\tilde{C}_\rmn{M}$ for large radii
     ($C_\rmn{vir}$), where each cross shows $\tilde{C}_\rmn{M}$ at $R_\rmn{vir}$
     for each cluster. The middle panel shows the transition radius
     $R_\rmn{trans}$, and the bottom panel shows the inverse transition width
     denoted by $\beta$.  \label{fig:tildeC_radial}}
   \end{minipage}
\end{figure}

In the top panel of Fig.~\ref{fig:tildeC_radial} we show the mean of the
$\tilde{C}_\rmn{M}$ profiles from the cluster simulation sample. We subdivide or
sample in two different mass intervals: large mass clusters (top red), and small
mass clusters (bottom blue), in the mass range $1\times 10^{15} <
\mvir/\mathrm{M_\odot} < 3\times 10^{15}$, and $7\times 10^{13} <
\mvir/\mathrm{M_\odot} < 4\times 10^{14}$, respectively.\footnote{Note that we
  exclude the intermediate mass cluster g1d, since it had a very recent merger
  which resulted in a double-cored system with a non-spherical CR profile.}  The
error-bars show the standard deviation from sample mean and the solid lines the
best fit that will be discussed in more detail in
Section~\ref{section:fitting_universal_CR}. The $\tilde{C}_\rmn{M}$ profile of
our large mass clusters is almost flat and shows only a very weak radial
dependence. In contrast, the $\tilde{C}$ profile of our small clusters has a
rather steep and long transition region and is increasing towards the
center. The difference in normalization, transition width, and transition radius
between low mass and large mass clusters indeed suggests that
$\tilde{C}_\rmn{M}$ scales with the cluster mass in a way that we
anticipated. We quantify the mass scaling of these shape parameters by a
power-law fit to the small and large clusters in the three lower panels of
Fig.~\ref{fig:tildeC_radial}: the value of $\tilde{C}_\rmn{M}$ in the
asymptotically flat regime in the cluster periphery, $C_\rmn{vir}$ (top); the
transition radius, $R_\rmn{trans}$ (middle); and the steepness of the
transition, $\beta$ (bottom). As expected, we find that all three quantities
increase slowly with mass. We additionally show the values of
$\tilde{C}_\rmn{M}$ at $R_\rmn{vir}$ for each cluster (black crosses). Within
the scatter, these values are consistent with the mass scaling found by the
profile fits in our two cluster mass bins. We have shown that there exists an
{\em almost universal spatial CR profile} after taking into account the weak
trends of the $\tilde{C}$ profile with cluster mass. Note that the
particle-by-particle variance of the spatial CR profile within a cluster (that
we address in Section~\ref{app:analytic_modeling} in the appendix) additionally
supports the conclusion of a universal spatial CR profile.

\subsection{A semi-analytic model for the spatial and spectral CR distribution}
\label{section:fitting_universal_CR}

In our simulations we use a CR spectral description which follows five different
CR proton populations; each being represented by a single power-law. The CR
populations are chosen such that we accurately can capture the total CR spectrum
in the entire momentum space (a convergence study on the number of CR
populations is presented in the appendix~\ref{app:convergence}). We want to
model this spectrum analytically with as few CR populations as possible,
but at the same time preserve the functional form of a power-law of each
population. In that way we can easily obtain the total CR spectrum by
superposition and apply a simple analytic formula to estimate the induced
radiative processes.

\begin{figure}
\begin{minipage}{1.0\columnwidth}
 \includegraphics[width=1.0\columnwidth]{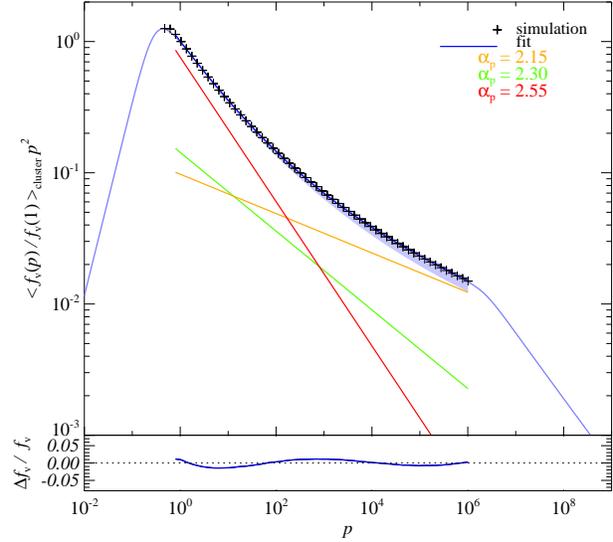}
 \caption{Cosmic ray proton spectrum as a function of dimensionless CR momentum
   $p=P_\p/\mpc$ for our sample of 14 galaxy clusters
   (Table~\ref{tab:cluster_sample}). The main panel shows median of the cosmic
   ray spectrum $f_\mathrm{v}(p)p^2$ across our cluster sample. The spectrum of
   each cluster has been normalized at $p=1$ (black crosses). The blue solid
   line shows the best fit triple power-law to the simulation data. The light
   blue area shows the 68 percentile of the data points across our cluster
   sample.  The bottom panel shows the difference between the relative fit and
   the simulation data (blue solid line) which amounts to less than 2
   percent. \label{fig:proton_spectrum_all_fit}}
\end{minipage}
\end{figure}

In detail, we use a total CR proton spectrum $f_\vv(p,R)$ that is obtained by
summing over the individual CR populations $f_i$
(equation~\ref{eq:single_CRp_spectrum}); each being a power-law in particle
momentum with the total CR amplitude $C_\vv(R)$,
\begin{eqnarray}
 \label{eq:fCRp_fit}
f_\vv(p,R) = C_\vv(R)\,g(\zeta_\rmn{p, max})\,D_\p(p,p_\rmn{break}, q)\sum_i\Delta_i\,p^{-\alpha_i}\,,
\end{eqnarray}
where we assume universal spectral shape throughout the cluster. The
normalization of $f_\vv$ depends on maximum shock acceleration efficiency
$\zeta_\rmn{p, max}$, where $g(0.5)=1$ and $g(\zeta_\rmn{p, max}<0.5)<1$
(functional will be studied in Pfrommer in prep.). Denoting the amplitude of
each CR population by $c_i(R)$, where the number of spectral bins might be
different from the five chosen in the simulations, we construct the relative
normalization for each CR population
\begin{eqnarray}
  \Delta_i \simeq  \frac{c_i(R)}{C_\vv(R)}\,.
\end{eqnarray}
Here we have assumed that $c_i(R)/C_\vv(R)$ is only a weak function of
radius. This is explicitly shown in Fig.~\ref{fig:proton_spectrum_radial}, where
the functional form is almost independent of the radius.  The two energy breaks
in the CR spectrum are represented by
\begin{eqnarray}
 \label{eq:fCRp_break}
  D_\p(p,p_\rmn{break}, q) =
 \left[ \frac{1 + q^\beta}{1+\left(\frac{p}{q}\right)^{-\beta}} \right]^{\frac{2}{\beta}}
 \left[ \frac{1+p_\rmn{break}^{-\beta}}{1+\left(\frac{p}{p_\rmn{break}}\right)^{\beta}}
   \right]^{\frac{\delta}{\beta}}\,.
\end{eqnarray}
The first term in equation~(\ref{eq:fCRp_break}) ensures the low-momentum slope $\sim
p^2$ as appropriate from the phase space volume that is populated by CRs
\citep{2007A&A...473...41E} and the last term accounts for diffusive losses of
CRs that steepen the spectrum by $\delta=1/3$ assuming Kolmogorov
turbulence. The shape parameter $\beta$ determines the size of the transition in
momentum space. Our choice of $\beta = 3$ ensures a fast transition from one
regime to the other. The low-momentum break $q$ is determined to $q\simeq0.3$
from Fig.~\ref{fig:proton_spectrum_all_fit}, and the high-momentum break
$p_\rmn{break}$ is derived from equation~(\ref{eq:break}).

To capture the spectral universality of our cluster sample, we fit in
Fig.~\ref{fig:proton_spectrum_all_fit} the median of our cluster sample of the
CR spectrum with a triple power-law function. Because we normalize the spectrum
at $p=1$, this ensures that the normalized spectrum $f_\rmn{v}(p, R) /
f_\rmn{v}(1 , R) = \sum_i \,\Delta_i\,p^{-\alpha_i}$ becomes independent of
radius (cf. equation~\ref{eq:fCRp_fit}). The triple power-law fit represented by the
blue line in the upper panel of Fig.~\ref{fig:proton_spectrum_all_fit} resulted
in
\begin{eqnarray}
\label{eq:Delta}
\boldsymbol{\Delta}=(0.767,0.143,
0.0975)\,,~\rmn{and}~~\boldsymbol{\alpha}=(2.55, 2.3, 2.15)\,.
\end{eqnarray}
The data from the simulation is denoted by black crosses, together with the 68
percentiles spread shown by the light blue area. In the lower panel, we show the
relative difference between the simulation and the fit which indicates a
difference of less than two percent from the GeV to PeV energy regime.

The spatial part of the CR spectrum is derived from the fit to the mean
$\tilde{C}_\rmn{M}$ in the top panel of Fig.~\ref{fig:tildeC_radial}. The solid
lines show the best fit to $\tilde{C}_\rmn{M}$ using
equation~(\ref{eq:tildeCM}). We find that the central value $C_{\rm center}=5\times10^{-7}$
is the most conservative value that still provides a good fit for both mass
intervals. Note that there is a large uncertainty in this value due to
insufficient data in the center\footnote{We cut the $\tilde{C}_\rmn{M}$ profiles
  at $R\simeq0.03\rvir$ due to the bias with the over-cooled center. The
  increased scatter in the center for the small clusters is caused by the low
  statistics of the fewer number of clusters that can contribute on these small
  scales. With both of these considerations in mind, we conclude that there is a
  large scatter in $C_{\rm center}$.}, implying that it should be treated as a
lower limit. However, the gamma-ray flux depends only weakly on the exact value
of $C_{\rm center}$ since most of the flux reside from the region outside the
transition region. The mass dependence of $\tilde{C}_\rmn{M}$ is quantified in
the three lower panels of Fig.~\ref{fig:tildeC_radial}, where we fit a
power-law in mass for the normalization $C_{\rm vir}$ (top), the transition
radius $R_{\rm trans}$ (middle), and the steepness of the transition region
$\beta$ (bottom).  We obtain the following relations,
\begin{eqnarray}
  \label{eq:C-scaling1}
  C_{\rm vir}   &=& 1.7\times 10^{-7}~~\times(\mvir / 10^{15}\,M_\odot)^{0.51}, \\
  \label{eq:C-scaling2}
  R_{\rm trans} &=& 0.021\,R_\rmn{vir}~~\times(\mvir / 10^{15}\,M_\odot)^{0.39}, \\
  \label{eq:C-scaling3}
  \beta       &=& 1.04~~\times(\mvir/10^{15}\,M_\odot)^{0.15}.
\end{eqnarray}

\section{Semi-analytic model for the $\gamma$-ray emission}
\label{sect:analytic_model}
In this section we derive a semi-analytic formula for the integrated
$\gamma$-ray source function that is based on our semi-analytic CR distribution.
Using the gas density profile of the cluster along with its virial radius and
mass, this formula enables us to predict the dominating $\pi^0$-decay induced
$\gamma$-ray emission from that cluster. The density profiles can either be
inferred from simulations or from X-ray data. To test the semi-analytic source
function, we use density profiles from our cluster simulations. We also make
$\gamma$-ray flux predictions for the Coma and Perseus cluster using their true
density profiles as inferred from X-ray observations.

\subsection{Schematic overview and semi-analytic formulae}
\label{sect:schematic_overview}
\begin{figure}
  \begin{minipage}{1.0\columnwidth}
    \includegraphics[width=1.0\columnwidth]{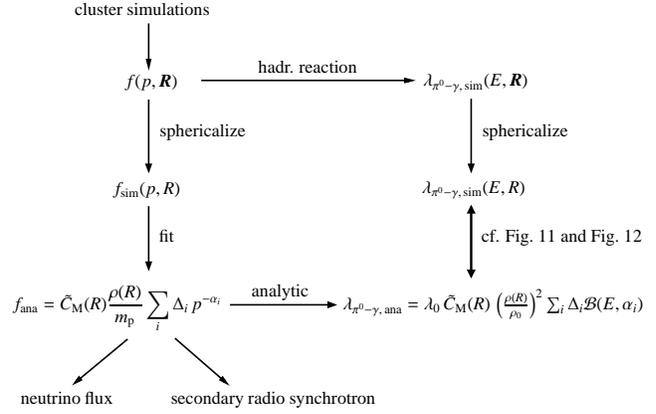}
    \caption{Schematic overview of how our semi-analytic framework relates to
      the simulated CR and $\gamma$-ray quantities. From the cluster simulations
      we derive the CR spectrum for each SPH particle, $f(p,\vr)$. Going
      clockwise, the integrated $\gamma$-ray production rate for each SPH
      particle $\lambda_\rmn{\pig,sim}(E,\vr)$ is derived within the framework
      of \citet{2008MNRAS.385.1211P}.  The radial profile of the integrated
      $\gamma$-ray production rate, $\lambda_\rmn{\pig,sim}(E,R)$, is obtained
      by volume weighting this quantity of individual SPH particles. Returning
      to $f(p,\vr)$ and going counter-clockwise instead, we can directly perform
      a volume-weighting of the CR spectrum in each radial bin, yielding
      $f_\rmn{sim}(p,R)$. Fitting both the spatial and spectral part provides us
      with a semi-analytic model for the spectrum, $f_\rmn{ana}$. The
      semi-analytic model for the integrated $\gamma$-ray production rate,
      $\lambda_\rmn{\pig,ana}$, is derived within the framework of
      \citet{2008MNRAS.385.1211P}. The explicit form of $\lambda_\rmn{\pig,ana}$
      can be found in equations~(\ref{eq:gammaray_analytic})-(\ref{eq:gamma_break}),
      while $\lambda_0$ and $\mathcal{B}(E,\alpha_i)$ are implicitly given by
      equation~(\ref{eq:gammaray_analytic_energy}).  Comparing the flux from the
      simulations and our semi-analytic model shows good agreement for our
      cluster sample.}
      \label{fig:schematics_formalism}
  \end{minipage}
\end{figure}

In Fig.~\ref{fig:schematics_formalism} we show a schematic overview of the
simulated CR spectrum and the integrated source function together with the
mapping to our semi-analytic model. From the cluster simulations we derive the
CR spectrum $f(p,\vr)$ describing the phase space density of CRs. When taking
the spherical (volume weighted) average of $f(p,\vr)$, we obtain
$f_\vv(p,R)$. We fit the spectral and spatial part of this function
separately. The semi-analytic model for the CR distribution in clusters that we
provide in equation~(\ref{eq:fCRp_fit}) can now be used to predict the secondary
radio synchrotron emission and the hadronically induced neutrino and
$\gamma$-ray emission from galaxy clusters. Following the formalism in
\citet{2008MNRAS.385.1211P}, we obtain the volume weighted and energy integrated
and omnidirectional (i.e integrated over the $4\pi$ solid angle) $\gamma$-ray
source function due to pion decay\footnote{From here on we always imply the
  volume weighted and energy integrated source function when talking about
  $\lambda$ if nothing else is stated.},
\begin{eqnarray}
\label{eq:gammaray_analytic}
\lambda_{\pig}(R,E) &=& A_\lambda(R)\,\tilde{\lambda}_{\pig}(E), \\
\label{eq:gammaray_analytic_radius}
A_\lambda(R) &=& \tilde{C}_\rmn{M}(R)\,\frac{\rho(R)^2}{\rho_0^2}\,,\\
\tilde{\lambda}_{\pig}(E) &=& g(\zeta_\rmn{p, max})\,D_\gamma(E_\gamma, E_\rmn{\gamma,~break})
\frac{4m_{\pi^0} c}{3m_\p^3}\,\rho_0^2 \nonumber\\
  &\times& \sum_{i=1}^3\frac{\sigma_{\rmn{pp},\,i}}{\alpha_i\,\delta_i}
\left(\frac{m_\p}{2m_{\pi^0}}\right)^{\alpha_i}
\Delta_i\left[\mathcal{B}_x\left(\frac{\alpha_i+1}{2\delta_i},
\frac{\alpha_i-1}{2\delta_i}\right)\right]_ {x_1}^{x_2} \, ,\nonumber\\
\label{eq:gammaray_analytic_energy}
& &\rmn{and} \qquad  
x_j=\left[1+\left(\frac{m_{\pi^0}c^2}{2E_{\gamma,j}}\right) ^{2\delta_i}\right]^{-1}\,,
\end{eqnarray}
where the sum extends over our three CR populations, $\tilde{C}_\rmn{M}(R)$ is
given by equations~(\ref{eq:tildeCM}), (\ref{eq:C-scaling1}) -
(\ref{eq:C-scaling3}), while $\boldsymbol{\Delta}$ and $\boldsymbol{\alpha}$ are
provided by equation~(\ref{eq:Delta}) and we introduced an auxiliary variable
$\rho_0$ for dimensional reasons to ensure that $A_\lambda$ is dimensionless.
In equation~(\ref{eq:gammaray_analytic_energy}) we have also have introduced the
abbreviation
\begin{equation}
\left[\mathcal{B}_x\left(a,b\right)\right]_{x_1}^{x_2}=
\mathcal{B}_{x_2}\left(a,b\right)-\mathcal{B}_{x_1}\left(a,b\right)\,,
\end{equation}
where $\mathcal{B}_x\left(a,b\right)$ denotes the incomplete Beta-function, we
have used that the number density of target nucleons is the sum of hydrogen
$n_\rmn{H}$ and helium $n_\rmn{He}$ number densities,
$n_\rmn{N}=n_\rmn{H}+4n_\rmn{He}=\rho/m_\p$. The shape parameter $\delta_i
\simeq 0.14\,\alpha_i^{-1.6}+0.44$ allows us to accurately predict the emission
close to the pion bump in combination with the effective inelastic cross-section
for proton-proton interactions, $\sigma_{\rmn{pp},\,i} \simeq
32\,(0.96+\e^{4.42-2.4\alpha_i})\,\rmn{mbarn}$. In addition we have a term,
similar to equation~(\ref{eq:fCRp_break}) that describes the diffusive CR losses
due to escaping protons from the cluster at the equivalent photon energy for the
break, $E_\rmn{\gamma,~break}$. It is derived from $E_\rmn{p,~break}$ of
equation~(\ref{eq:break}),
\begin{eqnarray}
\label{eq:gamma_break}
D_\gamma(E_\gamma, E_\rmn{\gamma,~break}) = 
\left[1 + 
    \left(\frac{E_\gamma}{E_\rmn{\gamma,~break}}\right)^\beta\right]^{-1/(3\beta)}\,,
\end{eqnarray}
where $\beta=3$. Finally we note that the semi-analytic $\gamma$-ray model is
derived within $\rvir$. Applying the model to the region outside $\rvir$, but
within $3\rvir$, would increase the $\gamma$-ray flux by less than $10\%$ for
small mass clusters and less than $30\%$ for large mass clusters
(cf. Table~\ref{tab:scaling}).

For convenience and completeness, we provide here the differential $\gamma$-ray
source function for pion decay \citep[for further details
  see][]{2004A&A...413...17P},
\begin{eqnarray}
\label{eq:diff_gammaray_analytic}
s_{\pig}(R,E) &=& A_\rmn{s}(R)\,\tilde{s}_{\pig}(E), \\
\tilde{s}_{\pig}(E) &=& g(\zeta_\rmn{p, max})\,D_\gamma(E_\gamma, E_\rmn{\gamma,~break})
\frac{16}{3m_\p^3c}\,\rho_0^2 \nonumber\\
  \lefteqn{\times \sum_{i=1}^3\frac{\sigma_{\rmn{pp},\,i}}{\alpha_i}
\left(\frac{m_\p}{2m_{\pi^0}}\right)^{\alpha_i}\Delta_i\,
\left[\left(\frac{2E_\gamma}{m_{\pi^0}c^2}\right)^{\delta_i}+
\left(\frac{2E_\gamma}{m_{\pi^0}c^2}\right)^{-\delta_i}\right]^{-\frac{\alpha_i}{\delta_i}}\!\!}
\end{eqnarray}
where $A_\rmn{s}(R) = A_\lambda(R)$ (see
equation~\ref{eq:gammaray_analytic_radius}).  The exact shape of
$D_\gamma(E_\gamma, E_\rmn{\gamma,~break})$ depends on the detailed physics of
CR diffusion and the characteristics of turbulence and is subject to future
studies. We note, however, that the break does not interfere with the energy
range of current $\gamma$-ray observatories.

\subsection{Comparing our semi-analytic model to simulations}

\begin{figure}
\begin{minipage}{1.0\columnwidth}
 \includegraphics[width=1.0\columnwidth]{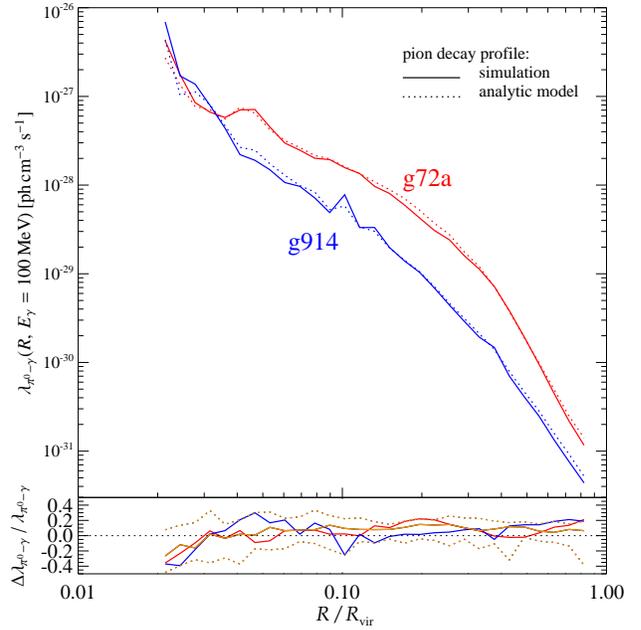}
 \caption{Comparison of the spatial profile of the $\gamma$-ray source function
   in our simulations and our semi-analytic framework. The main panel shows the
   profile of the $\gamma$-ray source function $\lambda_{\pig}$ that results
   from pion decay emission. We compare $\lambda_{\pig}$ at 100 MeV for two
   clusters, the large merging g72a (red) cluster and the the small CC g914
   (blue) cluster, both as a function of $R/\rvir$. The solid lines show the
   simulated integrated source function and the dotted lines the semi-analytic
   model. The lower panel shows the difference between the semi-analytic and the
   simulated source functions normalized by the simulated source
   function. The orange lines show the median (solid) difference between the
     semi-analytic and the simulated source functions across our cluster sample
     together with the 68 percentiles (dotted). \label{fig:lambda_comparison}}
\end{minipage}
\end{figure}

\begin{figure}
\begin{minipage}{1.0\columnwidth}
 \includegraphics[width=1.0\columnwidth]{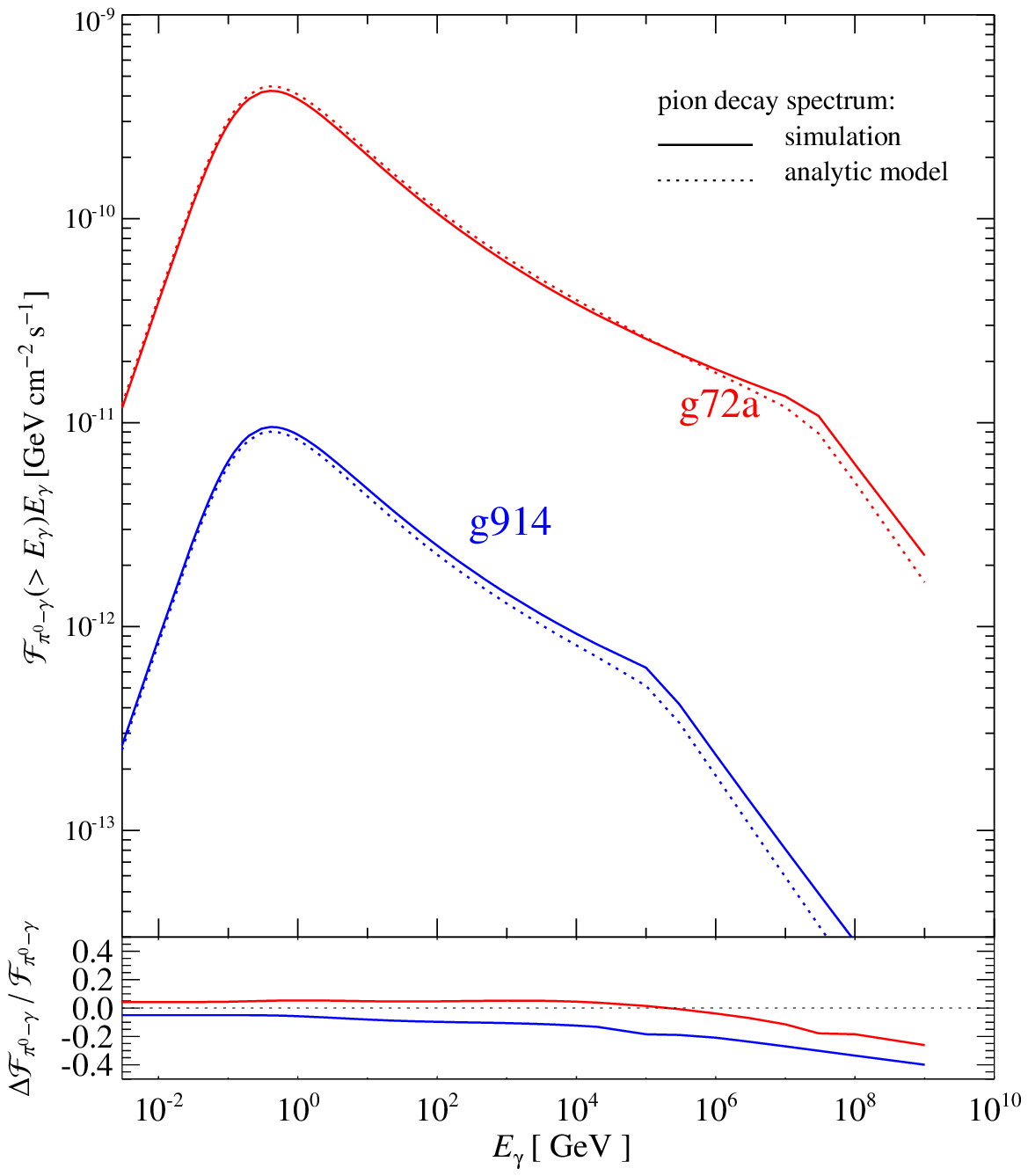}
 \caption{Comparison of the $\gamma$-ray spectrum in our simulations and our
   semi-analytic framework.  The main panel shows the $\gamma$-ray number flux
   from pion decay weighted by the photon energy.  We compare two clusters, the
   large merging g72a (red) cluster and the small CC g914 (blue) cluster, both
   as a function of photon energy. The solid lines show the simulated flux and
   the dotted lines the semi-analytic model. The lower panel shows the difference
   between the semi-analytic and the simulated flux normalized by the simulated
   flux. \label{fig:flux_comparison}}
\end{minipage}
\end{figure}

To test our semi-analytic $\gamma$-ray model we contrast it with numerically
calculated radial profiles and spectra.  In the upper panel of
Fig.~\ref{fig:lambda_comparison}, we compare the radial profile of our
semi-analytic $\gamma$-ray source function (equation~\ref{eq:gammaray_analytic})
at 100~MeV to a numerical emission profile for two representative clusters, a
large post-merging cluster g72a, and a small CC cluster g914. The numerical
profile has been obtained by means of calculating the $\gamma$-ray source
function $\lambda_{\pig}$ of every SPH particle and volume weighting the
resulting radial profile.  The overall shape of the radial profiles of
$\lambda_{\pig}$ for different clusters are quite similar. This is because the
main spatial dependence originates from the gas density, that enters with a
square in $\lambda_{\pig}$ and has a small scatter across the cluster
sample. The different behavior in the cluster centers stems from the steeper
profile of $\tilde{C}_\rmn{M}(R)$ for low mass clusters.  We show the difference
between the integrated source functions from our simulations and the
semi-analytic model in more detail in the lower panel of
Fig.~\ref{fig:lambda_comparison}.  In both clusters, we see an excellent
agreement with differences amounting to less than 20-30 percent at any
radius. These differences are representative for our cluster sample, which is
  indicated by the median difference across our cluster sample together with the
  68 percentiles that show a spread similar to the difference in these two
  clusters. Since these are fluctuating differences, they average partly away
when we perform the volume integral. Hence we obtain an agreement of the total
flux within the virial radius -- obtained directly from the simulations and in
our semi-analytic model -- that is better than 5 per cent for the two
representative clusters that are shown. This indicates that our semi-analytic
description manages to capture the CR physics, that is important for
$\gamma$-ray emission from clusters, surprisingly well.

\begin{table*}
\caption{\label{tab:real_cluster_data}Properties of the Coma and Perseus galaxy cluster.}
\begin{tabular}{cccccccc}
\hline\hline
\phantom{\Big|} 
 cluster & $z^{(1)}$ & $D_\rmn{lum}^{(1)}$  & $\rvir^{(1)}$ & $\mvir^{(1)}$ & 
  $L_{X,0.1-2.4}^{(1)}$  & $n_\e^{(1)}$\\

& & [Mpc] & [Mpc] &  [$10^{14}\,\rmn{M}_{\odot}$] &
[$10^{44}\,$erg~s$^{-1}$] &  [$10^{-3}\,$electrons cm$^{-3}$] \\

\hline 
\phantom{\Big|}
Coma & 0.0232 & 101 & 2.3 & $13.8 $ & $4.0$ & 
  $3.4\times\left[1+(R/294\,\rmn{kpc})^2\right]^{-1.125}$ \\
Perseus & 0.0183 & 77.7 & 1.9 & $7.71$ & $8.3$ & 
$46\times\left[1+(R/57\,\rmn{kpc})^2\right]^{-1.8} + 
4.79\times\left[1 + (R/200\,\rmn{kpc})^2\right]^{-0.87}$ \\
\hline\hline
\end{tabular}
  \begin{quote}
    Notes:\\
    (1) Data for redshift ($z$), luminosity distance ($D_\rmn{lum}$), $\rvir$,
    $\mvir$, and X-ray 0.1-2.4 keV luminosity ($L_{X,0.1-2.4}$) are taken from
    \cite{2002ApJ...567..716R} and rescaled to our assumed cosmology
    (Section~\ref{sec:cosmology}). The electron number density, $n_\e$, for Coma
    and Perseus are inferred from X-ray observations by
    \cite{1992A&A...259L..31B} and \cite{2003ApJ...590..225C}, respectively.
  \end{quote}
\end{table*}

\begin{table*}
  \caption{\label{tab:GRcomp}$\gamma$-ray flux comparison  of the Coma and Perseus galaxy cluster.}
\begin{tabular}{llcccccc}
\hline\hline
\phantom{\Big|} 
&  & $E_\gamma$ & Experiment & $\theta$ & $\mathcal{F}_{\gamma,\rmn{UL}}(> E_\gamma)$ & 
  $\mathcal{F}_{\pig}(> E_\gamma, \zeta_{\p,\rmn{max}})^{(5)}$ & 
  $\frac{\mathcal{F}_{\gamma,\rmn{UL}}(> E_\gamma)}{\mathcal{F}_{\pig}(> E_\gamma, \zeta_{\p,\rmn{max}})}$ \\  
&  & [GeV] &  & [deg] & [$\rmn{ph}\,\rmn{cm}^{-2}\,\rmn{s}^{-1}$] &  [$\rmn{ph}\,\rmn{cm}^{-2}\,\rmn{s}^{-1}$] &\\
\hline 
\phantom{\tiny a}\\
GeV-regime & Coma    & 100~MeV & EGRET$^{(1)}$ & 5.8$^{(4)}$ & $3.81\times10^{-8~~}$ & $4.20\times10^{-9~}$ & ~~9.1 \\
           & Perseus & 100~MeV & EGRET$^{(1)}$ & 5.8$^{(4)}$ & $3.7~~ \times10^{-8~~}$ & $1.46\times10^{-8~}$ & ~~2.5 \\
TeV-regime & Coma    & \phantom{00}1~TeV   & HESS$^{(2)}$  & 0.2~~  & $1.1~~\times10^{-12}$  & $4.86\times10^{-14}$ & 23.0  \\
           & Perseus & \phantom{00}1~TeV   & Magic$^{(3)}$ & 0.15 & $4.7~~ \times10^{-13}$ & $1.75\times10^{-13}$ & ~~2.7 \\
           & Perseus & 100~GeV & Magic$^{(3)}$ & 0.15 & $6.55\times10^{-12}$ & $3.2~~\times10^{-12}$  & ~~2.0 \\
\hline\hline
\end{tabular}
\begin{quote}
  {Notes:}\\
  (1) \citet{2003ApJ...588..155R}, (2) \cite{2009A&A...502..437A}, (3)
  \cite{2009arXiv0909.3267T}, (4) Since the flux is dominated by the region
  inside $\rvir$ we use $R_\theta=\rvir$. (5) $\gamma$-ray fluxes are obtained
  within our semi-analytic model that is based on our conservative model. The
  $\gamma$-ray flux level depends on the maximum shock acceleration efficiency of CRs,
  $\zeta_{\p,\rmn{max}}$, for which we assume an optimistic value of
  0.5. Smaller efficiencies imply smaller fluxes.
\end{quote}
\end{table*}

We show the $\gamma$-ray spectrum of the $\pi^0$-decay emission weighted by
energy for g72a and g914 in the upper panel of Fig.~\ref{fig:flux_comparison}.
Both clusters have very similar spectra with the exception of the diffusive
steepening that is inherited from the proton spectrum. Since the break of the
proton spectrum scales as $E_\rmn{p,~break} \propto M_\rmn{vir}^2$ with the
virial mass of the cluster (equation~\ref{eq:break}), the break in the pion
decay spectrum is reduced by a factor of $10^3$ for the smaller cluster g914.
The solid and dotted lines contrast the simulated spectrum to that obtained from
our semi-analytic model.  The difference between these two approaches amounts to
less than 20 percent for both clusters in the GeV and TeV band (shown in the
lower panel). The flux differences between our semi-analytic model and the
simulations for individual clusters are about a factor two smaller compared to
the scatter in the the mass-luminosity scaling relations for a given cluster
(see e.g. Fig.~\ref{fig:ML_energy}). The reason for the more accurate
predictions within our semi-analytic formalism is a direct consequence of the
essential additional spatial information of the gas and CR density that we
account for.

\subsection{Predicting the $\gamma$-ray emission from Perseus and Coma}

Here we demonstrate how our semi-analytic formalism can be applied to predict
the $\gamma$-ray flux and surface brightness from real clusters using their
electron density profile as inferred from X-ray measurements. The predicted flux
and surface brightness are then compared to current upper limits and previous
work.

The two clusters that we investigate are two of the brightest X-ray clusters in
the extended HIFLUGCS catalogue \citep{2002ApJ...567..716R} -- a sample of the
brightest X-ray clusters observed by ROSAT -- namely Coma and Perseus. Both Coma
and Perseus are well studied clusters, where Coma is a large post-merging
cluster while Perseus is a somewhat smaller cluster that hosts a massive cooling
flow and is the brightest X-ray cluster known \citep{1992MNRAS.258..177E}.  In
Table~\ref{tab:real_cluster_data} we show the data for respective cluster. Using
the electron number density profile, we can calculate the gas density profile of the
cluster through $\rho=m_\p\,n_\e/(X_\rmn{H}\,X_\e)$. Here denote $X_\rmn{H}=0.76$
the primordial hydrogen (H) mass fraction, and the ratio of electron and
hydrogen number densities in the fully ionized ICM is given by $X_\e =
1.157$. For sufficiently small angular scales, the $\gamma$-ray flux within the
radius $R_\theta=\theta\, D_\rmn{lum}$ of a disk of angular radius $\theta$
(measured in radians and centered at $R=0$) is calculated by
\begin{eqnarray}
\mathcal{F}_{\pig}(>E_\gamma) &=&
\frac{1}{D_\rmn{lum}^2}\int_0^{R_\theta}2\pi\,\dd R_\perp
\,R_\perp\, S_{\pig}(R_\perp, E_\gamma), \\
S_{\pig}(R_\perp, E_\gamma) &=&2\frac{\tilde{\lambda}_{\pig}(E_\gamma)}{4\pi}\int_{R_\perp}^{\infty}\dd R
\,\frac{\rho^2(R)}{\rho_0^2} \frac{\tilde{C}_\rmn{M}(R)\,R}{\sqrt{R^2-R_\perp^2}}\,,
\end{eqnarray}
where $\tilde{\lambda}_{\pig}(E_\gamma)$ is given by
equation~(\ref{eq:gammaray_analytic_energy}) and we introduced the definition
for the $\gamma$-ray surface brightness $S_{\pig}$ in the last step.  As a
consistency check, we compare the flux ratios from X-rays
$\mathcal{F}_\rmn{X-ray, Perseus} / \mathcal{F}_\rmn{X-ray, coma} \simeq 3.51$
to that of $\gamma$-rays as predicted by our semi-analytic model,
$\mathcal{F}_\rmn{\pig,Perseus}(>100\,\mev) /
\mathcal{F}_\rmn{\pig,coma}(>100\,\mev) \simeq 3.47$ and find excellent
agreement within one percent.

\subsubsection{Comparison to upper limits in the GeV/TeV regime}

We calculate the $\gamma$-ray fluxes for Coma and Perseus above 100~MeV and 100
GeV that is emitted within $R_\theta=\rvir$ (see Table~\ref{tab:GRcomp}).
Comparing the 100~MeV-flux in our model to the EGRET upper limit on the Coma
(Perseus) cluster \citep{2003ApJ...588..155R}, we find that it falls short of
our semi-analytic prediction by a factor of about 9 (2.5). With the two-year
data by Fermi, this upper limit on Coma will improve considerably and is
expected to become competitive with our predictions.  With this at hand, we will
be able to put important constraints on the adopted CR physics in our
simulations. In particular, we can test our assumptions about the maximum shock
acceleration efficiency at structure formation shocks.  Since Fermi detected
$\gamma$-rays from the central cD galaxy in Perseus, NGC1275, at a level that is
about five times higher than the EGRET upper limits, this indicates that the
source is variable on time scales of years to decades
\citep{2009ApJ...699...31A}. Hence it restricts and complicates the
detectability of the extended pion-day emission that might be buried
underneath.

In the TeV regime, we integrate our model prediction within a solid angle that
is comparable to the point-spread function of IACTs.  The best current upper
limit for Coma (\citealt{2009A&A...502..437A}) falls short of our semi-analytic
prediction by a factor 20.  The best current upper flux limit for Perseus (using
a spectral index of -2.2, \citealt{2009arXiv0909.3267T}) is only a factor of two
larger than our flux prediction and clearly within reach of future deeper TeV
observations. This demonstrates the huge potential of nearby CC clusters to
detect non-thermal $\gamma$-ray emission as suggested by
\citet{2004A&A...413...17P}.

\subsubsection{Surface brightness profile}
\begin{figure}
  \begin{minipage}{1.0\columnwidth}
    \includegraphics[width=1.0\columnwidth]{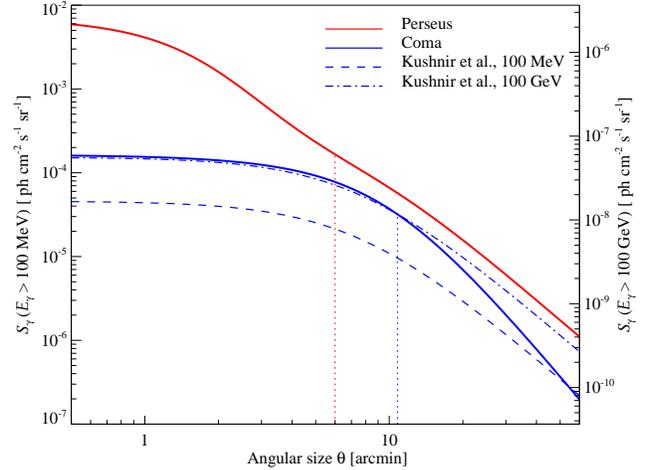}
    \caption{The $\gamma$-ray brightness profiles derived with our semi-analytic
      formalism (solid line) as a function of the angular size on the sky
      $\theta$. The left axis indicates the emission for $E_\gamma > 100$~MeV,
      and the right axis shows the emission for $E_\gamma > 100$~GeV. We compare
      the emission from the Perseus galaxy cluster (red solid line) to the Coma
      galaxy cluster (blue solid line) while employing the X-ray density profile
      for each cluster, respectively. The high surface brightness in the cool
      core cluster Perseus is a result of the high central gas densities in this
      cluster. The dotted line shows the radius from within half the flux
      originates. It is clearly smaller for Perseus due to the steep central
      density profile and the larger transition region in $\tilde{C}_\rmn{M}$
      for smaller clusters. The dashed and dash-dotted lines show the Coma
      surface brightness as predicted from \citet{Kushnir:2009vm} using their
      assumed parameters above the energy 100~MeV and 100~GeV, respectively.}
    \label{fig:coma.perseus_comparison}
  \end{minipage}
\end{figure}

To explain the large difference in flux between Coma and Perseus we show the
surface brightness as a function of the viewing angle $\theta$ in
Fig.~\ref{fig:coma.perseus_comparison}. The dense cooling core region of Perseus
provides ample target material for two-body interactions such as bremsstrahlung
or hadronic CR interactions which boosts the luminosities likewise.  This
results in a large increase in flux compared to the average cluster of similar
mass that are characterized by mass-luminosity scaling relations (see
e.g. Table~\ref{tab:scaling}). Assigning a flux to a cluster that is consistent
with the scaling relations should be a very conservative approach for CC
clusters. The half flux radius for Perseus (Coma) is $\theta_\rmn{HF}=0.11$~deg
(0.18 deg). It is shown with dotted lines in
Fig.~\ref{fig:coma.perseus_comparison}. Since these $\theta_\rmn{HF}$ values are
comparable to the angular scale of the point spread functions of IACTs (0.1-0.2
degrees) both clusters are suitable candidate sources.  The dashed and
dotted-dashed lines are obtained by using the semi-analytic formalism from
\citet{Kushnir:2009vm} to predict the surface brightness from the Coma cluster
above the energy 100~MeV and 100~GeV, respectively. See
Section~\ref{sect:comparison} for a detailed comparison to their result.

\subsection{CR-to-thermal pressure and temperature profile in Perseus and Coma}

\begin{figure}
  \begin{minipage}{1.0\columnwidth}
    \includegraphics[width=0.99\columnwidth]{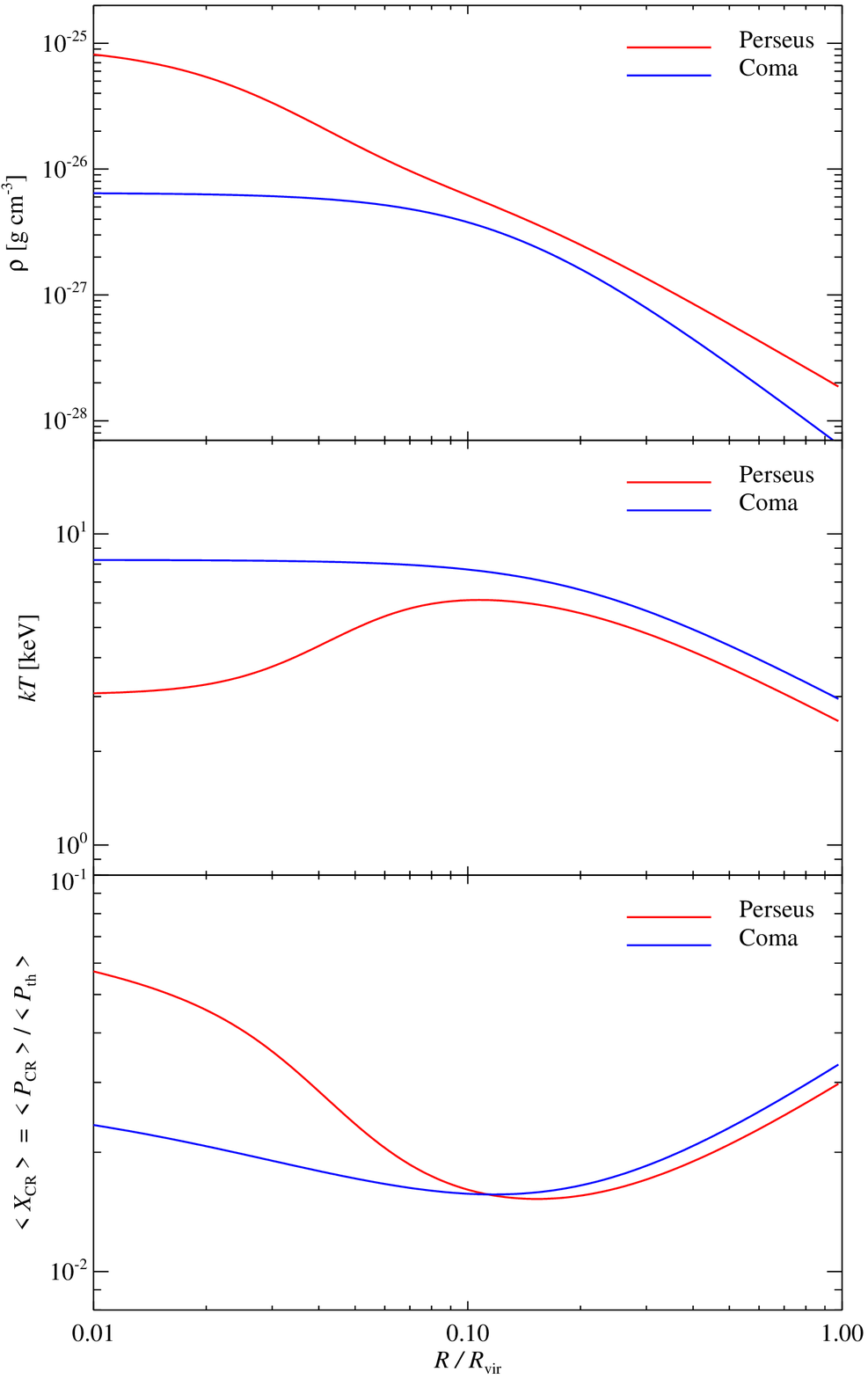}
    \caption{Radial profiles of the gas density $\rho$, the temperature $kT$,
      and the CR-to-thermal pressure $\bra X_\CR\ket = \bra P_\CR\ket/ \bra
      P_\rmn{th}\ket$. We compare the cool core cluster Perseus (red) to the
      non-cool core cluster Coma (blue). The density and and temperature
      profiles are taken from the X-ray data \citep{2003ApJ...590..225C,
        1992A&A...259L..31B} while we remodel the external temperature profile
      to bring it into agreement with cosmological cluster simulations as well
      as higher redshift X-ray observations. We obtain the CR-to-thermal
      pressure profile from our semi-analytic modeling in combinations with the
      presented gas profiles.  }
    \label{fig:coma.perseus_XCR}
  \end{minipage}
\end{figure}

A quantity that is of great theoretical interest is the CR pressure relative to
the thermal pressure, $X_\CR = P_\CR / P_\rmn{th}$ as it directly assesses the
CR bias of hydrostatic cluster masses where the CR pressure enters in the
equation of motion. Since $X_\CR \propto n_\rmn{th}/P_\rmn{th}=1/kT$ in the
external cluster regions, we have to accurately model the temperature profile of
our clusters. Note, however, that this relation should only hold for regions
with long thermal cooling times compared to the dynamical time scale. In
particular it breaks down towards the center of a cooling flow cluster where the
thermal gas cools on a shorter time scale such that the forming cooling flow
causes adiabatic contraction of the CR population. We model the central regions
of Coma and Perseus according to X-ray observations by
\cite{1992A&A...259L..31B} and \cite{2003ApJ...590..225C}, respectively.  These
observations are not sensitive to the outer temperature profile due to the high
particle background for XMM-Newton and Chandra.\footnote{We note that Suzaku has
  an approximately ten times lower background due to its low-Earth orbit that
  could enable such observations in the X-rays. Alternatively, combining the
  X-ray surface brightness with future Sunyaev-Zel'dovich measurements of these
  nearby clusters should in principle allow the derivation of the temperature
  profile of the entire cluster.}  X-ray observations of somewhat more distant
cluster sample show a universal declining temperature profile outside the
cooling core region up to $R_{500}$ \citep{2005ApJ...628..655V}.  We model this
behavior of the temperature profiles towards the cluster periphery according to
cosmological cluster simulations by \citet{2007MNRAS.378..385P} and obtain a
function $\mathcal{T}_\rmn{ext}(R)$ that accounts for the decreasing temperature
profile outside the core region. It is unity in the center and then smoothly
decreases until the virial radius beyond which we expect the spherical
approximation to break down and where the cluster accretion shocks should
introduce breaks in the temperature profile.\footnote{The mean temperature
  profile of the radiative simulations by \citet{2007MNRAS.378..385P} increases
  towards smaller radii until $R \sim 0.1 \rvir$ due to adiabatic compression of
  the gas and starts to drop sharply towards smaller radii where radiative
  cooling causes the temperature to decline. This behavior qualitatively matches
  the results of low mass clusters from a Chandra sample of nearby relaxed
  galaxy clusters \citep{2006ApJ...640..691V} whereas the temperature maximum
  for more massive clusters seems to shift to somewhat larger radii around $R
  \sim 0.2 \rvir$.  Hence we adopted this larger value as the transition radius
  for Perseus and Coma. We note that the resulting profiles are consistent with
  the observed central temperature profiles \citep{2001A&A...365L..60B,
    2003ApJ...590..225C} as well as the mosaiced entire temperature profile of
  Perseus beyond $R_{200}$ (S. Allen, private communication). } This function
can be multiplied to existing central temperature profiles to yield the
temperature profiles for Coma and Perseus,
\begin{eqnarray}
  \label{eq:kT(r)}
  kT_\rmn{Perseus}(R) &=& 7\,\rmn{keV} \times
  \frac{1 + (R/71\,\rmn{kpc})^3}{2.3 + (R/71\,\rmn{kpc})^3}\,
  \mathcal{T}_\rmn{ext}(R), \\
  kT_\rmn{Coma}(R)    &=& 8.25 \,\rmn{keV} \times \mathcal{T}_\rmn{ext}(R), \\
  \mathcal{T}_\rmn{ext}(R) &=& \left[1 + \left(\frac{R}{0.2\,\rvir}\right)^2\right]^{-0.32},
\end{eqnarray}
where $\rvir$ can be obtained from Table~\ref{tab:real_cluster_data}. The
density and temperature profiles for Coma and Perseus are shown in the first two
panels of Fig.~\ref{fig:coma.perseus_XCR}.

In the previous sections, we have seen that the $\gamma$-ray surface brightness
is a radially declining function and so is the CR pressure. In contrast, outside
the central cooling core regions, the CR-to-thermal pressure $X_\CR$ increases
with radius as can be seen in the bottom panel of
Fig.~\ref{fig:coma.perseus_XCR}. This increase is entirely driven by the
decreasing temperature profile.  The two $X_\CR$-profiles for Coma and Perseus
show our expectations for a typical non-CC and CC cluster.  The $X_\CR$-profile
in a CC cluster shows an additional enhancement towards the cluster center which
results from the centrally enhanced CR number density due to adiabatic
contraction during the formation of the cooling flow. During this process, the
thermal gas cools on a short time scale compared to that of the CRs which causes
an increase in density and hence adiabatic compression of the CRs.  We note that
the overall normalization of $X_\CR$ depends on the normalization of the CR
distribution that itself is set by the maximum shock acceleration
efficiency. The overall shape of $X_\CR$, however, should remain invariant since
CRs are adiabatically transported into the cluster
(\citealt{2007MNRAS.378..385P}, Pfrommer in prep.).

While the overall characteristics of $X_\CR$ in our semi-analytical model is
similar to that obtained in cosmological simulations
\citep{2007MNRAS.378..385P}, there are some noticeable differences particularly
in the central cooling core region around the cD galaxy of these clusters. Again
this can be traced back to known short-comings of modeling the physics in the
central regions correctly in current simulations such as to include AGN feedback
and anisotropic conduction in combination with magneto-hydrodynamics. This also
leads to different simulated temperature profiles in the center compared to
those inferred from X-ray observations and explains the discrepancy in the
$X_\CR$-profiles.  The volume average of the CR-to-thermal pressure for Coma and
Perseus is $\bra X_\CR\ket = \bra P_\CR \ket / \bra P_\rmn{th}\ket = 0.02$,
dominated by the region around the virial radius. These values assume an
optimistic saturation value of the shock acceleration efficiency of
$\zeta_\rmn{max}=0.5$ and decrease accordingly if this value is not realized at
the relevant structure formation shocks responsible for the CRs in clusters.

\section{High-energy scaling relations}
\label{sect:scaling}
We now discuss the scaling relations of the numerical $\gamma$-ray emission from clusters
and analyze their dependence on dynamical state, emission region and address the
bias of galaxies to the total luminosity.  The cluster scaling relations are
derived by integrating the surface brightness map of each cluster. By fitting
the total $\gamma$-ray emission of the 14 clusters in our cluster sample with a
power-law, we determine the mass-to-luminosity scaling. 

In the preceding sections we have shown that the pion decay emission dominates
the total $\gamma$-ray emission but we have not addressed the question, which
radii contribute most to the luminosity? To answer this, we have to consider the
$\gamma$-ray luminosity resulting from pion decay within a radius $R$,
\begin{equation}
\mathcal{L}_{\pig}(R) \propto \int_0^R \dd R'\, R'^2\, \tilde{C}_\rmn{M}(R')\, \rho(R')^2 
\sim \int_0^R \dd R'\, R'^2 \,\rho(R')^2.
  \label{sec:Lgamma}
\end{equation}
For the purpose of this simple argument, we neglected the very weak spatial
dependence of the CR distribution which is described by
$\tilde{C}_\rmn{M}(R)$. {\em The $\gamma$-ray luminosity $\mathcal{L}_{\pig}$ is
  dominated by the region around the scale radius $R_\rmn{s}$} which can be seen
by considering the contribution to $\mathcal{L}_{\pig}$ per logarithmic radius,
\begin{equation}
  \frac{\dd  \mathcal{L}_{\pig}}{\dd \log R}
  \propto R^3 \rho(R)^2
  \propto\left\{ \begin{array}{lr}
      R^3    & R < R_\rmn{s},\\
      R^{-3} & R\gg R_\rmn{s}.
    \end{array} \right. 
  \label{dLXdlogr}
\end{equation}
Here we assumed a central plateau of the density profile which steepens beyond
the scale radius $R_\rmn{s}$ and approaches the asymptotic slope of $R^{-3}$ of
the dark matter profile that shapes the gas distribution at large radii
\citep[][ and references therein]{2001MNRAS.327.1353K}.  This radial behavior
makes the simulated $\gamma$-ray luminosity only weakly dependent on
uncertainties from the incomplete physical modelling of feedback processes in
the cluster cores.

\subsection{Contribution of different  $\gamma$-ray emission processes}

\begin{figure}
\begin{minipage}{1.0\columnwidth}  
\includegraphics[width=1.0\columnwidth]{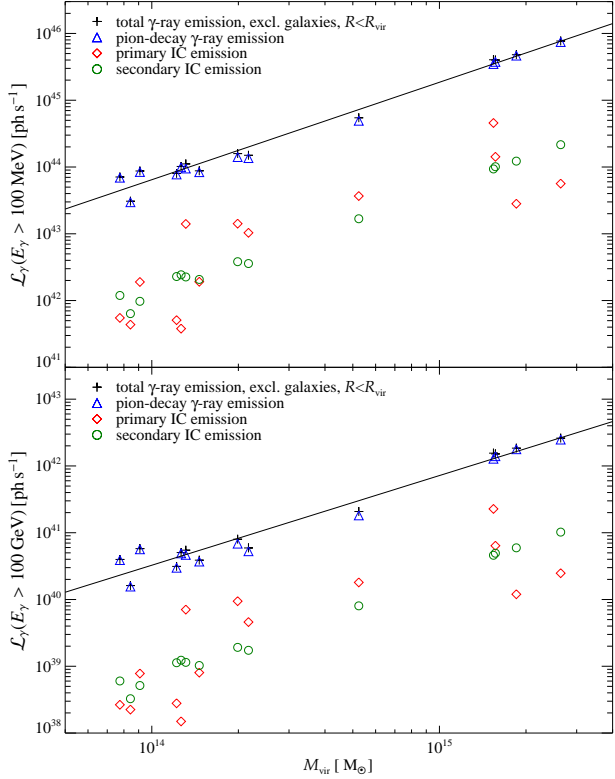}
\caption{Cluster scaling relations of the $\gamma$-ray luminosity for the energy
  regimes corresponding to the {\em Fermi $\gamma$-ray space telescope} and
  imaging air \v{C}erenkov telescopes.  Show is the contribution from individual
  $\gamma$-ray emission components within one virial radius (pIC - red squares,
  sIC - green circles, $\pi^0$ decay - blue triangles) to the total cluster
  luminosities (black crosses) and mass to luminosity scaling (black solid line)
  at the energies $E_\gamma > 100 \, \mev$ (upper panel) and $E_\gamma > 100 \,
  \gev$ (lower panel). \label{fig:ML_energy}}
 \end{minipage}
\end{figure}

Figure~\ref{fig:ML_energy} shows the scaling relations of the IC and pion decay
emission for two different energy scales of interest to the {\em Fermi
  $\gamma$-ray space telescope} and imaging air \v{C}erenkov telescopes. We
compare the total emission and the contribution from the individual emission
components of each cluster.  The very similar slopes of the mass-to-luminosity
scaling relation at both energies (Table~\ref{tab:scaling}) is a consequence of
the small variance in the proton spectrum (Fig~\ref{fig:proton_spectrum_all})
among different galaxy clusters. The individual emission processes also show
similar slopes, with small scatter for the pion decay emission and sIC
component. Contrary, the pIC emission has a larger scatter than the secondary
emission components due to the different dynamical states: the presence of
strong merger or accretion shocks is critical for the generation of primary CR
electrons and the associated radiative emission.

The ratio of the pion decay to the pIC emission in the 100~MeV and 100~GeV
regime is very similar. This is partly a coincidence and owed to our particular
choice of the two energy bands: the effective spectral index of the pion decay
between these two energies is flattened due the pion bump. It happens to be
similar to the power-law index of the pIC component which itself is unaffected
by the energy cutoff of the electron spectrum at these energies. If we chose a
smaller (larger) value than 100 MeV for the lower energy band, we would obtain a
lower (higher) pion-to-pIC ratio due to the steeper intrinsic spectrum of the CR
protons at lower energies (cf.~Fig.~\ref{fig:sketch_CR_gamma}).

\subsection{$\gamma$-ray emission from individual galaxies}

 \begin{figure}
   \begin{minipage}{1.0\columnwidth}  
 \includegraphics[width=1.0\columnwidth]{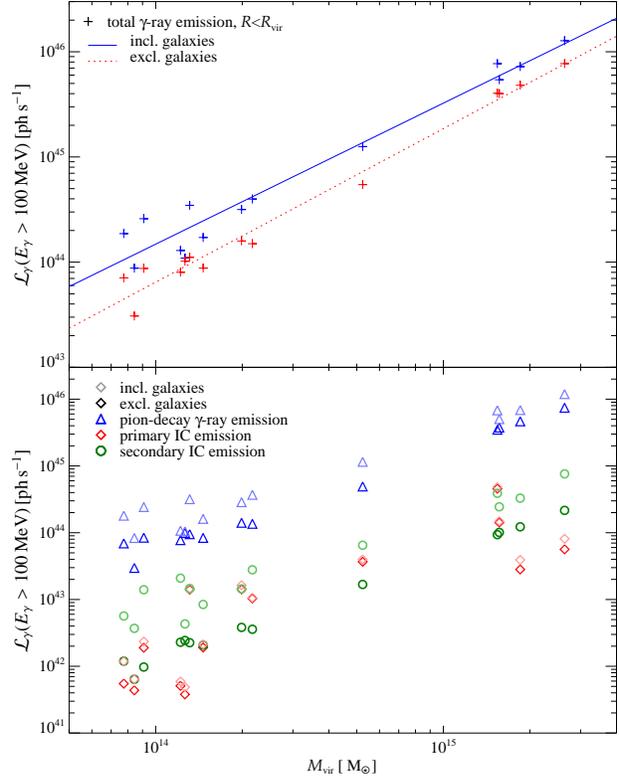}
 \caption{Studying the galaxy bias to cluster $\gamma$-ray scaling relations for
   energies $E_\gamma > 100 \, \mev$ and $R<R_\rmn{vir}$.  In the upper panel,
   the blue crosses show the cluster emission with galaxies and the red crosses
   show the cluster emission without galaxies, the solid lines show the mass to
   luminosity scaling (Table~\ref{tab:scaling}). The lower panel shows the
   contribution from individual components (primary IC -- red squares, secondary
   IC -- green circles, $\pi^0$ decay -- blue triangles) for emission including
   galaxies (light color) and excluding galaxies (dark color).}
       \label{fig:GalVsNoGal}
   \end{minipage}
 \end{figure}

We investigate the bias of galaxies to the scaling relation of the $\gamma$-ray
luminosity above 100~MeV in Fig.~\ref{fig:GalVsNoGal}. The top panel shows the
total $\gamma$-ray emission, where the presence of galaxies biases smaller mass
clusters slightly more compared to their larger analogues, with an average bias
of about a factor two across our cluster sample.  Masking galaxies reduces the
overall scatter in the scaling relations, particularly at low masses.  In the
lower panel, we show the contribution from individual emission components. The
largest bias originates from the pion decay (blue triangles) and the sIC
emission (green circles), while the pIC component (red diamonds) is barely
affected by this masking procedure since it does not scale with density.

\subsection{$\gamma$-ray emission from the cluster periphery and WHIM}

\begin{figure}
\begin{minipage}{1.0\columnwidth}
  \includegraphics[width=1.0\columnwidth]{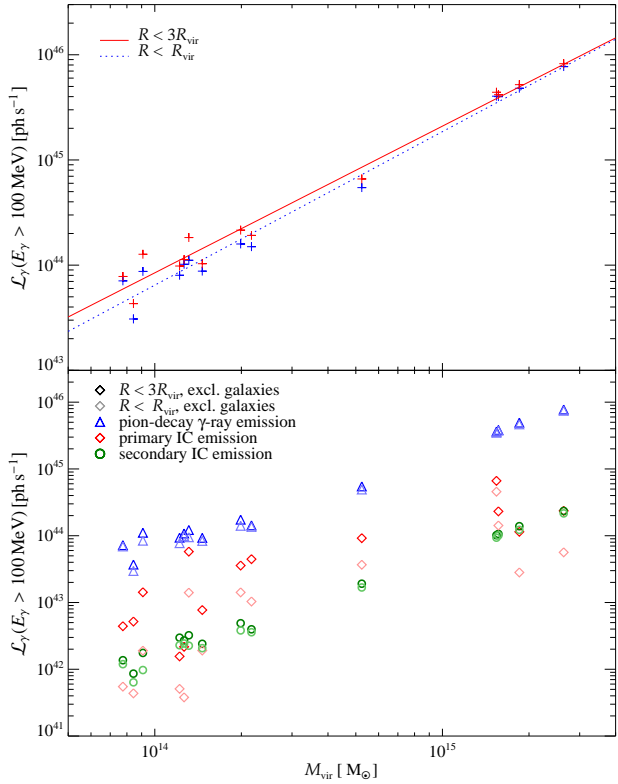}
  \caption{Influence of the emission from the accretion region around clusters
    on the mass to luminosity cluster scaling relations for $E_\gamma > 100 \,
    \mev$. In the upper panel, we show the luminosity integrated out to
    $3\,\rvir$ (red crosses) as well as the luminosity integrated out to $\rvir$
    (blue crosses). The solid lines indicate the mass to luminosity scaling
    relation found in Table~\ref{tab:scaling}. The lower panel shows the
    contribution from individual components (primary - red squares, secondary -
    green circles, $\pi^0$ decay - blue triangles) for luminosity integrated out
    to $\rvir$ (light color) and luminosity integrated out to $3\,\rvir$ (dark
    color). \label{fig:100gev3rvir}}
\end{minipage}
\end{figure}

In this section we study the dependence of the accretion region around clusters
on $\gamma$-ray scaling relations. Specifically, we compare the total
$\gamma$-ray emission within $\rvir$ to the emission within $3\,\rvir$ which
hosts the WHIM and individual satellite galaxies (and groups) that have not yet
accreted onto the cluster.

From Fig.~\ref{fig:100gev3rvir} it is clear that the total luminosity is
dominated from the region inside $R_\rmn{vir}$. For small clusters, the
flux-correction from the WHIM is of the order of 30 percent which the correction
is smaller for massive systems -- below 10 percent. This stems from the fact
that the emission of low mass systems with smaller potential wells is easier to
perturb -- through accreting clumps of matter or nearby satellite systems. The
pIC component contributes a factor $2-10$ more in the WHIM than within
$\rvir$. The reason being that especially for merging systems, the pIC profile
is rather flat (see Fig.~\ref{fig:emission_components_profiles}). Hence it
contributes substantially to WHIM luminosity, whereas the density dependent
secondary components in the WHIM are negligible due to the low gas
densities. Only satellite systems within the WHIM contribute at a low level to
the secondary components. This effect is especially pronounced when galaxies are
excluded and implies that the pIC component becomes comparable to the pion decay
emission for a few low mass clusters. Note that in the TeV regime, the pIC is
considerably suppressed due to the limited maximum energy of the primary CR
electrons which reduces that effect in the WHIM.

\begin{table*}
\caption{Cluster $\gamma$-ray scaling relations$^{(1)}$.}
\begin{tabular}{l c c  c c  c c}
\hline
\hline
&
\multicolumn{2}{c}{$\gamma$-rays ($E_\gamma > 100$~MeV):} &
\multicolumn{2}{c}{$\gamma$-rays ($E_\gamma > 1$~GeV):} &
\multicolumn{2}{c}{$\gamma$-rays ($E_\gamma > 100$~GeV):}\\
\hline
model$^{(2)}$
& $\LL_{\gamma,0}^{(3)}$ & $\beta_\gamma$
& $\LL_{\gamma,0}^{(4)}$ & $\beta_\gamma$
& $\LL_{\gamma,0}^{(5)}$ & $\beta_\gamma$ 
\\
\hline
Including galaxies, $1\,\rvir$  & $5.24 \pm 0.99$ & $1.34 \pm 0.09$ & $7.83 \pm 1.70$  
& $1.35 \pm 0.10$ & $2.78 \pm 0.71$ & $1.33 \pm 0.13$ \\ 
Including galaxies, $2\,\rvir$  & $5.73 \pm 0.97$ & $1.24 \pm 0.09$ & $8.48 \pm 1.64$  
& $1.24 \pm 0.10$ & $3.00 \pm 0.71$ & $1.23 \pm 0.12$ \\ 
Including galaxies, $3\,\rvir$  & $6.53 \pm 1.20$ & $1.14 \pm 0.08$ & $9.63 \pm 1.87$ 
& $1.15 \pm 0.09$ & $3.47 \pm 0.76$ & $1.14 \pm 0.11$ \\
Excluding galaxies, $1\,\rvir$  & $3.13 \pm 0.37$ & $1.46 \pm 0.06$ & $4.02 \pm 0.42$  
& $1.43 \pm 0.07$ & $1.16 \pm 0.11$ & $1.34 \pm 0.08$ \\ 
Excluding galaxies, $2\,\rvir$  & $3.32 \pm 0.54$ & $1.41 \pm 0.08$ & $4.22 \pm 0.53$  
& $1.39 \pm 0.06$ & $1.24 \pm 0.19$ & $1.31 \pm 0.08$ \\ 
Excluding galaxies, $3\,\rvir$  & $3.44 \pm 0.43$ & $1.39 \pm 0.06$ & $4.37 \pm 0.57$ 
& $1.37 \pm 0.07$ & $1.30 \pm 0.20$ & $1.29 \pm 0.08$ \\
\hline
\end{tabular}
\begin{quote} 
  Notes:\\ (1) The cluster $\gamma$-ray scaling relations are defined by $A =
  A_0\,M_{15}^{\beta}$, where $M_{15} = \mvir / (10^{15} \rmn{M}_\odot)$.
  \\ (2) The $\gamma$-ray luminosity from respective cluster is obtained by
  integrating over the region within an integer times $\rvir$. For scaling
  relations that include galaxies, we apply a central core cut-out and exclude a
  spherical region with radius $r < 0.025\, \rvir$ that is centered on the
  brightest central $\gamma$-ray point-source. \\ (3) The normalization of the
  $\gamma$-ray scaling relations ($E_\gamma>100$~MeV) is given in units of
  $10^{45}\,\rmn{ph}\mbox{ s}^{-1}$. \\ (4) The normalization of the
  $\gamma$-ray scaling relations ($E_\gamma>1$~GeV) is given in units of
  $10^{44}\,\rmn{ph}\mbox{ s}^{-1}$. \\ (5) The normalization of the
  $\gamma$-ray scaling relations ($E_\gamma>100$~GeV) is given in units of
  $10^{42}\,\rmn{ph}\mbox{ s}^{-1}$. \\
\label{tab:scaling}
\end{quote}
\end{table*}

\section{Prediction of the $\gamma$-ray emission from nearby galaxy clusters}
\label{sect:prediction}

\begin{figure*}
\begin{minipage}{2.0\columnwidth}
  \includegraphics[width=1.0\columnwidth]{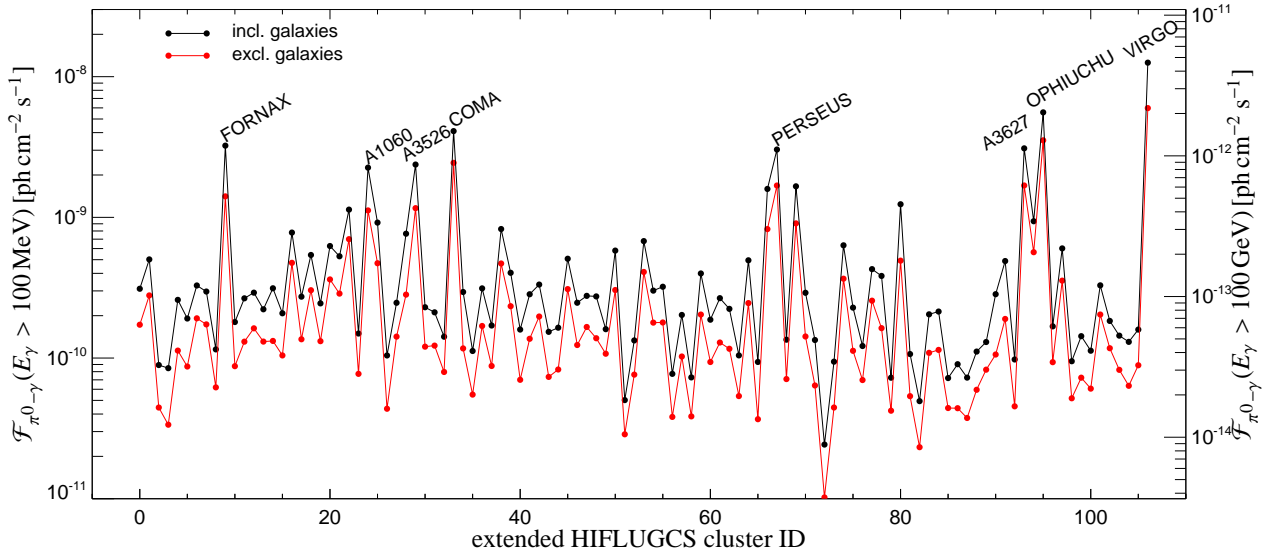}
  \caption{Predicted $\gamma$-ray flux in clusters and groups in the extended
    HIFLUGCS catalog to which we also add the Virgo cluster. For each cluster
    and group, we account for the flux from the region within one virial
    radius. The left axis shows the flux above 100 MeV while the right axis
    accounts for the flux above 100 GeV. The black line refers to our optimistic
    model where we include the flux contribution from galaxies and the red line
    shows the flux without galaxies (cf. Table~\ref{tab:scaling}). We name the
    clusters and groups with $\mathcal{F}_{\pig}(E_\gamma > 100\,\rmn{MeV}) >
    2\times10^{-9}\, \rmn{ph}\,\rmn{cm}^{-2}\,\rmn{s}^{-1}$ in our optimistic
    model which roughly corresponds to the sensitivity of the Fermi all-sky
    survey after two years of data taking.}
    \label{fig:Flux_HIFLUGCS}
\end{minipage}
\end{figure*}

\begin{figure*}
\begin{minipage}{2.0\columnwidth}
  \includegraphics[width=1.0\columnwidth]{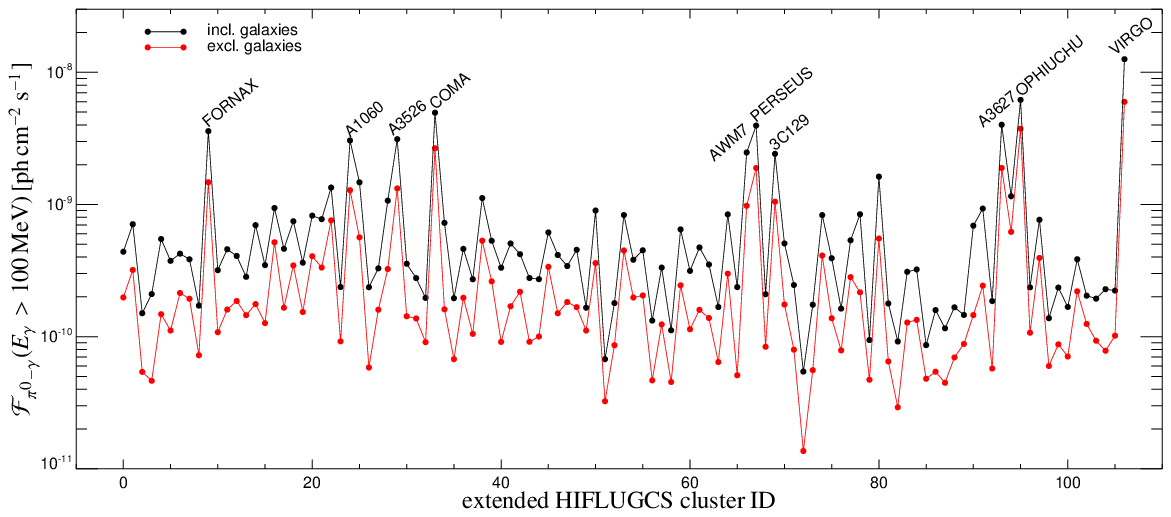}
  \caption{Predicted $\gamma$-ray flux above 100 MeV in clusters and groups in
    the extended HIFLUGCS catalog to which we also add the Virgo cluster. The
    flux comes from the region within the Fermi angular resolution at 100 MeV,
    i.e.{\ }a circular region of radius 3.5 degree that contains 68 per cent of
    the PSF, but with the limit at $3\,\rvir$ for each cluster and group. The
    black line refers to our optimistic model where we include the flux
    contribution from galaxies and the red line shows the flux without galaxies
    (cf. Table~\ref{tab:scaling}). We name the clusters and groups with
    $\mathcal{F}_{\pig}(E_\gamma > 100\,\rmn{MeV}) > 2\times10^{-9}\,
    \rmn{ph}\,\rmn{cm}^{-2}\,\rmn{s}^{-1}$ in our optimistic model which
    roughly corresponds to the sensitivity of the Fermi all-sky survey after two
    years of data taking.}
    \label{fig:Flux_Fermi_HIFLUGCS}
\end{minipage}
\end{figure*}

We use the mass-to-luminosity scaling relations as derived in
Section~\ref{sect:scaling} in combinations with the virial masses of galaxy
clusters of the extended HIFLUGCS catalogue \citep{2002ApJ...567..716R} -- the
``HIghest X-ray FLUx Galaxy Cluster Sample'' from the ROSAT all-sky survey -- to
predict their $\gamma$-ray emission.\footnote{We have also added the Virgo
  cluster to the sample that we nevertheless refer to as extended HIFLUGCS
  catalogue in the following.}  In Fig.~\ref{fig:Flux_HIFLUGCS}, we show the
$\gamma$-ray flux for radii $R<\rvir$ and energies above 100~MeV and 100~GeV, as
a function of the identifier (ID) in the extended HIFLUGCS catalogue. We
specifically name those clusters that have a flux (in our optimistic model)
which is larger than the sensitivity of the Fermi all-sky survey after two years
of data taking.  We find that the {\em brightest clusters in $\gamma$-rays are
  Virgo, Ophiuchus, Coma, Perseus and Fornax}. This result agrees well with
previous $\gamma$-ray studies using galaxy clusters from the HIFLUGCS sample
\citep{2008MNRAS.385.1242P, 2009PhRvD..80b3005J}.

Figure~\ref{fig:Flux_HIFLUGCS} should serve as a starting point to identify
promising sources for the $\gamma$-ray experiment in question. In a realistic
setting, we would have to include the instrumental response and the point-spread
function to obtain the predicted detection significance for the model in
question. We note that this procedure of using scaling relations does not take
into account deviations of individual systems from the mean $\gamma$-ray flux at
a given cluster mass. One would rather have to model each system separately
along the lines presented in Section~\ref{sect:analytic_model}.

To address the effect of source extension on the detection significance of
Fermi, we compute the $\gamma$-ray flux of each cluster within the Fermi
equivalent angular resolution at 100~MeV in
Fig.~\ref{fig:Flux_Fermi_HIFLUGCS}. To this end, we interpolate the scaling
relations in Table~\ref{tab:scaling} to the radius corresponding to the angular
resolution of 3.5~deg.  We limit the size of each source to $3\,\rvir$ since
there is negligible additional flux beyond this radius.  For most clusters the
flux is very similar to what we found in Fig.~\ref{fig:Flux_HIFLUGCS} because of
the similar mass to luminosity scaling relations for $\rvir$ and $3\,\rvir$.

\section{Discussion and comparison to previous work}
\label{sect:discussion_comparison}
\subsection{Comparison to previous work on $\gamma$-ray emission from clusters}
\label{sect:comparison}

In support of the new instrumental capabilities in $\gamma$-ray astronomy,
several pioneering papers have appeared that simulate the high-energy
$\gamma$-ray emission from clusters. Here we make a comparison to some of those
papers.

{\bf Comparison to Pfrommer et al.} -- In the series of papers
\citet{2007MNRAS.378..385P, 2008MNRAS.385.1211P} and \citet{2008MNRAS.385.1242P}
simulate the same cluster sample as we do. In fact, our work represents an
extension of these earlier works.  Overall, our results are in continuity with
their results. The differences emerge from the details of the CR physics, where
they adopted a simplified description using a single CR population with a
spectral index of 2.3. Hence they concentrated on the non-thermal radio emission
as well as the $\gamma$-ray emission at energies $E_\gamma>100$~MeV that depend
only weakly on the particular value or even a running of the CR spectral index
(as long as it is close to the true one).  For the primary electron populations,
a maximum electron injection efficiency of $\zeta_\e=0.05$ was assumed. In
addition, the bias from anomalous galaxies was not addressed.
 
Comparing profiles of the total surface brightness above 100~MeV in our
optimistic model with galaxies, shown in grey in
Fig.~\ref{fig:emission_components_profiles}, to the brightness profiles in
\citet{2008MNRAS.385.1211P}, we find only very small differences. These
differences are caused by a combination of different binning and Poisson noise
in the galaxies' spatial distribution which have a different realization due to
our slightly modified CR description that changes the hydrodynamics.  The
largest difference is seen for the pIC component in the periphery of merging
clusters and a consequence of the different values for the injection efficiency
$\zeta_\e$ adopted.  In addition, we compare the mass to luminosity scaling
relation above 100~MeV in our optimistic model (see Table~\ref{tab:scaling}) and
find that they agree to the percent level with what was found in
\citet{2008MNRAS.385.1242P}.

{\bf Comparison to Miniati et al.} -- There have been a series of pioneering
papers simulating the non-thermal emission from clusters by numerically
modelling discretized CR proton and electron spectra on top of Eulerian
grid-based cosmological simulations \citep{2001CoPhC.141...17M,
  2001ApJ...559...59M, 2001ApJ...562..233M, 2002MNRAS.337..199M,
  2003MNRAS.342.1009M}. In contrast to our approach, these models neglected the
hydrodynamic pressure of the CR component, were quite limited in their adaptive
resolution capability, and they neglected dissipative gas physics including
radiative cooling, star formation, and supernova feedback.  Comparing the
$\gamma$-ray emission characteristics of the IC emission from primary CR
electrons and hadronically generated secondary CR electrons as well as the pion
decay $\gamma$-rays, we confirm the qualitative picture of the emission
characteristics of the different $\gamma$-ray components put forward by these
authors. However, we find important differences on smaller scales especially in
cluster cores, the emission strength of the individual components and their
spectra.

We confirm that the high-energy $\gamma$-ray emission ($E_\gamma > 100$~MeV)
from cluster cores is dominated by pion decays while at lower energies, the IC
emission of secondary CR electrons takes over \citep{2003MNRAS.342.1009M}.  We
reproduce their finding that the $\gamma$-ray emission in the virial regions of
clusters and beyond in super-cluster regions is very inhomogeneous and stems in
part from the IC emission of primary shock accelerated electrons. Contrarily to
these authors, we find that the surface brightness of this emission component
remains sub-dominant in projection compared to the hadronically induced emission
components in the cluster core and that the pion decay completely dominates the
high-energy $\gamma$-ray emission of clusters above a few~MeV
(cf. Fig.~\ref{fig:sketch_CR_gamma}).  In addition we predict a pIC spectrum
that is somewhat steeper with a photon index of $\Gamma\simeq 1.15$ which
resembles a steeper primary electron spectrum in our simulations compared to
theirs. This points to on average weaker shocks that are responsible for the
acceleration of primary CR electrons that dominate the pIC emission. This
discrepancy of the pIC spectral index causes the discrepancy of the pIC flux at
high $\gamma$-ray energies.

In the WHIM we find that the pIC emission dominates the total $\gamma$-ray
spectrum below about 100~MeV, and a comparable flux level of $\pi^0$-decay and
pIC between 100~MeV and 1~TeV, where the $\pi^0$-decay takes over. This is in
stark contrast to the finding of \citet{2003MNRAS.342.1009M}, where the pIC
is dominating the pion decay emission by a factor of about 10 or more over the
entire $\gamma$-ray energy band.  We note that our $\gamma$-ray fluxes from
clusters are typically a factor of two smaller than the estimates given in
\citet{2001ApJ...559...59M} which has important implications for the
detectability of clusters by Fermi.

There are several factors contributing to the mentioned discrepancies. (1) Our
simulations are Lagrangian in nature and hence adaptively resolve denser
structures with a peak resolution of $5\,h^{-1}$~kpc. In contrast, the
cosmological simulations of \citet{2003MNRAS.342.1009M} have a fixed spatial
resolution of $\sim100\,h^{-1}$~kpc which is too coarse to resolve the
observationally accessible, dense central regions of clusters in this grid-based
approach and underestimates CR cooling processes such as Coulomb and hadronic
losses.  It also cannot resolve the adiabatic compression of CRs into the core.
(2) \citet{2000ApJ...542..608M} identified shocks with Mach numbers in the range
$4 \lesssim \M \lesssim 5$ as the most important in thermalizing the plasma. In
contrast, \citet{2003ApJ...593..599R}, \citet{2006MNRAS.367..113P}, and
\citet{2008ApJ...689.1063S} found that the Mach number distribution peaks in the
range $1 \lesssim \M \lesssim 3$. This finding seems to be robust as different
computational methods have been used which range from fixed and adaptive
Eulerian grid codes to Lagrangian Tree-SPH codes.  Since diffusive shock
acceleration of CRs depends sensitively on the Mach number, this implies a more
efficient CR injection in the simulations by \citet{2001ApJ...559...59M}.  It
also results in a flatter CR electron and CR ion spectrum compared to ours shown
in Fig.~\ref{fig:sketch_CR_gamma}. Hence, the pIC emission of
\citet{2003MNRAS.342.1009M} has a flatter photon index and a boosted flux.  (3)
For the CR ion spectrum, \citet{2003MNRAS.342.1009M} uses only four momentum
bins which is not enough to resolve the pion bump accurately. The large pion
decay plateau which he found indicates a constant CR ion spectral index in this
energy range. This is in contradiction to the concavely shaped CR spectrum that
our cluster simulations show, where the shape is a consequence of the Mach
number statistics and the adiabatic transport. The difference in the CR spectral
shape is especially important for CR energies above 1~GeV, since those CRs give
rise to the $\pi^0$-decay emission at energies above the pion bump.

{\bf Comparison to Kushnir et al.} -- \citet{Kushnir:2009vm} use a simple
analytic model to follow the evolution of ICM CRs, accelerated in strong
accretion shocks. Interestingly, their approach predicts similar characteristics
for the pion decay emission, in particular its flux agrees with our prediction
within a factor two. In contrast, their model predicts a high-energy
$\gamma$-ray flux of the pIC component that is approximately a factor of $600$
larger than ours due to a different spectral description where they adopted
a spectral index of 2. This, however, is in conflict with our simulated average
spectral index of 2.3 that is a consequence of cluster assembly history (compare
our Fig.~\ref{fig:emission_components_profiles} with Fig. 3 in
\citealt{Kushnir:2009vm}). This finding, together with different adopted values
for the electron injection efficiency and CR-to-thermal ratio, led them to the
contradicting conclusion that the expected overall $\gamma$-ray emission would
be much more extended. We believe that their model would over-predict the amount
of observed radio relic emission in clusters, in particular when considering
magnetic field amplification at accretion shocks on a level that is only a
fraction of what is observed in supernova remnant shocks
\citep{2007Natur.449..576U}. It can be easily seen that the different power-law
indices are indeed the reason for the flux discrepancies by comparing the energy
flux of electrons at 170 GeV which are responsible for IC photons at 100 MeV
(assuming the up-scattering of CMB photons). Adopting our effective spectral
injection index of primary electrons of $\alpha_{\e,\rmn{inj}}\simeq 2.3$ and
assuming a post-shock temperature at a typical accretion shock of $kT\simeq
0.1$~keV, we find a flux ratio between their model and ours of $(0.1
\,\rmn{keV}/170 \,\rmn{GeV})^{-0.3}\simeq 600$.

We would like to compare their model predictions for the pion decay emission in
more detail. To this end, we use a proton injection efficiency in their model of
$\eta_\p\simeq 0.2$ for the comparison. In our simulations we use a maximum
proton injection efficiency of $\eta_\rmn{max,p}=0.5$. The proton injection
efficiency in our simulations is dynamical, and depends on the strength of the
shocks. Since the cluster evolves with time and the majority of CRs are injected
at higher redshift during the formation of clusters, the final CR pressure
depends on the interesting interplay of the actual value of the shock injection
efficiency and the successive CR transport. Using results from
\citep{2007A&A...473...41E}, we estimate an effective injection efficiency of
approximately $\eta_\p\simeq0.2$. We note, however, that their baseline model
assumes $\eta_\p=0.02$, which will suppress the flux by a factor 10 in their
model.  We now contrast the $\gamma$-ray flux predictions of the Coma and
Perseus cluster of our semi-analytic model to the one worked out in
\citet{Kushnir:2009vm}.  First we study the $\gamma$-ray flux from Coma within
$R_\theta=\rvir$ above 100~MeV where we find with their model a flux
$\mathcal{F}_\rmn{\pig,k\&w}(>100\,\mev) = (0.70-1.7)\times10^{-9}\,
\rmn{ph}\,\rmn{cm}^{-2}\,\rmn{s}^{-1}$. This flux is only a factor two lower
than what we predict for the pion decay emission from Coma. Turning to pIC,
which play a much more important role for the total $\gamma$-ray flux than the
$\pi^0$-decay emission in their model, results in the flux
$\mathcal{F}_\rmn{pIC, k\&w}(>100\,\mev)\simeq
1.5\times10^{-8}\,\rmn{ph}\,\rmn{cm}^{-2}\,\rmn{s}^{-1}$. This is only a factor
two below the EGRET upper limit and can be readily tested with the one-year data
from Fermi.

Studying fluxes in the TeV $\gamma$-ray regime is also of great importance since
it compares the spectral representation of the models. Using the analytic model
of \citet{Kushnir:2009vm}, results in the flux $\mathcal{F}_\rmn{\pig,
  coma}(>1\,\tev)=9.3\times10^{-14}\,\rmn{ph}\,\rmn{cm}^{-2}\,\rmn{s}^{-1}$
within $\theta=0.2$~deg that is about a factor 10 below the upper limit for Coma
set by HESS and only a factor two larger than the flux we predict for Coma.
However, their result is most probably flawed by their too simplistic assumption
for the CR spectral index of $\alpha=2$. If we use our universal concave shaped
spectrum instead of their flat CR spectrum, we can show that their flux would
decrease by a factor
$\sim\left(1\,\tev/100\,\mev\right)^{(\Gamma_{100\,\mev}^{1\,\tev}-1)}\sim10^{4\times0.24}\sim9$.

Finally we study the surface brightness profiles from Coma and Perseus predicted
by our semi-analytic model and compare it to theirs. The dashed and
dotted-dashed lines in Fig.~\ref{fig:coma.perseus_comparison} show their
predictions for the Coma cluster above the energy 100~MeV and 100~GeV,
respectively. Using their formalism we find a surface brightness above 100~MeV
that is about a factor two smaller than what we predict. However, above 100~GeV
our predictions are in better agreement.

\subsection{Limitations and future work}
\label{sect:limitations}

The ideal CR formalism would trace the spectral energy evolution, as well as the
spatial evolution of CRs, and at the same time time keep track of the dynamical
non-liner coupling with magneto-hydrodynamics. In order to make cosmological
simulations less expensive in computational power, we are forced to make
compromises. The simplifying assumptions chosen, enable us to run cosmological
simulations of the formation of galaxy clusters with the necessary resolution to
resolve their cores. At the same time, these assumptions enable us to follow the
CR physics self-consistently on top of the radiative gas physics. Here we
outline our most severe limitations for computing the $\gamma$-ray emission from
clusters.

\begin{enumerate}
\item In our simulations we neglect the effect of microscopic CR diffusion and
  CR streaming. The collisionless plasma forces CRs to stay predominantly on a
  given field line and to diffuse along it. The random walk of field lines cause
  initially closely confined CRs to be transported to larger scales which can be
  described as a diffusion process. In our model we assume the magnetic field to
  be tangled on scales smaller than those we are interested in, $\lambda
  \sim10$~kpc in the center and even larger scales outside. Hence, CRs are
  magnetically coupled to the thermal gas and advected alongside it. The
  diffusivity can be rewritten into a macroscopic advection term that we fully
  resolve in our Lagrangian SPH simulations by construction and a microscopic
  diffusivity.  The advection term dominates over microscopic term, as the
  following estimate for the diffusivities shows: $\kappa_\rmn{adv} \simeq 100
  \mbox{ kpc}\times 1000 \mbox{ km/s} \simeq 10^{31.5} \mbox{ cm}^2/\mbox{s} \gg
  \kappa_\rmn{diff} \simeq 10^{29} \mbox{ cm}^2/\mbox{s}$. Further work is
  needed to study microscopic anisotropic diffusion, in combination with
  self-consistent modelling of the magnetic fields.
\item We also did not account for the injection of CRs by AGN or supernova
  remnants where the additional CRs would diffuses out of AGN-inflated bubbles
  or drive starburst winds that enrich the IGM. In addition we do not account
  for the feedback processes by AGN despite their importance for understanding
  the nature of the very X-ray luminous cool cores found in many clusters of
  galaxies. For further details we refer the reader to
  \citet{2008MNRAS.387.1403S}.
\item We postpone the study of the potential contribution of a population of
  re-accelerated electrons to the IC $\gamma$-ray emission throughout this work:
  strong merger shocks and shear motions at the cluster periphery might inject
  hydrodynamic turbulence that cascades to smaller scales, feeds the MHD
  turbulence and eventually might be able to re-accelerate an aged CR electron
  population. Due to non-locality and intermittency of turbulence, this could
  partly smooth the very inhomogeneous primary emission component predominantly
  in the virial regions of clusters where simulations indicate a higher energy
  density in random motions. However, to study these effects, high-resolution
  AMR simulations are required that refine not only on the mass but also on some
  tracer for turbulence such as the dimensionless vorticity parameter
  \citep{2008MNRAS.388.1079I, 2008MNRAS.388.1089I, 2009ApJ...707...40M}.
\item Our model for the diffusive shock acceleration assumes a featureless
  power-law for both, the proton and the electron acceleration, that is injected
  from the thermal distribution. The complete theoretical understanding of this
  mechanism is currently an active research topic that includes non-linear
  effects and magnetic field amplification \citep{2006ApJ...652.1246V}.
  Phenomenologically, we believe that there are strong indications for the
  diffusive shock acceleration mechanism to be at work which come from
  observations of supernova remnants over a wide range of wavelengths from the
  radio, X-rays into the TeV $\gamma$-rays \citep[e.g.,][]{2000AIPC..528..383E,
    2000ApJ...543L..61H, 2005ICRC....3..261E, 2005ApJ...634..376W,
    2004Natur.432...75A, 2006Natur.439..695A} as well as the bow shock of the
  Earth \citep{1990ApJ...352..376E, 1999Ap&SS.264..481S}. Theoretical work
  suggests that the spectrum of CRs which is injected at strong shocks shows an
  intrinsic concave curvature: the feedback of the freshly accelerated and
  dynamically important CR pressure to the shock structure results in a weaker
  sub-shock that is proceeded by a smooth CR precursor extending into the
  upstream. Hence low-energy protons are only shock-compressed at the weaker
  sub-shock and experience a smaller density jump which results in a steeper
  low-energy spectrum (compared to the canonical value $\alpha=2$ from linear
  theory). In contrast, the Larmor radii of high-energy protons also sample the
  CR precursor and experience a much larger density contrast that results in a
  flatter high-energy spectrum with $\alpha<2$ \citep{1984ApJ...286..691E,
    2008MNRAS.385.1946A, 2009arXiv0912.2714C}. The low-energy part of the CR
  spectrum in clusters (as found in this work) should be unaffected since a
  softer population of CRs dominate there with $\alpha \sim 2.5$ and
  non-linear effects are presumably negligible in this regime. However the
  high-energy part of the CR spectrum in clusters could become harder compared
  to what we found due to these non-linear effects. Future work will be
  dedicated to improve our model and to incorporate more elaborate plasma
  physical models and to study the uncertainty of our results with respect to
  the saturated value of our CR acceleration efficiency
  \citep[e.g.,][]{2007APh....28..232K, 2007arXiv0706.0587E}. 
  \item An artificial surface tension effect limits the ability of SPH (in its
    standard conservative form) to follow the growth of boundary instabilities
    such as the Kelvin-Helmholtz instability at the interface of a dense and
    under-dense phase accurately, i.e. on the predicted linear growth time
    \citep{2007MNRAS.380..963A}. In the context of galaxy cluster simulations,
    this only occurs at the interface of the ISM of individual galaxies and the
    ICM. This causes an unphysically long survival time of dense
    gaseous point sources after they got ram pressure/tidally stripped from
    their galactic halo -- for simplicity, we call them "galaxies" and describe
    the physics in detail in Section~\ref{sect:galaxies}. In our paper, we
    decided to show our result for an optimistic model that includes all
    galaxies and one conservative model that cuts all galaxies. This is meant to
    bracket the realistic case. Also, the main result is based on our
    conservative model where we cut out the galaxies. In this way we circumvent
    these issues.
\end{enumerate}

\subsection{Impact of these limitations on our results}

Generally, we acknowledge that the CR spatial distribution is more uncertain
than the spectrum due to the details outlined in
Section~\ref{sect:limitations}. Below, we detail our considerations why we
believe that the spectrum that we found is robust even when considering
uncertainties such as additional CRs injected from AGN, re-acceleration of CRs
at MHD turbulence, CR diffusion, and non-linear shock acceleration. We outline
here the reasons in detail:

{\bf Additional CRs injected from AGN.} It is very uncertain whether AGN jets
are powered hadronically or though Poynting flux
\citep[e.g.,][]{1998MNRAS.293..288C, 1998ApJ...497..563H, 2000ApJ...534..109S}.
Irrespectively, the energetics of AGN are insufficient to account for a majority
of CRs in clusters -- in particular for the most massive systems (Thompson \&
Pfrommer in prep.).
  
{\bf CR re-acceleration through MHD turbulence.} The involved physics is
currently very uncertain such as the level and nature of turbulence in the ICM,
how CRs exactly interact with plasma waves, how efficient this accelerates CRs,
and whether a power-law extrapolation between the gyro radius of a CR, $R \sim
10^{-5} \,\rmn{pc}\,(B/\mug)^{-1}\,(E/10\,\gev)$, and the scales accessible to
current simulations with peak resolution of a few kpc is justified. Hence it
appears that is is impossible to constrain the impact of turbulent
re-acceleration on the CR spectrum in clusters self-consistently from first
principles. However, in our Milky Way, we are able to understand the observed CR
spectrum on Earth with $f_\CR \sim p^{-2.7}$ fairly well in terms of injection
and transport. CRs with an injected spectrum of $p^{-(2.3-2.4)}$ experience
momentum dependent diffusion so that the more energetic particles can leave the
system in a so-called `leaky-box model'; an effect that accounts for the
observed steepening \citep{2002cra..book.....S}. This leaves little room for
spectral modifications through turbulent re-acceleration. Hypothesizing that the
fundamental interactions of plasma waves with particles should not be very
different in the ISM and ICM, we believe that we are safe to neglect this
process to first order. Note, however, that this argument neglects possible
important CR transport processes that might become important in the cluster
environment due to the much longer CR life time compared to the ISM in our
Galaxy.

{\bf Spatial diffusion of CRs.} If spatial CR diffusion is momentum dependent it
will introduce a radial dependence in the shape of the CR spectrum since high
energy CRs can diffuse out of the central regions faster.  This, in turn, will
affect the observed morphology and spectrum of all relevant non-thermal emission
components, including gamma-rays from pion decay, IC from secondaries, and
synchrotron emission.

{\bf Non-linear shock acceleration.} Since the CR spectrum at GeV-to-TeV
energies is sufficiently steep, intermediate Mach number shocks are responsible
for the acceleration of these CRs -- with efficiencies there are modest (not in
the saturated regime) so that the non-linear back-reaction is expected to be
small or even negligible.

{\bf Over-cooling problem.} Due to the over-cooling problem, the modeling of
cluster cores with radiative cooling but without any significant feedback
process produces an unphysically high stellar mass fraction. This is not
expected to impact the CR spectrum significantly. However, the effect on the CR
morphology might be substantial and might depend on the details of the required
feedback process. This can potentially impact the non-thermal cluster
observables. Future work is needed to solve this problem.

\section{Conclusions}
\label{sect:conclusions}

In this paper we have simulated 14 galaxy clusters spanning two orders of
magnitude in mass and a broad range of dynamical stages. The simulations follow
self-consistent CR physics on top of the dissipative gas physics including
radiative cooling and star formation. We have simulated high-energy $\gamma$-ray
emission maps, profiles and spectra of various emission components. These
include the inverse Compton emission from primary, shock-accelerated electrons
(pIC) and secondary electrons that result from hadronic interactions of CR
protons with ambient gas protons (sIC), as well as $\gamma$-rays from neutral
pion decay that are also generated in these hadronic reactions. 

We would like to emphasize that we focus on the intrinsic spectrum emitted at
the cluster position without taking into account photon propagation effects to
highlight the various physical process that shape the emission
spectra. Depending on the cluster redshift, these spectra attain a high-energy
cutoff due to $e^+e^-$-pair production on IR and optical photons which can be
easily derived from the photon-photon opacity
\citep[e.g.,][]{2008A&A...487..837F}. We also caution the reader that we assume
an optimistic value for the maximum shock injection efficiency (based on data
from supernova remnant studies by \citealt{2009Sci...325..719H}); smaller values
would reduce the resulting $\gamma$-ray emission accordingly. To date it is not
clear whether these high efficiencies apply in an average sense to strong
collisionless shocks or whether they are realized for structure formation shocks
at higher redshifts. Hence the goal of this work is to establish a thorough
framework and to predict the level of $\gamma$-ray emission that we expect for
this efficiency. We note that one cannot lower the acceleration efficiency
infinitely if one wants to explain radio (mini-)halos in the hadronic model of
CR interactions.  For clusters that host such a large, unpolarized, and
centrally peaked radio halo emission that resembles the thermal X-ray surface
brightness, one can derive a minimum $\gamma$-ray flux. The idea is based on the
fact that a steady state distribution of CR electrons loses all its energy to
synchrotron radiation for strong magnetic fields ($B \gg B_\rmn{CMB} \simeq 3.2
\mu\rmn{G}\times(1+z)^2$) so that the ratio of $\gamma$-ray to synchrotron flux
becomes independent of the spatial distribution of CRs and thermal gas.
Lowering the magnetic field would require an increase in the energy density of
CR electrons to reproduce the observed synchrotron luminosity and thus increase
the associated $\gamma$-ray flux \citep[for applications to Coma and Perseus,
see][ respectively]{2008MNRAS.385.1242P, 2009arXiv0909.3267T}.

According to our simulations, clusters have very similar morphology in the
$\mbox{100 \, MeV - 100 \, GeV}$ Fermi band, and in the $\mbox{100 \, GeV - 10
  \, TeV}$ \v{C}erenkov band.  This is due to the power-law spectra of the
dominating pion decay emission (which show a slowly running spectral index) and
ultimately inherited by the parent CR proton distribution.  The emission from
the central parts of clusters are dominated by $\gamma$-rays from pion decay,
while the periphery of the ICM and the WHIM have a considerable contribution
from pIC, which is especially pronounced in merging clusters.  The energy
dependent photon index for 100~MeV to 1~GeV has a median value of $\Gamma = 0.9$
due to pion decay induced emission in the central parts of the clusters, while
that in the periphery shows a slightly higher value of $\Gamma = 1.1$ which is
due to the substantial contribution from pIC. In the energy range from 100~GeV
to 1~TeV, the photon index steepens to $\Gamma = 1.25$ in the central
regions. This spectral steepening in the cluster center is due to the convex
curvature of the pion bump around 100~MeV causing a steepening in the asymptotic
$\gamma$-ray spectrum at higher energies. The small concave curvature at higher
energies is not able to compensate for this effect. At energies $E_\gamma\gtrsim
1$~TeV, the photon index in the cluster outskirts attain a much higher value due
to a super-exponential cutoff of the primary IC spectrum. This emission
component contributes substantially to the total $\gamma$-ray emission there. At
these energies, the electron cooling time is smaller than the time scale for
diffusive shock acceleration which causes this cutoff in the electron
spectrum. We used a semi-analytic formula for the injected electrons from which
we derive the cooled electron distribution with the characteristic
super-exponential cutoff. The shape of this spectrum is passed on to the pIC
spectrum and we capture this shape with a fit that is valid both in the
low-energy Thomson regime as well as in the high-energy Klein-Nishina regime.

The simulated CR proton spectra show an approximate power-law in momentum with a
few additional features; a cutoff at $p=P_\p/\mpcc \simeq 0.1$, a concave shape
between $p \simeq1-10^6$, and a steepening by $p^{1/3}$ between
$p\simeq10^6-10^8$. The overall shape of the spectrum shows only little variance
between the clusters, indicating a universal CR spectrum of galaxy clusters. The
radial dependence of the spectrum within the virial radius is negligible to
first order.  This allowed us to construct a semi-analytic model of the median
CR proton spectrum across our cluster sample. Using the semi-analytic CR
spectrum we derive a semi-analytic formula for the $\gamma$-ray flux from the
pion decay induced emission that dominate the total $\gamma$-ray spectrum above
100~MeV. We apply this formalism to the Perseus and Coma clusters, using their
density profiles as inferred from X-ray measurements and predict that the flux
from Perseus is close to the recent upper limits obtained by the MAGIC
collaboration \citep{2009arXiv0909.3267T}.

The mass-to-luminosity scaling for the 100~MeV, 1~GeV, and 100~GeV regimes show
very similar slopes for both the total $\gamma$-ray luminosity and all the
components, which is due to the small variance in the CR spectrum.  Masking
galaxies decreases the total $\gamma$-ray emission by a factor of 2-3.  The cut
has a larger effect on smaller mass clusters since the emission of low mass
systems with smaller potential wells are easier to perturb -- through accreting
clumps of matter or nearby satellite systems.  We also found that the presence
of galaxies considerably increases the scatter in the $\gamma$-ray scaling
relation. The region outside $R_\rmn{vir}$ only contributes marginally (of order
ten per cent) to the total $\gamma$-ray emission for massive clusters while it
contributes significantly to the total $\gamma$-ray luminosity of low mass
clusters with a factor $\lesssim 1.5$. This is again mostly due to the pion decay
emission from satellite systems that have not yet accreted on the cluster.  The
flux of the pIC component is increased by a factor of $2-10$ when the WHIM is
included. This can be explained by the rather flat spatial profiles of the pIC
emission.

Combining our $\gamma$-ray scaling relations with the virial masses of galaxy
clusters of the extended HIFLUGCS catalogue, we predict a detection of a few
galaxy clusters above 100~MeV with {\em Fermi} after two years, where Virgo,
Ophiuchus, Coma, Perseus and Fornax are expected to be the brightest clusters in
$\gamma$-rays (barring uncertainties in the injection efficiency). Since Fermi
already discovered the central AGN in Virgo/M87 and Perseus/NGC1275 the
detection of the somewhat more extended and dimmer pion decay component will be
very challenging in these clusters and requires careful variability studies to
subtract the AGN component.  For energies above 100~GeV, the flux of these
clusters as determined by our scaling relation is more than $5 \times 10^3$
times lower. This provides a challenge for current \v{C}erenkov telescopes as it
is almost an order magnitude lower than the 50~h sensitivities. However, future
upgrades of IACTs or the CTA telescope might considerably change the
expectations. We note however that these estimates are too conservative for cool
core clusters, which are known to show enhanced X-ray fluxes by a factor of up
to ten relative to clusters on the X-ray luminosity scaling relation. Since we
expect the X-ray luminosity to tightly correlate with the $\gamma$-ray
luminosity, this sub-class of clusters should provide very rewarding targets due
to the ample target matter for inelastic collisions of relativistic protons
leading to $\gamma$-rays. Applying our semi-analytic model for the $\gamma$-ray
emission, we identify Perseus among the best suited clusters to target for the
current IACT experiments.

\section*{Acknowledgments}
We thank our referee for a constructive report. 
We gratefully acknowledge the great atmosphere at the Kavli Institute for
Theoretical Physics program on Particle Acceleration in Astrophysical Plasmas,
in Santa Barbara (2009, July 26-October 3) where this paper was finished.  That
program was supported in part by the National Science Foundation under Grant No.
PHY05-51164.  A.P. is grateful the Swedish National Allocations Committee (SNAC)
for the resources granted at HPC2N.  C.P. gratefully acknowledges the financial
support of the National Science and Engineering Research Council of Canada. Some
computations were also performed on CITA’s McKenzie and Sunnyvale clusters which
are funded by the Canada Foundation for Innovation, the Ontario Innovation
Trust, and the Ontario Research Fund.

\bibliography{bibtex/gadget}

\begin{thebibliography}{}

\bibitem[\protect\citeauthoryear{{Abdo}, {Ackermann}, {Ajello}, {Asano},
  {Baldini}, {Ballet}, {Barbiellini}, {Bastieri}, {Baughman}, {Bechtol},
  {Bellazzini}, {Blandford} \& {et al.}}{{Abdo}
  et~al.}{2009}]{2009ApJ...699...31A}
{Abdo} A.~A.,  {Ackermann} M.,  {Ajello} M.,  {Asano} K.,  {Baldini} L.,
  {Ballet} J.,  {Barbiellini} G.,  {Bastieri} D.,  {Baughman} B.~M.,  {Bechtol}
  K.,  {Bellazzini} R.,  {Blandford} R.~D.,    {et al.} 2009, \apj, 699, 31

\bibitem[\protect\citeauthoryear{{Abramowitz} \& {Stegun}}{{Abramowitz} \&
  {Stegun}}{1965}]{1965hmfw.book.....A}
{Abramowitz} M.,  {Stegun} I.~A.,  1965, {Handbook of mathematical functions}.
New York: Dover

\bibitem[\protect\citeauthoryear{{Acciari}, {Aliu}, {Arlen}, {Aune},
  {Bautista}, {Beilicke}, {Benbow}, {Boltuch}, {Bradbury}, {Buckley}, {Bugaev},
  {Byrum} \& {et al.}}{{Acciari} et~al.}{2009}]{2009Natur.462..770V}
{Acciari} V.~A.,  {Aliu} E.,  {Arlen} T.,  {Aune} T.,  {Bautista} M.,
  {Beilicke} M.,  {Benbow} W.,  {Boltuch} D.,  {Bradbury} S.~M.,  {Buckley}
  J.~H.,  {Bugaev} V.,  {Byrum} K.,    {et al.} 2009, \nat, 462, 770

\bibitem[\protect\citeauthoryear{{Agertz}, {Moore}, {Stadel}, {Potter},
  {Miniati}, {Read}, {Mayer}, {Gawryszczak}, {Kravtsov}, {Nordlund}, {Pearce},
  {Quilis}, {Rudd}, {Springel}, {Stone}, {Tasker}, {Teyssier}, {Wadsley} \&
  {Walder}}{{Agertz} et~al.}{2007}]{2007MNRAS.380..963A}
{Agertz} O.,  {Moore} B.,  {Stadel} J.,  {Potter} D.,  {Miniati} F.,  {Read}
  J.,  {Mayer} L.,  {Gawryszczak} A.,  {Kravtsov} A.,  {Nordlund} {\AA}.,
  {Pearce} F.,  {Quilis} V.,  {Rudd} D.,  {Springel} V.,  {Stone} J.,  {Tasker}
  E.,  {Teyssier} R.,  {Wadsley} J.,    {Walder} R.,  2007, \mnras, 380, 963

\bibitem[\protect\citeauthoryear{{Aharonian}, {Akhperjanian}, {Anton}, {Barres
  de Almeida}, {Bazer-Bachi}, {Becherini}, {Behera}, {Bernl{\"o}hr}, {Boisson},
  {Bochow} \& {et al.}}{{Aharonian} et~al.}{2009}]{2009A&A...502..437A}
{Aharonian} F.,  {Akhperjanian} A.~G.,  {Anton} G.,  {Barres de Almeida} U.,
  {Bazer-Bachi} A.~R.,  {Becherini} Y.,  {Behera} B.,  {Bernl{\"o}hr} K.,
  {Boisson} C.,  {Bochow} A.,    {et al.} 2009, \aap, 502, 437

\bibitem[\protect\citeauthoryear{{Aharonian}, {Akhperjanian}, {Bazer-Bachi},
  {Beilicke}, {Benbow}, {Berge}, {Bernl{\"o}hr}, {Boisson}, {Bolz}, {Borrel},
  {Braun}, {Breitling} \& {et al.}}{{Aharonian}
  et~al.}{2006}]{2006Natur.439..695A}
{Aharonian} F.,  {Akhperjanian} A.~G.,  {Bazer-Bachi} A.~R.,  {Beilicke} M.,
  {Benbow} W.,  {Berge} D.,  {Bernl{\"o}hr} K.,  {Boisson} C.,  {Bolz} O.,
  {Borrel} V.,  {Braun} I.,  {Breitling} F.,    {et al.} 2006, \nat, 439, 695

\bibitem[\protect\citeauthoryear{{Aharonian}, {Akhperjanian}, {Aye},
  {Bazer-Bachi}, {Beilicke}, {Benbow}, {Berge}, {Berghaus}, {Bernl{\"o}hr},
  {Bolz}, {Boisson}, {Borgmeier} \& {et al.}}{{Aharonian}
  et~al.}{2004}]{2004Natur.432...75A}
{Aharonian} F.~A.,  {Akhperjanian} A.~G.,  {Aye} K.-M.,  {Bazer-Bachi} A.~R.,
  {Beilicke} M.,  {Benbow} W.,  {Berge} D.,  {Berghaus} P.,  {Bernl{\"o}hr} K.,
   {Bolz} O.,  {Boisson} C.,  {Borgmeier} C.,    {et al.} 2004, \nat, 432, 75

\bibitem[\protect\citeauthoryear{{Aleksi{\'c}}, {Antonelli}, {Antoranz},
  {Backes}, {Baixeras}, {Balestra}, {Barrio}, {Bastieri}, {Becerra
  Gonz{\'a}lez}, {Bednarek}, {Berdyugin}, {Berger} \& {et al.}}{{Aleksi{\'c}}
  et~al.}{2010}]{2009arXiv0909.3267T}
{Aleksi{\'c}} J.,  {Antonelli} L.~A.,  {Antoranz} P.,  {Backes} M.,  {Baixeras}
  C.,  {Balestra} S.,  {Barrio} J.~A.,  {Bastieri} D.,  {Becerra Gonz{\'a}lez}
  J.,  {Bednarek} W.,  {Berdyugin} A.,  {Berger} K.,    {et al.} 2010, \apj,
  710, 634

\bibitem[\protect\citeauthoryear{{Amato}, {Blasi} \& {Gabici}}{{Amato}
  et~al.}{2008}]{2008MNRAS.385.1946A}
{Amato} E.,  {Blasi} P.,    {Gabici} S.,  2008, \mnras, 385, 1946

\bibitem[\protect\citeauthoryear{{Berezinsky}, {Blasi} \&
  {Ptuskin}}{{Berezinsky} et~al.}{1997}]{1997ApJ...487..529B}
{Berezinsky} V.~S.,  {Blasi} P.,    {Ptuskin} V.~S.,  1997, \apj, 487, 529

\bibitem[\protect\citeauthoryear{{Blandford} \& {Eichler}}{{Blandford} \&
  {Eichler}}{1987}]{1987PhR...154....1B}
{Blandford} R.,  {Eichler} D.,  1987, \physrep, 154, 1

\bibitem[\protect\citeauthoryear{{Blasi} \& {Colafrancesco}}{{Blasi} \&
  {Colafrancesco}}{1999}]{1999APh....12..169B}
{Blasi} P.,  {Colafrancesco} S.,  1999, Astroparticle Physics, 12, 169

\bibitem[\protect\citeauthoryear{{Blumenthal} \& {Gould}}{{Blumenthal} \&
  {Gould}}{1970}]{1970RvMP...42..237B}
{Blumenthal} G.~R.,  {Gould} R.~J.,  1970, Reviews of Modern Physics, 42, 237

\bibitem[\protect\citeauthoryear{{Briel}, {Henry} \& {Boehringer}}{{Briel}
  et~al.}{1992}]{1992A&A...259L..31B}
{Briel} U.~G.,  {Henry} J.~P.,    {Boehringer} H.,  1992, \aap, 259, L31

\bibitem[\protect\citeauthoryear{{Briel}, {Henry}, {Lumb}, {Arnaud}, {Neumann},
  {Aghanim}, {Gastaud}, {Mittaz}, {Sasseen} \& {Vestrand}}{{Briel}
  et~al.}{2001}]{2001A&A...365L..60B}
{Briel} U.~G.,  {Henry} J.~P.,  {Lumb} D.~H.,  {Arnaud} M.,  {Neumann} D.,
  {Aghanim} N.,  {Gastaud} R.,  {Mittaz} J.~P.~D.,  {Sasseen} T.~P.,
  {Vestrand} W.~T.,  2001, \aap, 365, L60

\bibitem[\protect\citeauthoryear{{Caprioli}, {Amato} \& {Blasi}}{{Caprioli}
  et~al.}{2009}]{2009arXiv0912.2714C}
{Caprioli} D.,  {Amato} E.,    {Blasi} P.,  2009, ArXiv e-prints

\bibitem[\protect\citeauthoryear{{Celotti}, {Kuncic}, {Rees} \&
  {Wardle}}{{Celotti} et~al.}{1998}]{1998MNRAS.293..288C}
{Celotti} A.,  {Kuncic} Z.,  {Rees} M.~J.,    {Wardle} J.~F.~C.,  1998, \mnras,
  293, 288

\bibitem[\protect\citeauthoryear{{Churazov}, {Forman}, {Jones} \& {B{\"
  o}hringer}}{{Churazov} et~al.}{2003}]{2003ApJ...590..225C}
{Churazov} E.,  {Forman} W.,  {Jones} C.,    {B{\" o}hringer} H.,  2003, \apj,
  590, 225

\bibitem[\protect\citeauthoryear{{Dav{\' e}}, {Cen}, {Ostriker}, {Bryan},
  {Hernquist}, {Katz}, {Weinberg}, {Norman} \& {O'Shea}}{{Dav{\' e}}
  et~al.}{2001}]{2001ApJ...552..473D}
{Dav{\' e}} R.,  {Cen} R.,  {Ostriker} J.~P.,  {Bryan} G.~L.,  {Hernquist} L.,
  {Katz} N.,  {Weinberg} D.~H.,  {Norman} M.~L.,    {O'Shea} B.,  2001, \apj,
  552, 473

\bibitem[\protect\citeauthoryear{{Dennison}}{{Dennison}}{1980}]{1980ApJ...239L%
..93D}
{Dennison} B.,  1980, \apjl, 239, L93

\bibitem[\protect\citeauthoryear{{Dolag}, {Bartelmann} \& {Lesch}}{{Dolag}
  et~al.}{1999}]{1999A&A...348..351D}
{Dolag} K.,  {Bartelmann} M.,    {Lesch} H.,  1999, \aap, 348, 351

\bibitem[\protect\citeauthoryear{{Dolag}, {Borgani}, {Murante} \&
  {Springel}}{{Dolag} et~al.}{2009}]{2009MNRAS.399..497D}
{Dolag} K.,  {Borgani} S.,  {Murante} G.,    {Springel} V.,  2009, \mnras, 399,
  497

\bibitem[\protect\citeauthoryear{{Dolag} \& {En{\ss}lin}}{{Dolag} \&
  {En{\ss}lin}}{2000}]{2000A&A...362..151D}
{Dolag} K.,  {En{\ss}lin} T.~A.,  2000, \aap, 362, 151

\bibitem[\protect\citeauthoryear{{Drury}}{{Drury}}{1983}]{1983RPPh...46..973D}
{Drury} L.~O.,  1983, Reports of Progress in Physics, 46, 973

\bibitem[\protect\citeauthoryear{Dubois \& Teyssier}{Dubois \&
  Teyssier}{2008}]{Dubois:2008mz}
Dubois Y.,  Teyssier R.,  2008

\bibitem[\protect\citeauthoryear{{Dursi} \& {Pfrommer}}{{Dursi} \&
  {Pfrommer}}{2008}]{2008ApJ...677..993D}
{Dursi} L.~J.,  {Pfrommer} C.,  2008, \apj, 677, 993

\bibitem[\protect\citeauthoryear{{Edge}, {Stewart} \& {Fabian}}{{Edge}
  et~al.}{1992}]{1992MNRAS.258..177E}
{Edge} A.~C.,  {Stewart} G.~C.,    {Fabian} A.~C.,  1992, \mnras, 258, 177

\bibitem[\protect\citeauthoryear{{Edmon}, {Jones} \& {Kang}}{{Edmon}
  et~al.}{2007}]{2007arXiv0706.0587E}
{Edmon} P.~P.,  {Jones} T.~W.,    {Kang} H.,  2007, ArXiv:0706.0587

\bibitem[\protect\citeauthoryear{{Ellison}}{{Ellison}}{2000}]{2000AIPC..528..3%
83E}
{Ellison} D.~C.,  2000, in {Mewaldt} R.~A.,  {Jokipii} J.~R.,  {Lee} M.~A.,
  {M{\"o}bius} E.,   {Zurbuchen} T.~H.,  eds, Acceleration and Transport of
  Energetic Particles Observed in the Heliosphere Vol.~528 of American
  Institute of Physics Conference Series, {The Cosmic Ray-X-ray Connection:
  Effects of Nonlinear Shock Acceleration on Photon Production in SNRs}.
p.~383

\bibitem[\protect\citeauthoryear{{Ellison} \& {Eichler}}{{Ellison} \&
  {Eichler}}{1984}]{1984ApJ...286..691E}
{Ellison} D.~C.,  {Eichler} D.,  1984, \apj, 286, 691

\bibitem[\protect\citeauthoryear{{Ellison} \& {et al.}}{{Ellison} \& {et
  al.}}{2005}]{2005ICRC....3..261E}
{Ellison} D.~C.,  {et al.} 2005, in International Cosmic Ray Conference Vol.~3
  of International Cosmic Ray Conference, {Thermal Particle Injection in
  Nonlinear Diffusive Shock Acceleration}.
p.~261

\bibitem[\protect\citeauthoryear{{Ellison}, {Jones} \& {Eichler}}{{Ellison}
  et~al.}{1981}]{1981JGZG...50..110E}
{Ellison} D.~C.,  {Jones} F.~C.,    {Eichler} D.,  1981, Journal of Geophysics
  Zeitschrift Geophysik, 50, 110

\bibitem[\protect\citeauthoryear{{Ellison}, {Moebius} \& {Paschmann}}{{Ellison}
  et~al.}{1990}]{1990ApJ...352..376E}
{Ellison} D.~C.,  {Moebius} E.,    {Paschmann} G.,  1990, \apj, 352, 376

\bibitem[\protect\citeauthoryear{{En{\ss}lin}}{{En{\ss}lin}}{2003}]{2003A&A...%
399..409E}
{En{\ss}lin} T.~A.,  2003, \aap, 399, 409

\bibitem[\protect\citeauthoryear{{En{\ss}lin}, {Biermann}, {Klein} \&
  {Kohle}}{{En{\ss}lin} et~al.}{1998}]{1998A&A...332..395E}
{En{\ss}lin} T.~A.,  {Biermann} P.~L.,  {Klein} U.,    {Kohle} S.,  1998, \aap,
  332, 395

\bibitem[\protect\citeauthoryear{{En{\ss}lin}, {Biermann}, {Kronberg} \&
  {Wu}}{{En{\ss}lin} et~al.}{1997}]{1997ApJ...477..560E}
{En{\ss}lin} T.~A.,  {Biermann} P.~L.,  {Kronberg} P.~P.,    {Wu} X.-P.,  1997,
  \apj, 477, 560

\bibitem[\protect\citeauthoryear{{En{\ss}lin}, {Pfrommer}, {Springel} \&
  {Jubelgas}}{{En{\ss}lin} et~al.}{2007}]{2007A&A...473...41E}
{En{\ss}lin} T.~A.,  {Pfrommer} C.,  {Springel} V.,    {Jubelgas} M.,  2007,
  \aap, 473, 41

\bibitem[\protect\citeauthoryear{{Feretti}}{{Feretti}}{2003}]{2003ASPC..301..1%
43F}
{Feretti} L.,  2003, in {Bowyer} S.,  {Hwang} C.-Y.,  eds, Astronomical Society
  of the Pacific Conference Series Vol.~301 of Astronomical Society of the
  Pacific Conference Series, {Radio Observations of Clusters of Galaxies}.
pp 143--+

\bibitem[\protect\citeauthoryear{{Ferrari}, {Govoni}, {Schindler}, {Bykov} \&
  {Rephaeli}}{{Ferrari} et~al.}{2008}]{2008SSRv..134...93F}
{Ferrari} C.,  {Govoni} F.,  {Schindler} S.,  {Bykov} A.~M.,    {Rephaeli} Y.,
  2008, Space Science Reviews, 134, 93

\bibitem[\protect\citeauthoryear{{Franceschini}, {Rodighiero} \&
  {Vaccari}}{{Franceschini} et~al.}{2008}]{2008A&A...487..837F}
{Franceschini} A.,  {Rodighiero} G.,    {Vaccari} M.,  2008, \aap, 487, 837

\bibitem[\protect\citeauthoryear{{Freedman}, {Madore}, {Gibson}, {Ferrarese},
  {Kelson}, {Sakai}, {Mould}, {Kennicutt} Jr., {Ford}, {Graham}, {Huchra},
  {Hughes}, {Illingworth}, {Macri} \& {Stetson}}{{Freedman}
  et~al.}{2001}]{2001ApJ...553...47F}
{Freedman} W.~L.,  {Madore} B.~F.,  {Gibson} B.~K.,  {Ferrarese} L.,  {Kelson}
  D.~D.,  {Sakai} S.,  {Mould} J.~R.,  {Kennicutt} Jr. R.~C.,  {Ford} H.~C.,
  {Graham} J.~A.,  {Huchra} J.~P.,  {Hughes} S.~M.~G.,  {Illingworth} G.~D.,
  {Macri} L.~M.,    {Stetson} P.~B.,  2001, \apj, 553, 47

\bibitem[\protect\citeauthoryear{{Gould}}{{Gould}}{1972}]{1972Phy....58..379G}
{Gould} R.~J.,  1972, Physica, 58, 379

\bibitem[\protect\citeauthoryear{{Hayashi} \& {White}}{{Hayashi} \&
  {White}}{2008}]{2008MNRAS.388....2H}
{Hayashi} E.,  {White} S.~D.~M.,  2008, \mnras, 388, 2

\bibitem[\protect\citeauthoryear{{Helder}, {Vink}, {Bassa}, {Bamba}, {Bleeker},
  {Funk}, {Ghavamian}, {van der Heyden}, {Verbunt} \& {Yamazaki}}{{Helder}
  et~al.}{2009}]{2009Sci...325..719H}
{Helder} E.~A.,  {Vink} J.,  {Bassa} C.~G.,  {Bamba} A.,  {Bleeker} J.~A.~M.,
  {Funk} S.,  {Ghavamian} P.,  {van der Heyden} K.~J.,  {Verbunt} F.,
  {Yamazaki} R.,  2009, Science, 325, 719

\bibitem[\protect\citeauthoryear{{Hirotani} \& {Okamoto}}{{Hirotani} \&
  {Okamoto}}{1998}]{1998ApJ...497..563H}
{Hirotani} K.,  {Okamoto} I.,  1998, \apj, 497, 563

\bibitem[\protect\citeauthoryear{{Hughes}, {Rakowski} \&
  {Decourchelle}}{{Hughes} et~al.}{2000}]{2000ApJ...543L..61H}
{Hughes} J.~P.,  {Rakowski} C.~E.,    {Decourchelle} A.,  2000, \apjl, 543, L61

\bibitem[\protect\citeauthoryear{{Iapichino}, {Adamek}, {Schmidt} \&
  {Niemeyer}}{{Iapichino} et~al.}{2008}]{2008MNRAS.388.1079I}
{Iapichino} L.,  {Adamek} J.,  {Schmidt} W.,    {Niemeyer} J.~C.,  2008,
  \mnras, 388, 1079

\bibitem[\protect\citeauthoryear{{Iapichino} \& {Niemeyer}}{{Iapichino} \&
  {Niemeyer}}{2008}]{2008MNRAS.388.1089I}
{Iapichino} L.,  {Niemeyer} J.~C.,  2008, \mnras, 388, 1089

\bibitem[\protect\citeauthoryear{{Inoue} \& {Nagashima}}{{Inoue} \&
  {Nagashima}}{2005}]{2005AIPC..745..567I}
{Inoue} S.,  {Nagashima} M.,  2005, in {Aharonian} F.~A.,  {V{\"o}lk} H.~J.,
  {Horns} D.,  eds, High Energy Gamma-Ray Astronomy Vol.~745 of American
  Institute of Physics Conference Series, {Gamma-Rays from Large Scale
  Structure Formation and the Warm-Hot Intergalactic Medium: Cosmic Baryometry
  with Gamma-Rays}.
pp 567--572

\bibitem[\protect\citeauthoryear{{Jeltema}, {Kehayias} \& {Profumo}}{{Jeltema}
  et~al.}{2009}]{2009PhRvD..80b3005J}
{Jeltema} T.~E.,  {Kehayias} J.,    {Profumo} S.,  2009, \prd, 80, 023005

\bibitem[\protect\citeauthoryear{{Jubelgas}, {Springel}, {En{\ss}lin} \&
  {Pfrommer}}{{Jubelgas} et~al.}{2008}]{2008A&A...481...33J}
{Jubelgas} M.,  {Springel} V.,  {En{\ss}lin} T.,    {Pfrommer} C.,  2008, \aap,
  481, 33

\bibitem[\protect\citeauthoryear{{Kang} \& {Jones}}{{Kang} \&
  {Jones}}{2005}]{2005ApJ...620...44K}
{Kang} H.,  {Jones} T.~W.,  2005, \apj, 620, 44

\bibitem[\protect\citeauthoryear{{Kang} \& {Jones}}{{Kang} \&
  {Jones}}{2007}]{2007APh....28..232K}
{Kang} H.,  {Jones} T.~W.,  2007, Astroparticle Physics, 28, 232

\bibitem[\protect\citeauthoryear{{Katz}, {Weinberg} \& {Hernquist}}{{Katz}
  et~al.}{1996}]{1996ApJS..105...19K}
{Katz} N.,  {Weinberg} D.~H.,    {Hernquist} L.,  1996, \apjs, 105, 19

\bibitem[\protect\citeauthoryear{{Katz} \& {White}}{{Katz} \&
  {White}}{1993}]{1993ApJ...412..455K}
{Katz} N.,  {White} S.~D.~M.,  1993, \apj, 412, 455

\bibitem[\protect\citeauthoryear{{Keshet}, {Waxman}, {Loeb}, {Springel} \&
  {Hernquist}}{{Keshet} et~al.}{2003}]{2003ApJ...585..128K}
{Keshet} U.,  {Waxman} E.,  {Loeb} A.,  {Springel} V.,    {Hernquist} L.,
  2003, \apj, 585, 128

\bibitem[\protect\citeauthoryear{{Komatsu} \& {Seljak}}{{Komatsu} \&
  {Seljak}}{2001}]{2001MNRAS.327.1353K}
{Komatsu} E.,  {Seljak} U.,  2001, \mnras, 327, 1353

\bibitem[\protect\citeauthoryear{{Kushnir}, {Katz} \& {Waxman}}{{Kushnir}
  et~al.}{2009}]{2009JCAP...09..024K}
{Kushnir} D.,  {Katz} B.,    {Waxman} E.,  2009, Journal of Cosmology and
  Astro-Particle Physics, 9, 24

\bibitem[\protect\citeauthoryear{Kushnir \& Waxman}{Kushnir \&
  Waxman}{2009}]{Kushnir:2009vm}
Kushnir D.,  Waxman E.,  2009, JCAP, 0908, 002

\bibitem[\protect\citeauthoryear{{Loeb} \& {Waxman}}{{Loeb} \&
  {Waxman}}{2000}]{2000Natur.405..156L}
{Loeb} A.,  {Waxman} E.,  2000, \nat, 405, 156

\bibitem[\protect\citeauthoryear{{Lu}, {Mo}, {Katz} \& {Weinberg}}{{Lu}
  et~al.}{2006}]{2006MNRAS.368.1931L}
{Lu} Y.,  {Mo} H.~J.,  {Katz} N.,    {Weinberg} M.~D.,  2006, \mnras, 368, 1931

\bibitem[\protect\citeauthoryear{{Maier}, {Iapichino}, {Schmidt} \&
  {Niemeyer}}{{Maier} et~al.}{2009}]{2009ApJ...707...40M}
{Maier} A.,  {Iapichino} L.,  {Schmidt} W.,    {Niemeyer} J.~C.,  2009, \apj,
  707, 40

\bibitem[\protect\citeauthoryear{{Malkov} \& {O'C Drury}}{{Malkov} \& {O'C
  Drury}}{2001}]{2001RPPh...64..429M}
{Malkov} M.~A.,  {O'C Drury} L.,  2001, Reports of Progress in Physics, 64, 429

\bibitem[\protect\citeauthoryear{{Markevitch} \& {Vikhlinin}}{{Markevitch} \&
  {Vikhlinin}}{2007}]{2007PhR...443....1M}
{Markevitch} M.,  {Vikhlinin} A.,  2007, \physrep, 443, 1

\bibitem[\protect\citeauthoryear{{Miniati}}{{Miniati}}{2001}]{2001CoPhC.141...%
17M}
{Miniati} F.,  2001, Computer Physics Communications, 141, 17

\bibitem[\protect\citeauthoryear{{Miniati}}{{Miniati}}{2002}]{2002MNRAS.337..1%
99M}
{Miniati} F.,  2002, \mnras, 337, 199

\bibitem[\protect\citeauthoryear{{Miniati}}{{Miniati}}{2003}]{2003MNRAS.342.10%
09M}
{Miniati} F.,  2003, \mnras, 342, 1009

\bibitem[\protect\citeauthoryear{{Miniati}, {Jones}, {Kang} \& {Ryu}}{{Miniati}
  et~al.}{2001b}]{2001ApJ...562..233M}
{Miniati} F.,  {Jones} T.~W.,  {Kang} H.,    {Ryu} D.,  2001b, \apj, 562, 233

\bibitem[\protect\citeauthoryear{{Miniati}, {Ryu}, {Kang} \& {Jones}}{{Miniati}
  et~al.}{2001a}]{2001ApJ...559...59M}
{Miniati} F.,  {Ryu} D.,  {Kang} H.,    {Jones} T.~W.,  2001a, \apj, 559, 59

\bibitem[\protect\citeauthoryear{{Miniati}, {Ryu}, {Kang}, {Jones}, {Cen} \&
  {Ostriker}}{{Miniati} et~al.}{2000}]{2000ApJ...542..608M}
{Miniati} F.,  {Ryu} D.,  {Kang} H.,  {Jones} T.~W.,  {Cen} R.,    {Ostriker}
  J.~P.,  2000, \apj, 542, 608

\bibitem[\protect\citeauthoryear{Molnar et~al.,}{Molnar
  et~al.}{2009}]{Molnar:2009rk}
Molnar S.~M.,  et~al., 2009, Astrophys. J., 696, 1640

\bibitem[\protect\citeauthoryear{{Percival}, {Reid}, {Eisenstein}, {Bahcall},
  {Budavari}, {Fukugita}, {Gunn}, {Ivezic}, {Knapp}, {Kron}, {Loveday},
  {Lupton}, {McKay}, {Meiksin}, {Nichol}, {Pope}, {Schlegel}, {Schneider} \&
  {et al.}}{{Percival} et~al.}{2009}]{2009arXiv0907.1660P}
{Percival} W.~J.,  {Reid} B.~A.,  {Eisenstein} D.~J.,  {Bahcall} N.~A.,
  {Budavari} T.,  {Fukugita} M.,  {Gunn} J.~E.,  {Ivezic} Z.,  {Knapp} G.~R.,
  {Kron} R.~G.,  {Loveday} J.,  {Lupton} R.~H.,  {McKay} T.~A.,  {Meiksin} A.,
  {Nichol} R.~C.,  {Pope} A.~C.,  {Schlegel} D.~J.,  {Schneider} D.~P.,    {et
  al.} 2009, arXiv:0907.1660

\bibitem[\protect\citeauthoryear{{Pfrommer}}{{Pfrommer}}{2008}]{2008MNRAS.385.%
1242P}
{Pfrommer} C.,  2008, \mnras, 385, 1242

\bibitem[\protect\citeauthoryear{{Pfrommer} \& {Dursi}}{{Pfrommer} \&
  {Dursi}}{2009}]{2009arXiv0911.2476P}
{Pfrommer} C.,  {Dursi} L.~J.,  2009, arXiv:0911.2476

\bibitem[\protect\citeauthoryear{{Pfrommer} \& {En{\ss}lin}}{{Pfrommer} \&
  {En{\ss}lin}}{2003}]{2003A&A...407L..73P}
{Pfrommer} C.,  {En{\ss}lin} T.~A.,  2003, \aap, 407, L73

\bibitem[\protect\citeauthoryear{{Pfrommer} \& {En{\ss}lin}}{{Pfrommer} \&
  {En{\ss}lin}}{2004}]{2004A&A...413...17P}
{Pfrommer} C.,  {En{\ss}lin} T.~A.,  2004, \aap, 413, 17

\bibitem[\protect\citeauthoryear{{Pfrommer}, {En{\ss}lin} \&
  {Springel}}{{Pfrommer} et~al.}{2008}]{2008MNRAS.385.1211P}
{Pfrommer} C.,  {En{\ss}lin} T.~A.,    {Springel} V.,  2008, \mnras, 385, 1211

\bibitem[\protect\citeauthoryear{{Pfrommer}, {En{\ss}lin}, {Springel},
  {Jubelgas} \& {Dolag}}{{Pfrommer} et~al.}{2007}]{2007MNRAS.378..385P}
{Pfrommer} C.,  {En{\ss}lin} T.~A.,  {Springel} V.,  {Jubelgas} M.,    {Dolag}
  K.,  2007, \mnras, 378, 385

\bibitem[\protect\citeauthoryear{{Pfrommer}, {Springel}, {En{\ss}lin} \&
  {Jubelgas}}{{Pfrommer} et~al.}{2006}]{2006MNRAS.367..113P}
{Pfrommer} C.,  {Springel} V.,  {En{\ss}lin} T.~A.,    {Jubelgas} M.,  2006,
  \mnras, 367, 113

\bibitem[\protect\citeauthoryear{{Quilis}, {Ibanez} \& {Saez}}{{Quilis}
  et~al.}{1998}]{1998ApJ...502..518Q}
{Quilis} V.,  {Ibanez} J.~M.~A.,    {Saez} D.,  1998, \apj, 502, 518

\bibitem[\protect\citeauthoryear{{Reimer}, {Pohl}, {Sreekumar} \&
  {Mattox}}{{Reimer} et~al.}{2003}]{2003ApJ...588..155R}
{Reimer} O.,  {Pohl} M.,  {Sreekumar} P.,    {Mattox} J.~R.,  2003, \apj, 588,
  155

\bibitem[\protect\citeauthoryear{{Reiprich} \& {B{\"o}hringer}}{{Reiprich} \&
  {B{\"o}hringer}}{2002}]{2002ApJ...567..716R}
{Reiprich} T.~H.,  {B{\"o}hringer} H.,  2002, \apj, 567, 716

\bibitem[\protect\citeauthoryear{{Rosati}, {Borgani} \& {Norman}}{{Rosati}
  et~al.}{2002}]{2002ARA&A..40..539R}
{Rosati} P.,  {Borgani} S.,    {Norman} C.,  2002, \araa, 40, 539

\bibitem[\protect\citeauthoryear{{Rybicki} \& {Lightman}}{{Rybicki} \&
  {Lightman}}{1979}]{1979rpa..book.....R}
{Rybicki} G.~B.,  {Lightman} A.~P.,  1979, Radiative processes in astrophysics.
New York, Wiley-Interscience

\bibitem[\protect\citeauthoryear{{Ryu}, {Kang}, {Hallman} \& {Jones}}{{Ryu}
  et~al.}{2003}]{2003ApJ...593..599R}
{Ryu} D.,  {Kang} H.,  {Hallman} E.,    {Jones} T.~W.,  2003, \apj, 593, 599

\bibitem[\protect\citeauthoryear{{Sarazin}}{{Sarazin}}{1999}]{1999ApJ...520..5%
29S}
{Sarazin} C.~L.,  1999, \apj, 520, 529

\bibitem[\protect\citeauthoryear{{Schlickeiser}}{{Schlickeiser}}{2002}]{2002cr%
a..book.....S}
{Schlickeiser} R.,  2002, {Cosmic ray astrophysics}.
Springer

\bibitem[\protect\citeauthoryear{{Shimada}, {Terasawa}, {Hoshino}, {Naito},
  {Matsui}, {Koi} \& {Maezawa}}{{Shimada} et~al.}{1999}]{1999Ap&SS.264..481S}
{Shimada} N.,  {Terasawa} T.,  {Hoshino} M.,  {Naito} T.,  {Matsui} H.,  {Koi}
  T.,    {Maezawa} K.,  1999, \apss, 264, 481

\bibitem[\protect\citeauthoryear{{Sijacki}, {Pfrommer}, {Springel} \&
  {En{\ss}lin}}{{Sijacki} et~al.}{2008}]{2008MNRAS.387.1403S}
{Sijacki} D.,  {Pfrommer} C.,  {Springel} V.,    {En{\ss}lin} T.~A.,  2008,
  \mnras, 387, 1403

\bibitem[\protect\citeauthoryear{{Sikora} \& {Madejski}}{{Sikora} \&
  {Madejski}}{2000}]{2000ApJ...534..109S}
{Sikora} M.,  {Madejski} G.,  2000, \apj, 534, 109

\bibitem[\protect\citeauthoryear{{Skillman}, {O'Shea}, {Hallman}, {Burns} \&
  {Norman}}{{Skillman} et~al.}{2008}]{2008ApJ...689.1063S}
{Skillman} S.~W.,  {O'Shea} B.~W.,  {Hallman} E.~J.,  {Burns} J.~O.,
  {Norman} M.~L.,  2008, \apj, 689, 1063

\bibitem[\protect\citeauthoryear{{Slane}, {Gaensler}, {Dame}, {Hughes},
  {Plucinsky} \& {Green}}{{Slane} et~al.}{1999}]{1999ApJ...525..357S}
{Slane} P.,  {Gaensler} B.~M.,  {Dame} T.~M.,  {Hughes} J.~P.,  {Plucinsky}
  P.~P.,    {Green} A.,  1999, \apj, 525, 357

\bibitem[\protect\citeauthoryear{{Springel}}{{Springel}}{2005}]{2005MNRAS.364.%
1105S}
{Springel} V.,  2005, \mnras, 364, 1105

\bibitem[\protect\citeauthoryear{{Springel} \& {Hernquist}}{{Springel} \&
  {Hernquist}}{2002}]{2002MNRAS.333..649S}
{Springel} V.,  {Hernquist} L.,  2002, \mnras, 333, 649

\bibitem[\protect\citeauthoryear{{Springel}, {Yoshida} \& {White}}{{Springel}
  et~al.}{2001}]{2001NewA....6...79S}
{Springel} V.,  {Yoshida} N.,    {White} S.~D.~M.,  2001, New Astronomy, 6, 79

\bibitem[\protect\citeauthoryear{{Totani} \& {Kitayama}}{{Totani} \&
  {Kitayama}}{2000}]{2000ApJ...545..572T}
{Totani} T.,  {Kitayama} T.,  2000, \apj, 545, 572

\bibitem[\protect\citeauthoryear{{Uchiyama}, {Aharonian}, {Tanaka}, {Takahashi}
  \& {Maeda}}{{Uchiyama} et~al.}{2007}]{2007Natur.449..576U}
{Uchiyama} Y.,  {Aharonian} F.~A.,  {Tanaka} T.,  {Takahashi} T.,    {Maeda}
  Y.,  2007, \nat, 449, 576

\bibitem[\protect\citeauthoryear{{Vestrand}}{{Vestrand}}{1982}]{1982AJ.....87.%
1266V}
{Vestrand} W.~T.,  1982, \aj, 87, 1266

\bibitem[\protect\citeauthoryear{{Vikhlinin}, {Kravtsov}, {Forman}, {Jones},
  {Markevitch}, {Murray} \& {Van Speybroeck}}{{Vikhlinin}
  et~al.}{2006}]{2006ApJ...640..691V}
{Vikhlinin} A.,  {Kravtsov} A.,  {Forman} W.,  {Jones} C.,  {Markevitch} M.,
  {Murray} S.~S.,    {Van Speybroeck} L.,  2006, \apj, 640, 691

\bibitem[\protect\citeauthoryear{{Vikhlinin}, {Markevitch}, {Murray}, {Jones},
  {Forman} \& {Van Speybroeck}}{{Vikhlinin} et~al.}{2005}]{2005ApJ...628..655V}
{Vikhlinin} A.,  {Markevitch} M.,  {Murray} S.~S.,  {Jones} C.,  {Forman} W.,
   {Van Speybroeck} L.,  2005, \apj, 628, 655

\bibitem[\protect\citeauthoryear{{Vink}, {Bleeker}, {van der Heyden}, {Bykov},
  {Bamba} \& {Yamazaki}}{{Vink} et~al.}{2006}]{2006ApJ...648L..33V}
{Vink} J.,  {Bleeker} J.,  {van der Heyden} K.,  {Bykov} A.,  {Bamba} A.,
  {Yamazaki} R.,  2006, \apjl, 648, L33

\bibitem[\protect\citeauthoryear{{Vladimirov}, {Ellison} \&
  {Bykov}}{{Vladimirov} et~al.}{2006}]{2006ApJ...652.1246V}
{Vladimirov} A.,  {Ellison} D.~C.,    {Bykov} A.,  2006, \apj, 652, 1246

\bibitem[\protect\citeauthoryear{{Vogt} \& {En{\ss}lin}}{{Vogt} \&
  {En{\ss}lin}}{2005}]{2005A&A...434...67V}
{Vogt} C.,  {En{\ss}lin} T.~A.,  2005, \aap, 434, 67

\bibitem[\protect\citeauthoryear{{Voit}}{{Voit}}{2005}]{2005RvMP...77..207V}
{Voit} G.~M.,  2005, Reviews of Modern Physics, 77, 207

\bibitem[\protect\citeauthoryear{{V{\"o}lk}, {Aharonian} \&
  {Breitschwerdt}}{{V{\"o}lk} et~al.}{1996}]{1996SSRv...75..279V}
{V{\"o}lk} H.~J.,  {Aharonian} F.~A.,    {Breitschwerdt} D.,  1996, Space
  Science Reviews, 75, 279

\bibitem[\protect\citeauthoryear{{Warren}, {Hughes}, {Badenes}, {Ghavamian},
  {McKee}, {Moffett}, {Plucinsky}, {Rakowski}, {Reynoso} \& {Slane}}{{Warren}
  et~al.}{2005}]{2005ApJ...634..376W}
{Warren} J.~S.,  {Hughes} J.~P.,  {Badenes} C.,  {Ghavamian} P.,  {McKee}
  C.~F.,  {Moffett} D.,  {Plucinsky} P.~P.,  {Rakowski} C.,  {Reynoso} E.,
  {Slane} P.,  2005, \apj, 634, 376

\bibitem[\protect\citeauthoryear{Webb~G.M.
  Drury~L.O'C.}{Webb~G.M.}{1984}]{webb84}
Webb~G.M. Drury~L.O'C. B.~P.,  1984, \aap, 137, 185

\bibitem[\protect\citeauthoryear{{Yoshida}, {Sheth} \& {Diaferio}}{{Yoshida}
  et~al.}{2001}]{2001MNRAS.328..669Y}
{Yoshida} N.,  {Sheth} R.~K.,    {Diaferio} A.,  2001, \mnras, 328, 669

\bibitem[\protect\citeauthoryear{{Zhao}, {Jing}, {Mo} \& {B{\"o}rner}}{{Zhao}
  et~al.}{2009}]{2009ApJ...707..354Z}
{Zhao} D.~H.,  {Jing} Y.~P.,  {Mo} H.~J.,    {B{\"o}rner} G.,  2009, \apj, 707,
  354

\bibitem[\protect\citeauthoryear{{Zirakashvili} \& {Aharonian}}{{Zirakashvili}
  \& {Aharonian}}{2007}]{2007A&A...465..695Z}
{Zirakashvili} V.~N.,  {Aharonian} F.,  2007, \aap, 465, 695

\bibitem[\protect\citeauthoryear{{Zirakashvili} \& {Aharonian}}{{Zirakashvili}
  \& {Aharonian}}{2010}]{2010ApJ...708..965Z}
{Zirakashvili} V.~N.,  {Aharonian} F.~A.,  2010, \apj, 708, 965

\end{thebibliography}
\bibliographystyle{mn2e}

\clearpage
\appendix

\section{Supporting material for our semi-analytic model: spectrum and spatial distribution}
 \label{appendixA}
 Here we present additional details on the spectral and spatial distribution of
 cosmic rays (CRs) that are important for the consistency of our semi-analytic
 model.

\subsection{Formalism for the semi-analytic modelling}
\label{sect:formalism_analyic_model}
The CR spectrum for each SPH particle at the dimensionless proton momentum $p
=P_\p/\mpc = 1$ is given by $f(p=1,\vr)= C(\vr)$, where we assume that the
low-momentum cutoff $q < 1$. The CR normalization is denoted by
$C=\tilde{C}\rho/m_\p$, where $C$ has units of inverse volume. At each radial
bin, we use the the volume weighted $C$ to calculate the normalized spectrum
through
\begin{eqnarray}
  \langle f \rangle_\vv(p=1,R) &\equiv& f_{\vv,\rmn{sim}}(p=1,R)
  =  \frac{1}{V}\sum_{\rmn{SPH},i}\frac{M_i}{\rho_i}C_i  \nonumber\\
  &=& 
  \frac{1}{V}\sum_{\rmn{SPH},i}\tilde{C}_i\,\frac{M_i}{m_\p}\,,
\end{eqnarray}
where the sum extends over SPH particles labeled by $i$. In our semi-analytic
formalism, we provide a fit to the mass weighted $\tilde{C}$, denoted by
$\tilde{C}_\rmn{M}$, and use the gas density profile (which in fact is the
volume weighted density $\rho_\vv$, that we simply denoted by $\rho$ throughout
the paper). The two methods to calculate $f_\vv$ are equivalent for all radial
bins, since
\begin{eqnarray}
f_{\vv,\rmn{ana}}(p=1,R)
&=&\tilde{C}_\rmn{M} \frac{\rho_\rmn{v}}{m_\p} =
\frac{\sum_{\rmn{SPH},i}\tilde{C}_i\,M_i}{\sum_{\rmn{SPH},i}
  M_i}\frac{\sum_{\rmn{SPH},i}\frac{M_i}{\rho_i}\rho_i}{V\,m_\p} \nonumber\\
&=&\frac{1}{V}\sum_{\rmn{SPH},i}\tilde{C}_i\,\frac{M_i}{m_\p} = f_{\vv,\rmn{sim}}(p=1,R)\,.
\end{eqnarray}

We use the CR proton spectrum to calculate the integrated $\gamma$-ray source
density $\lambda$ by integrating the inverse Compton (IC) and pion decay
$\gamma$-ray source functions $s_\gamma(E_\gamma)$ given by equation~(43) and by
equation~(19) of \citet{2004A&A...413...17P}, respectively. The luminosity is
calculated through the volume integral of $\lambda$. In analogue to the CR
spectrum, we show here that the pion decay induced luminosity on an SPH basis
for any radial bin,
\begin{eqnarray}
\mathcal{L}_\rmn{sim} &=& \int\dd V \lambda_{\pig}(E,R) \simeq
\tilde{\lambda}_{\pig}(E)\sum_{\rmn{SPH},i}\tilde{C}_i\,\frac{\rho_i^2}{\rho_0^2}\,
\frac{M_i}{\rho_i} \nonumber\\
 &=& \tilde{\lambda}_{\pig}(E)\sum_{\rmn{SPH},i}\tilde{C}_i\,\frac{\rho_i\,M_i}{\rho_0^2}\,,
\end{eqnarray}
  is equivalent to our semi-analytic $\gamma$-ray luminosity
\begin{eqnarray}
\label{eq:sem_ana_L}
 \mathcal{L}_\rmn{ana}&=&\int\dd V \lambda_{\pig}(E,R) = \lambda_\vv V \simeq
 \tilde{\lambda}_{\pig}(E)\tilde{C}_\rmn{M}\frac{\rho_\vv^2}{\rho_0^2} V \nonumber\\ 
 &=&\tilde{\lambda}_{\pig}(E)\frac{\sum_{\rmn{SPH},i}
   M_i\,\tilde{C}_i}{\sum_{\rmn{SPH},i} M_i}\sum_{\rmn{SPH},i}\frac{\rho_i^2}{\rho_0^2}
\frac{M_i}{\rho_i} \nonumber\\ 
 &\simeq& \tilde{\lambda}_{\pig}(E)\sum_{\rmn{SPH},i} \tilde{C}_i\,\frac{\rho_i\,M_i}{\rho_0^2}
 = \mathcal{L}_\rmn{sim}\,.\qquad
\end{eqnarray}
In the semi-analytic expression in equation~(\ref{eq:sem_ana_L}), we have used
that the spectral part is separable from the spatial part in the first
approximation, and that $\tilde{C}$ is only a weak function of radius in the
second approximation. Both these assumptions are validated by the universal CR
spectrum and approximate spatial universal profile that we found across our
cluster sample where the details are shown in Section~\ref{sect:CRp_spec}.

\subsection{Spectral convergence test}
\label{app:convergence}
\begin{figure}
\begin{minipage}{1.0\columnwidth}
  \includegraphics[width=1.0\columnwidth]{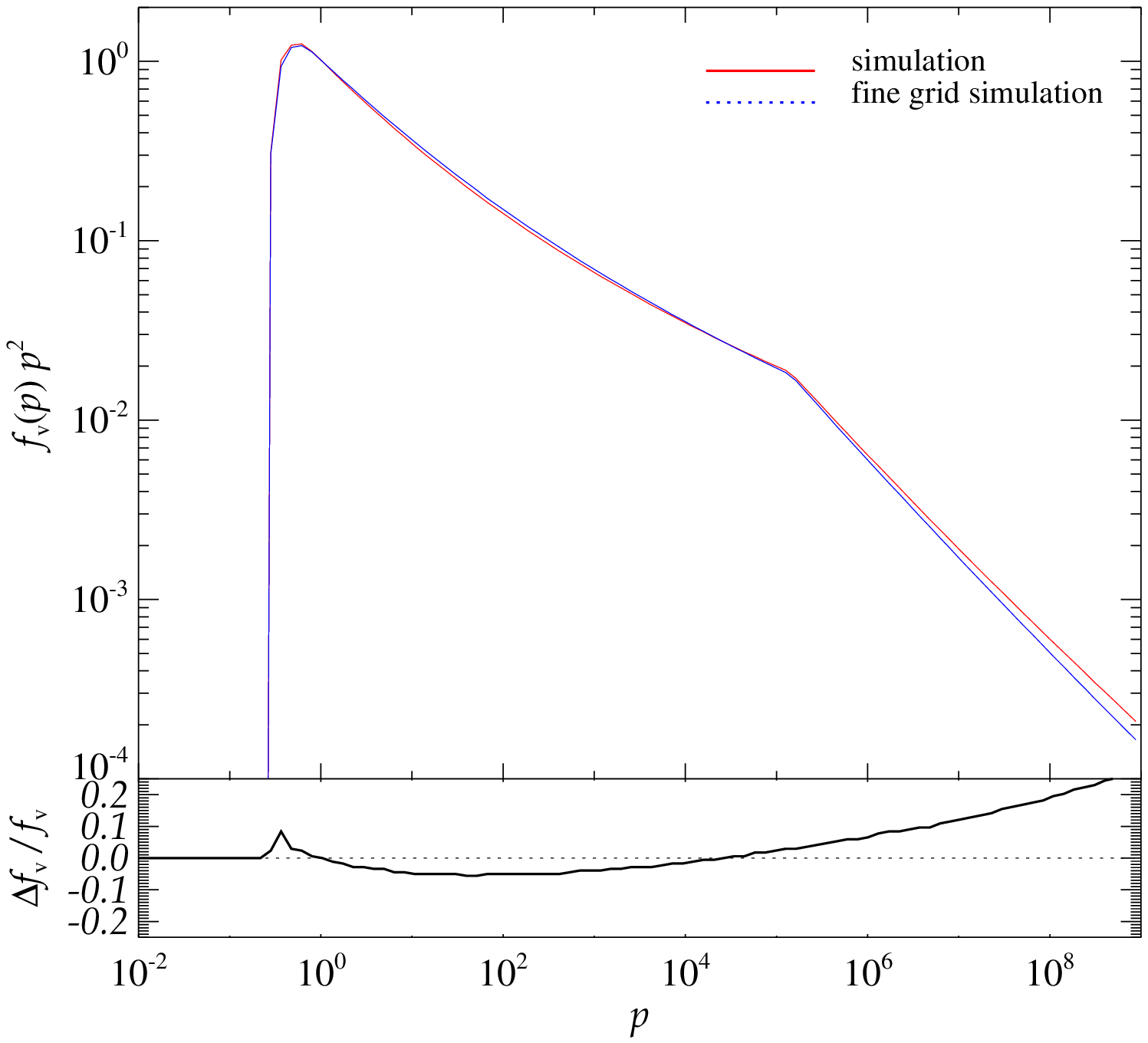}
  \caption{Convergence test for the CR spectrum: the main panel shows the volume
    weighted CR spectrum as a function of dimensionless CR momentum
    $p=P_\p/\mpc$ for the small CC cluster g676. The red line represents our
    standard spectral resolution using $\boldsymbol{\alpha}=(2.1, 2.3, 2.5, 2.7,
    2.9)$, and the blue line shows the finer spectral resolution given by
    $\boldsymbol{\alpha}=(2.1, 2.2, 2.3, 2.4, 2.5, 2.7)$. The bottom panel shows
    the difference between the two simulations normalized with the finer
    spectral resolution simulation.}
    \label{fig:conv.test}
\end{minipage}
\end{figure}

Here we test the convergence of our updated CR model that we use in our
simulations, where we account for both the spatial and spectral information of
the CRs. In particular we allow for multiple CR populations, where each
population is characterized by a fixed spectral index. For the test we run
simulations with a finer spacing between the spectral indices,
$\boldsymbol{\alpha}=(2.1, 2.2, 2.3, 2.4, 2.5, 2.7)$, where we omitted the last
bin $\alpha=2.9$ that was found to have negligible contribution to the
spectrum. In Fig.~\ref{fig:conv.test} we show the result of the convergence test
of the CR spectrum for the  small CC cluster g676. The red line shows g676
with our standard spectral resolution and the blue line shows the g676
simulation with a finer spectral resolution. The difference between the two
simulations normalized with the spectrum from the better spectral resolution
simulation is shown in the lower panel. In the GeV and TeV region the difference
is less than 10 percent, showing that we are accurately able to capture the CR
spectrum with our choice of $\alpha$ that has a wider spacing between the
spectral indices.

\subsection{Variance of the spatial CR distribution}
\label{app:analytic_modeling}
In this section we discuss the details of our spatial semi-analytic modelling
and address the particle-by-particle variance of the spatial CR profile within a
cluster. To this end, we plot the correlation space density of the dimensionless
CR normalization $\tilde{C}$ versus the overdensity of gas
$\delta_\rmn{gas}=\rho/(\Omega_\rmn{b}\rho_\rmn{cr})-1$ for individual
clusters. We over-plot the mean together with the 1-sigma standard deviation
$\tilde{C}_\rmn{M}$ as a function of overdensity. The scatter of 0.3 dex is
roughly constant with density. This variance is most probably caused by the
variance in shock strength and associated CR acceleration efficiency among
different fluid elements in the past history of a cluster.  If CRs are
adiabatically transported into a cluster, we expect a weak radial dependence of
the dimensionless normalization of the CR spectrum, $\tilde{C} \sim
\rho^{(\alpha-1)/3}$ \citep{2007A&A...473...41E}. However, we stress that the
scale dependence of DM halos and which seems to be inherited by the gas and CR
distribution could have an important effect in shaping $\tilde{C}$. Similarly,
non-adiabatic CR transport effects could have another important impact on
$\tilde{C}$. The outer cluster regions (particularly of mergers) typically host
weak to intermediate strength shocks that increase $\tilde{C}$ in that
region. In contrast, in the very central cluster regions, the hadronic losses
dominate and cause a suppression of $\tilde{C}$.

\begin{figure}
\begin{minipage}{1.0\columnwidth}
  \includegraphics[width=1.0\columnwidth]{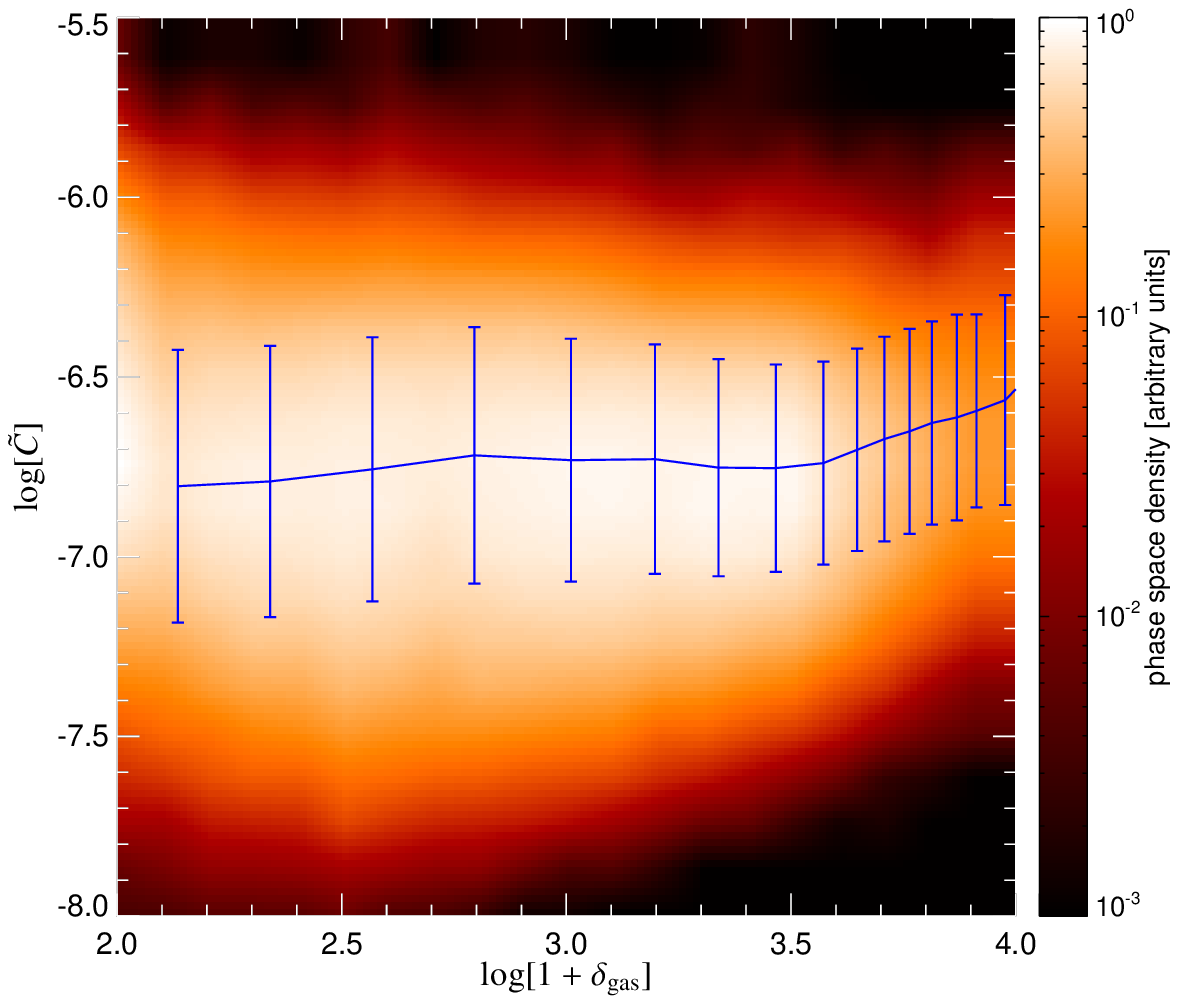}
  \caption{Correlation space density between the dimensionless normalization of
    the CR spectrum, $\tilde{C}$, and the overdensity $\delta_\rmn{gas}= \rho /
    (\Omega_\rmn{b} \rho_\rmn{cr})-1$, both in logarithmic representation. Shown
    is our simulated post-merging g72a cluster. The solid line shows the mean
    $\tilde{C}_\rmn{M}$ as a function of the overdensity and the error bars the
    1-sigma standard deviation.}
    \label{fig:PSfigure}
\end{minipage}
\end{figure}

\section{Primary, secondary electrons, and inverse Compton emission}
\label{appendixB}
In Section~\ref{sec:secondary_e} we introduce the steady state cosmic ray
electron (CRe) population for secondary electrons. These electrons are created
through CR proton-proton interactions via charged pions decaying. The primary
electrons in our simulations are accelerated through {\em diffusive shock
  acceleration} using the thermal leakage model originally proposed by
\citet{1981JGZG...50..110E}. In Section~\ref{sec:fe_inj} we start by deriving the
primary steady state CRe population under the assumption that the diffusive time
scale can be neglected. Then we continue the discussion for primary electrons
where we additionally account for an energy dependent diffusion that in
combination with the losses leads to a cutoff in the electron spectrum. In
\ref{sec:pIC} we use these CRe spectra to calculate the integrated secondary and
primary IC source functions. This description follows the approach of
\citet{2007A&A...473...41E}, but we refer the reader to the appendix of
\citet{2008MNRAS.385.1211P} for a detailed discussion about electron
acceleration, cooling, and IC emission within the framework of the SPH
formalism.

\subsection{Secondary electrons}
\label{sec:secondary_e}
Here we provide the equilibrium distribution of secondary CR electrons above a
GeV. It is shaped by the high-energy part of the CR proton population that
interacts with ambient gas and creates electrons through hadronic
reactions. These electrons cool through IC and synchrotron radiative losses. The
CRe equilibrium spectrum \citep{2008MNRAS.385.1211P} is derived through the
balance of injection and losses, and is steeper by one compared to the CR proton
spectrum ($\aei=\alpha_i + 1$). The volume weighted spectrum is given by
\begin{eqnarray}
\label{eq:fe_hadr}
f_\e (p_\e)\, \dd p_\e &=& \sum_{i=1}^{3}C_{\rmn{e},i} p_\e^{-\aei} \,\dd p_\e \\
C_{\rmn{e},i} &=& \rho^2_\rmn{v}(R)\,\tilde{C}_\rmn{M}(R)\,\frac{16^{2-\aei}
  \sigma_{\rmn{pp},\,i}\, \Delta_i\,c^2}
     {(\aei - 2)\,\sigma_\rmn{T} \,m_\p\,(\eps_B + \eps_\rmn{ph})}
\left(\frac{m_\p}{m_\e}\right)^{\aei-3} \nonumber\\
\label{eq:C_e}
&&
\end{eqnarray}
where $\tilde{C}_\rmn{M}(R)$ is the mass weighted CR proton normalization
(equation~\ref{eq:tildeCM}), $\Delta_i$ the relative normalization for each CR
population (equation~\ref{eq:Delta}), and $\sigma_{\rmn{pp},\,i}$ denote the
effective inelastic cross-section for proton-proton interactions and is defined
in Section~\ref{sect:schematic_overview}. In addition, $\sigma_\rmn{T}$
represents the Thompson cross-section, $\eps_B$ is the magnetic energy density,
and $\eps_\rmn{ph}$ denotes the photon energy density, taken to be that of CMB
photons.

\subsection{Spectrum of shock-accelerated electrons}
\label{sec:fe_inj}
In clusters of galaxies, the dynamical and diffusive time scales of electrons
are much longer compared to the shock injection and IC/synchrotron time
scales. The radio synchrotron emitting electron population cools on such a short
time scale $\tau_\rmn{sync} < 10^8$~yrs (compared to the very long dynamical
time scale $\tau_\rmn{dyn} \sim 1$~Gyr) that we can describe this by
instantaneous cooling at each timestep -- in contrast to the CR protons.  In
combination with the fact that the CRes have a negligible pressure contribution,
this enables us to account for the CRes in the post-processing.  In this
instantaneous cooling approximation, there is no steady-state electron
population and we would have to convert the energy from the electrons to inverse
Compton and synchrotron radiation. Instead, we introduce a virtual electron
population that lives in the SPH-broadened shock volume only, defined to be the
volume of energy dissipation. Within this volume, which is co-moving with the
shock, we can use the steady-state solution for the distribution function of CR
electrons and we assume no CR electrons in the post-shock volume, where no
energy dissipation occurs. Thus, the CR electron equilibrium spectrum can be
derived from balancing the shock injection with the IC/synchrotron cooling:
above a GeV and below 30~TeV, it is given by
\begin{eqnarray}
\label{eq:f_eq_full}
f_\e(p_\e)\,\dd p_\e &=& C_\e\, p_\e^{-\alpha_\e}\,\dd p_\e, \\
\label{eq:C_e_prim}
C_\e &=& \frac{3\, C_\rmn{inj}\, m_\e c}
{4\,(\alpha_\e - 2)\,\sigma_\rmn{T}\,\tau_\inj\,(\eps_B + \eps_\rmn{ph})}\,.
\end{eqnarray}
Here, we introduced the unit-less electron momentum $p_\e=P_\e/(m_\e c)$, where
$P_\e$ is the electron momentum.  The spectral index of the equilibrium electron
spectrum is denoted by $\alpha_\e = \alpha_\inj + 1$, where $\alpha_\inj$ is the
spectral index of the injected electron population in one-dimensional momentum
space given by equation~(\ref{eq:alpha_rc}). The CR electron normalization scales
linearly with the gas density $C_\e \propto \rho$ which we evolve dynamically in
our simulations and depends indirectly on $\alpha_\inj$ and the dissipated
energy rate per electron, $\dot{E}_\rmn{diss}$, through the normalization of the
injected CRes, $C_\rmn{inj}$ \citep[for further details
  see][]{2008MNRAS.385.1211P}. $\tau_\inj$ represent the electron injection time
scale, which depends on the time it takes for an electron to pass through the
broadened shock. 

The shape of the steady-state electron power-law spectrum in
equation~(\ref{eq:f_eq_full}) changes when the energy of the accelerated
electrons reach a maximum electron energy that is determined by the competition
between the diffusive acceleration, and radiative synchrotron and IC losses.  In
the cutoff region, the spectrum is proportional to the product of two terms -- a
power-law term ($p_\e^{4.5-\alpha_\rmn{inj}}$ which includes the phase space
volume) and a super-exponential term, $\exp(-p_\e^2/p_\rmn{max}^2)$. The first
term reflects a pile-up of electrons as their cooling time becomes comparable to
the acceleration time. The exponential term, however, effectively cancels this
pile-up feature which results in a prolonged power-law up to the electron cutoff
momentum $p_\e\sim p_\rmn{max}$, where a steeper super-exponential cutoff takes
over \citep{2007A&A...465..695Z}. Applying the theory of plane-parallel shock
acceleration that is justified because of the large curvature radius of the
shock, the equilibrium electron distribution function at each position in the
shock is given by
\begin{eqnarray}
\label{eq:fe_exp}
f_\e(x,p_\e) &=& C_\e p_\e^{-\alpha_\rmn{inj}}
\left[1+j(x, p_\e)\right]^{\delta_\rmn{e}}\exp\left[\frac{-p_\e^2}{p_\rmn{max}^2(x)}\right]\,.
\end{eqnarray}
Here, $x$ is a spatial coordinate along the shock normal, measured from the
shock position and the electron cutoff momentum is
\begin{eqnarray}
  p_\rmn{max}(x) &\equiv& (F+|x|G)^{-1}\,, \qquad \rmn{where}  \\
  G&\equiv&\frac{u}{4p_0a\,\kappa}\,, \qquad \rmn{and} \\
  F&\equiv&\frac{\alpha_\inj+2}{4p_0a_1}+\frac{\alpha_\inj-1}{4p_0a_2}\,.
\end{eqnarray}
The shape of the pile-up region of electrons in equation~(\ref{eq:fe_exp}) is given by
$\delta_\rmn{e}=9/5$ and the power-law function
\begin{eqnarray}
\label{eq:fe_pileup}
j(x, p_\e) = 0.66\left(\frac{p_\e}{p_\rmn{max}(x)}\right)^\frac{5}{2}\,.
\end{eqnarray}

The upstream quantities have the index 1 and the downstream quantities the index
2.  The flow velocity in the inertial frame of the shock is denoted by $u$
($u_1=\upsilon_\rmn{s}$, $u_2=u_1/r_\rmn{c}$ where $r_\rmn{c}=\rho_2/\rho_1$ is the
density compression factor at the shock), and $p_0$ represents the injection
momentum normalized with $m_\e c$. Note that the electron cutoff momentum
$p_\rmn{max}$ is independent of $p_0$, as expected.  The ratio between the
cooling and diffusive acceleration time scales given by
\begin{eqnarray}
  a=\frac{u^2\tau_\rmn{loss}}{4\kappa}\,.
\end{eqnarray}
Here, the inverse energy loss time scale of an electron dominated by synchrotron
and inverse-Compton losses is given by
\begin{eqnarray}
\tau_\rmn{loss}^{-1} = \frac{\dot{E}}{E}=
\frac{4 \sigma_\rmn{T}\,p_0}{3m_\e c} ({\eps_B + \eps_\rmn{ph}})\,.
\end{eqnarray}
The CRe acceleration depends on the type of diffusion which we assume to be
parallel to the magnetic field and to be in the Bohm-limit (as motivated by
young supernova remnants observations, see also
Section~\ref{section:diffusive_shock_acceleration}). The energy dependent
diffusion constant is given by
\begin{eqnarray}
 \kappa=\eta\frac{c\,r_g}{3}=\eta\frac{p_\e\,m_\e c^3}{3Z\,e\,B}\,,
\end{eqnarray}
which we assume to be the same across the shock, i.e.{\ }$\kappa\simeq \kappa_1
\simeq \kappa_2$. We also use that the magnetic field fluctuations $\delta B$
are of the same amplitude as the magnetic field $\eta = (B/\delta B)^2 \sim 1$.

\begin{figure}
\begin{minipage}{1.0\columnwidth}
  \includegraphics[width=1.0\columnwidth]{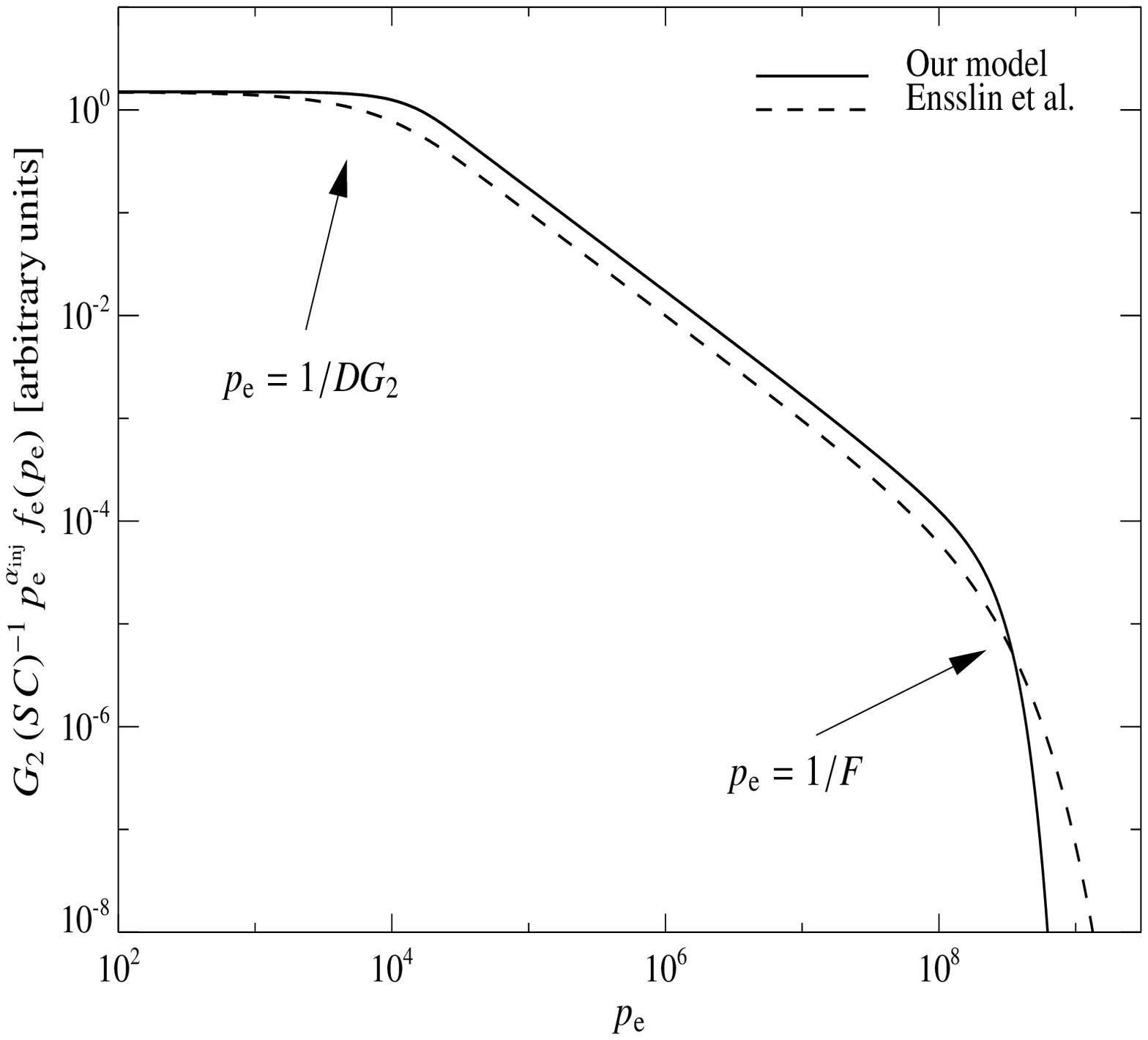}
  \caption{Comparison of the spectrum of primary electrons. We compare the
    electron spectrum in our model (equation~\ref{eq:fe_cool}) using results
    from \citet{2007A&A...465..695Z} of the high-energy cutoff region to a
    spectrum from \citet{1998A&A...332..395E} where an energy independent
    diffusion has been assumed. There is a break in the spectrum at
    $p_\e=1/D\tilde{G}_2$ and a cutoff at $p_\e=1/F$. The comparison shows
    similar spectral shape up to the cutoff region where the energy dependent
    Bohm diffusion in our model induces a much steeper cutoff. The offset of the
    two spectra between the break and cutoff regimes is caused by the integrated
    pile-up regime indicated by $J(p_\rmn{e})$ (equation~\ref{eq:fe_pileup})
    that effectively shifts the break by a factor of two to higher energies.}
    \label{fig:elcc_comp}
\end{minipage}
\end{figure}

Bohm diffusion becomes more effective at higher energies, which causes the
cooling induced cutoff in the electron spectrum to move to lower energies as the
electrons are transported advectively with the flow downstream. Integration over
the post-shock volume causes the cutoffs to add up to the power-law that is
steeper by unity compared to the injection power-law. Thus the steady-state
spectrum that balance losses through IC/synchrotron cooling with gains through
electron shock acceleration, is derived through the integral over the IC
radiating volume Vol = $SD$ (where $S$ is the surface area, and $D$ the
thickness). It is given by
\begin{eqnarray}
f_\e(p_\e) &=& S\int_0^D \dd x\,f(x,p_\e) \nonumber\\ &\simeq&
\frac{SC_\e}{\tilde{G}_2}\frac{\sqrt{\pi}}{2}p_\e^{-\left(\alpha_\rmn{inj}+1\right)}
\left[1+J(p_\e)\right]^{\delta_\rmn{e,cool}}\nonumber\\ &\times&
\left[\rmn{erf}\left(p_\e D\tilde{G}_2+p_\e F\right)-\rmn{erf}\left(p_\e
F\right)\right]\,,
\label{eq:fe_cool}
\end{eqnarray}
where erf is the standard Gaussian error function \citep{1965hmfw.book.....A}.
We do not expect the main characteristics of the pile-up region to change much
when it is integrated over the post-shock regime \citep{2007A&A...465..695Z}; to
obtain a consistent semi-analytic formula of the spectrum, we determine the
specific values of $J(p_\e)$, $\tilde{G}_2$, and $\delta_\rmn{e,cool}$ through
fits to numerically integrated spectra (see below).  The distribution function
in equation~(\ref{eq:fe_cool}) has a break in the spectrum at the electron
momentum $p_\e=p_\rmn{e,break}=1/D\tilde{G}_2$ and a cutoff at
$p_\e=p_\rmn{e,cut}=1/F$. The electron momentum cutoff is determined from the
strength of the magnetic field and the properties of electron diffusion in the
shock \citep{webb84}, and are given by
\begin{eqnarray}
p_\rmn{e,cut}=\frac{1}{F}=\frac{1}{\frac{\alpha_\rmn{inj}+2}{4p_0a_1}+\frac{\alpha_\rmn{inj}-1}{4p_0a_2}}=
\frac{3e\,B\,\tau_\rmn{loss}\,p_0}{p_\e \,m_\e c^3}
\frac{1}{\frac{\alpha_\rmn{inj}+2}{u_1^2}+\frac{\alpha_\rmn{inj}-1}{u_2^2}}\,.
\end{eqnarray}
The equivalent electron energy cutoff in the relativistic regime is given by  
\begin{eqnarray}
  \label{eq:Emax}
E_\rmn{max} = p_\rmn{e,cut}\,m_\e c^2 = u\left[\frac{6\pi e}{\sigma_\rmn{T}}\frac{B}{B^2+B_\rmn{CMB}^2}
\frac{r_\cc-1}{r_\cc\left(r_\cc+1\right)}\right]^{0.5}\,.
\end{eqnarray}
In Fig.~\ref{fig:elcc_comp} we show a comparison between our electron
distribution given by equation~(\ref{eq:fe_cool}) compared to the electron
spectrum in \citet{1998A&A...332..395E} where an energy independent diffusion
has been assumed. To plot the spectrum, we used typical values for the
temperature, Mach number and density in the virial region. These values are
found in Table~\ref{tab:diffusion}. In addition we assumed a magnetic field
amplification in the shock of a factor ten, resulting in a magnetic field of
about $6\,\mug$. The spectral comparison show that the spectral shapes are
similar up to the cutoff region where the energy dependent Bohm diffusion in our
model induces a steeper cutoff. Furthermore, the offset in the normalization of
the two spectra between the break and cutoff regime is caused by the integrated
pile-up regime (equation~\ref{eq:fe_pileup}) that effectively increases the
energy of the break from $1/D\,G_2$ to $1/D\,\tilde{G}_2$.

\begin{figure}
\begin{minipage}{1.0\columnwidth}
  \includegraphics[width=1.0\columnwidth]{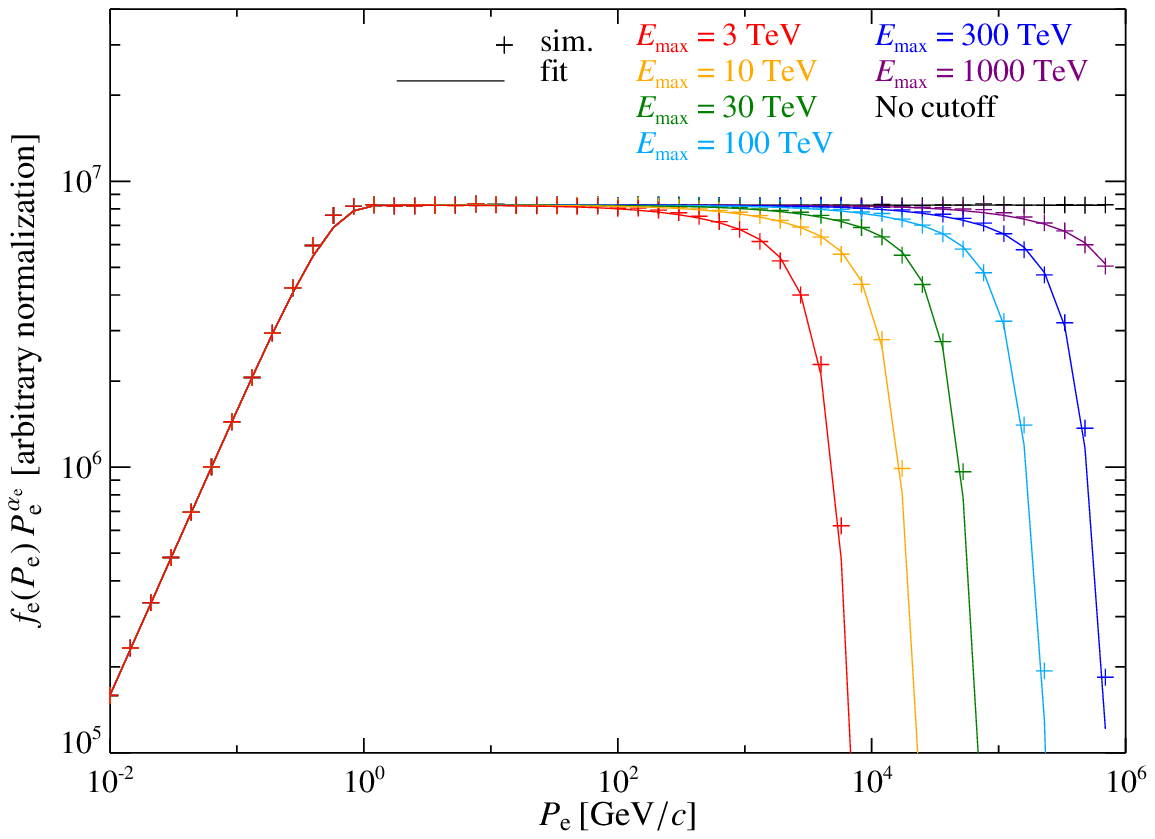}
  \caption{Cooled electron distribution of primary electrons for different
    cutoff energies.  We show the electron distribution for a fixed spectral
    index $\alpha_\e=3.1$ and varying electron cutoff energies $E_\rmn{max}$ in
    raising order; 3~TeV (red), 10~TeV (yellow), 30~TeV (green), 100~TeV (light
    blue), 300~TeV (blue), 1000~TeV (purple), and no cutoff (black). The data
    (crosses) is generated by numerically integrating
    equation~(\ref{eq:fe_cool}) for different cutoff energies. The fit (solid
    lines) are derived using three free parameters, where the fitted variables
    are shown in equation~(\ref{eq:fe_cool_fit}).}
    \label{fig:elec_cool_fit}
\end{minipage}
\end{figure}

In Fig.~\ref{fig:elec_cool_fit} we show both the cooled electron spectra that we
derive by numerically integrating $f_\e(x,p_\e)$ over the shock as well as the
fitted spectra. For $f_\e(p_\e)$ we use typical values for both the strength of
the shocks ($\alpha_\e=3.1$) and the break in the electron spectrum which we fix
at a constant momentum ($p_\rmn{e,break}=300\,\rmn{MeV}/m_\e c^2$). In our fit,
we allow for three independent fit variables, $A_1$, $A_2$, and $A_3$. We find
that,
\begin{eqnarray}
A_1&=&0.4\,,\qquad A_2=1.45\,, \qquad A_3=1.95\,,\qquad \rmn{where}\,, \nonumber \\
J(p_\e) &=& A_1\left(\frac{p_\e}{p_\rmn{e,cut}}\right)^{A_2}\,, \nonumber \\
\delta_{\rmn{e,cool}} &=&  4.5 - A_2 \,, \qquad \rmn{and}  \nonumber \\
\tilde{G}_2 &=& \frac{G_2}{A_3} \,.
\label{eq:fe_cool_fit}
\end{eqnarray}
We note that the factor 4.5 in $\delta_{\rmn{e,cool}}$ follows from
\citet{2007A&A...465..695Z} and represents the spectral flattening of the
power-law in the low-energy regime up to the regime where the exponential cutoff
dominates.

\subsection{Inverse Compton radiation}
\label{sec:pIC}
Inverse Compton scattering of CMB photons off ultra-relativistic electrons with
Lorentz factors of $\gamma_\e> 10^4$ redistributes these photons into the hard
X-ray/$\gamma$-ray regime according to equation~(\ref{eq:ICenergy}).  The
integrated IC source density $\lambda_\IC$ for an isotropic power-law
distribution of CR electrons, as described by equation~(\ref{eq:f_eq_full}), is
obtained by integrating the IC source function $s_\gamma(E_\gamma)$ in
equation~(43) of \citet{2004A&A...413...17P} (in the case of Thomson scattering)
over an energy interval between observed photon energies $E_1$ and $E_2$
yielding
\begin{eqnarray}
\label{eq:IC}
\lambda_\IC(E_1, E_2) &=& \int_{E_1}^{E_2} \dd E_\IC\,s_\IC(E_\IC) \\
&=& \tilde{\lambda}_0\,
f_\IC (\alpha_e)\, 
\left[\left(\frac{E_\IC}{k T_\mathrm{CMB}}\right)^{-\alpha_\nu}
  \right]_{E_2}^{E_1}, \\
\label{eq:IC_f_norm}
f_\mathrm{IC} (\alpha_e) &=& \frac{2^{\alpha_\e+3}\, 
  (\alpha_\e^2 + 4\, \alpha_\e + 11)}
  {(\alpha_\e + 3)^2\, (\alpha_e + 5)\, (\alpha_e + 1)} \nonumber\\
&\times&\Gamma\left(\frac{\alpha_e + 5}{2}\right)\,
  \zeta\left(\frac{\alpha_e + 5}{2}\right)\,,\\
\label{eq:IC_l_norm}
\mbox{and }\tilde{\lambda}_0 &=& 
\frac{16\, \pi^2\, r_\e^2\, C_\e\,
      \left(k T_\mathrm{CMB}\right)^3\,}{(\alpha_\e-1)\,h^3\, c^2}\,, 
\end{eqnarray}
where $\alpha_\nu = (\alpha_\e-1)/2$ denotes the photon spectral index, $r_\e =
e^2 /(m_\e\, c^2)$ the classical electron radius, and $\zeta(a)$ the Riemann
$\zeta$-function \citep{1965hmfw.book.....A}. The CRe normalization $C_\e$ is
given by equation~(\ref{eq:C_e}) and (\ref{eq:C_e_prim}) for the secondary and
primary electrons, respectively.

In the following, we provide a simple analytic formula that captures the primary
inverse-Compton emission from galaxy clusters for strong shocks and intermediate
strength shocks (i.e. $\alpha_\rmn{inj} \eqsim 2-2.7$) with an accuracy of five
percent or better. In addition, we want the analytic formula to be valid in the
full energy range of pIC emission, i.e. not limited by the Klein-Nishina (KN)
suppression where the center of mass energy of photons becomes comparable to the
electron mass and the less efficient energy transfer from electron to photon
causes a break in the photon spectrum.

\begin{figure}
\begin{minipage}{0.99\columnwidth}
  \includegraphics[width=0.99\columnwidth]{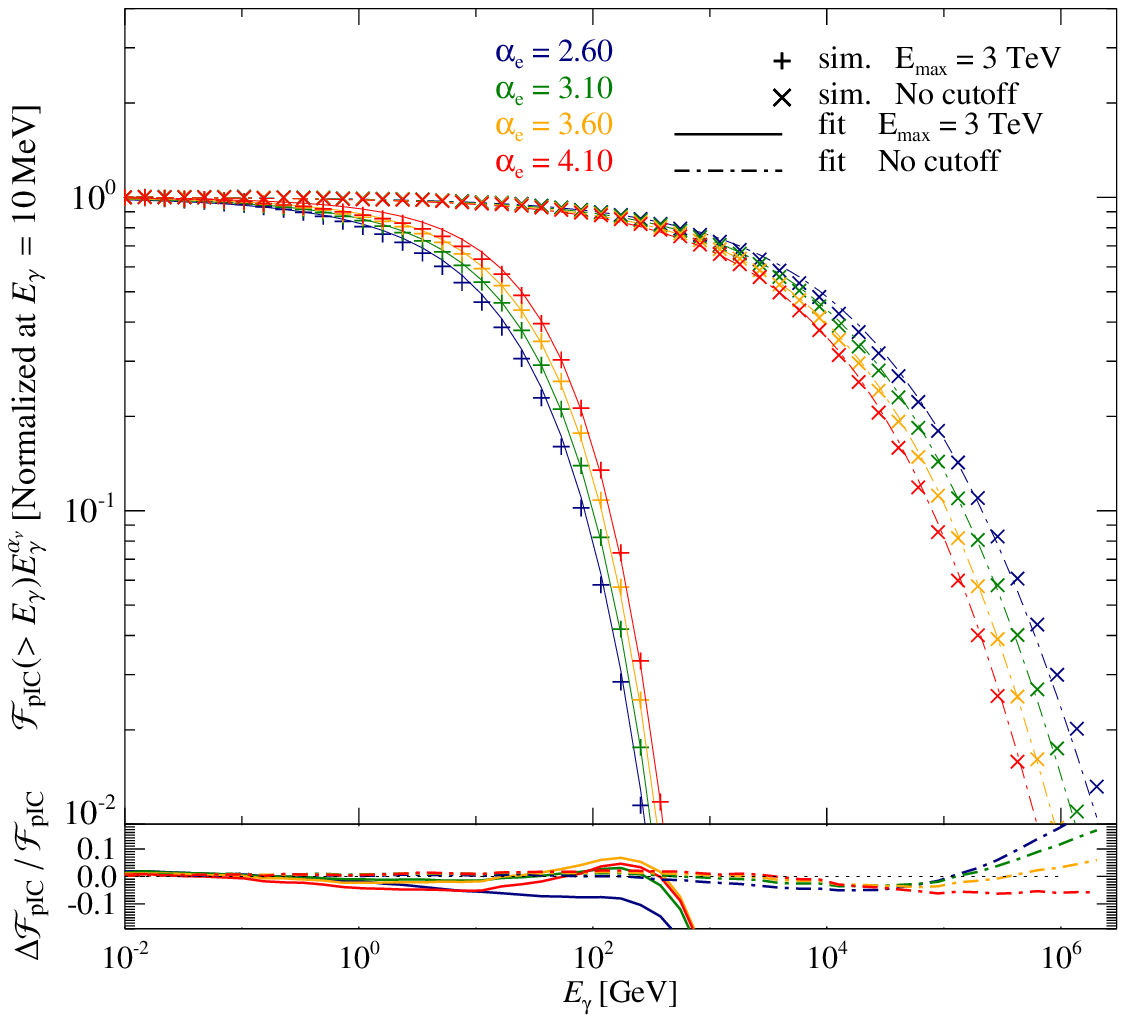}
  \caption{The primary inverse Compton $\gamma$-ray number flux weighted by
    photon energy for different electron spectral indices and electron cutoff
    energies. \emph{Main panel:} The data is numerically calculated using a
    cutoff energy $E_\rmn{max}=3\,\rmn{TeV}$ (crosses) and without a cutoff (X)
    for different electron spectral indices; $\alpha_\e=2.6$ (blue),
    $\alpha_\e=3.1$ (green), $\alpha_\e=3.6$ (yellow), and $\alpha_\e=4.1$
    (red). The fits to the data using $E_\rmn{max}=3\,\rmn{TeV}$ (no cutoff) are
    represented by the solid (dash-dotted) lines. \emph{Bottom panel:} The
    difference between the data and the fits. The flux from the fits using the
    dominating electron spectral indices $\alpha_\e=3.1$ and $\alpha_\e=3.6$,
    agree within a few percent with the data in the dominating flux regime. The
    flux from the fits using electron spectral indices of $\alpha_\e=2.6$ and
    $\alpha_\e=4.1$ have slightly lower precision, and agree within 10 percent
    with the data in the dominating flux regime.}
    \label{fig:pIC_alpha_fit}
\end{minipage}
\end{figure}

Using the exact spectra in the asymptotic low- and high-energy regime, together
with the result of numerical calculations at intermediate energies, we
parametrize the integrated source function for pIC emission by,
\begin{eqnarray}
\lambda_\pic &=& \tilde{\lambda}_0(\zeta_\rmn{e,max}, C_\rmn{e})\,f_\mathrm{IC}(\alpha_\e)\,
f_\mathrm{KN}(E_\IC,\alpha_\e)\,
\left(\frac{E_\IC}{k T_\mathrm{CMB}}\right)^{-\frac{\alpha_\e-1}{2}} \nonumber\\
&&
\times\,\left(1+0.84\sqrt{\frac{\,E_\IC}{E_{\IC, \rmn{cut}}}}\,\right)^
{\delta_{\rmn{IC}}(E_\IC, \alpha_\e)}
\rmn{exp}\left(-\sqrt{\frac{4.07\,E_\IC}{E_{\IC, \rmn{cut}}}}\right), \nonumber\\
&&
\label{eq:pIC_fit_appendix}
\end{eqnarray}
where Bohm diffusion has been assumed. The normalization constants
$\tilde{\lambda}_0(\zeta_\rmn{e,max}, C_\rmn{e})$ and $f_\mathrm{IC}(\alpha_\e)$
are derived in equations~(\ref{eq:IC_l_norm}) and (\ref{eq:IC_f_norm}),
respectively.  The shape of the IC spectrum without an exponential cutoff scales
as $E_\IC^{-\alpha_\nu}$ in the Thomson regime, and steepens to
$E_\IC^{-\alpha_\e}\rmn{log}(E_\IC)$ in the KN suppressed high-energy regime
\citep{1970RvMP...42..237B}. In Fig.~\ref{fig:pIC_alpha_fit} we fit the
numerically calculated intermediate energy regime using the following
parametrization,
\begin{eqnarray}
f_\KN(E_\IC,\alpha_\e) &=& \left\{1+\left\{\frac{E_\IC}{E_{\rmn{KN,1}}}
\left[A_{\KN}\,\rmn{log}\left(\frac{E_\IC}{E_{\rmn{KN,2}}}\right)+1\right]
^{\frac{1}{\alpha_\KN}}\right\}^{\beta_\KN}\right\}^{\frac{\alpha_\KN}{\beta_\KN}} \nonumber \\
&&
\label{eq:KN_fit}
\end{eqnarray}
which respects this asymptotic behavior of the IC spectrum. We use three free
independent fit variables, $E_{\rmn{KN,1}}$, $\beta_\KN$, and $A_{\KN}$ and find
that
\begin{eqnarray}
E_{\rmn{KN,1}} &=& 2\times 10^5\,\rmn{GeV}\,, \nonumber \\
E_{\rmn{KN,2}} &=& \frac{\rmn{GeV}^2}{E_{\rmn{KN,1}}} = 5\times 10^{-6}\,\rmn{GeV}\,,  \nonumber \\
A_{\KN} &=& 0.1 \,\alpha_\nu = 0.025 \,\alpha_\e - 0.025\,, \nonumber \\
\alpha_\KN &=& -\alpha_\e+\alpha_\nu=-\frac{\alpha_\e+1}{2}\,, \nonumber \\
\beta_\KN &=& 0.452\,.
\label{eq:fit_param_KN}
\end{eqnarray}
The resulting spectra accurately describe the pIC emission without an
exponential cutoff for $\alpha_\e \sim 2.5 - 4.5$. In addition, we capture the
shape of the integrated source function for pIC in
equation~(\ref{eq:pIC_fit_appendix}), where we include the super-exponential
cutoff. We find that the shape of the transition region is well approximated by
\begin{eqnarray}
\label{eq:fit_trans_KN}
\delta_{\rmn{IC}}(E_\IC, \alpha_\e) = 
0.529\alpha_\e-0.134\,\rmn{log}_{10}\left(\frac{E_{\IC, \rmn{cut}}}{30\,\rmn{GeV}}\right)\,,
\end{eqnarray}
which depends on both the photon energy and the electron spectral index. In the
process of finding a good fit, we allowed for multiple free parameters in
equation~(\ref{eq:pIC_fit_appendix}). In the end, however, we only allowed
$\delta_{\rmn{IC}}$ to vary and fixed all the other free parameters at
typical values to keep the formula as simple as possible.

\begin{figure}
\begin{minipage}{0.99\columnwidth}
  \includegraphics[width=0.99\columnwidth]{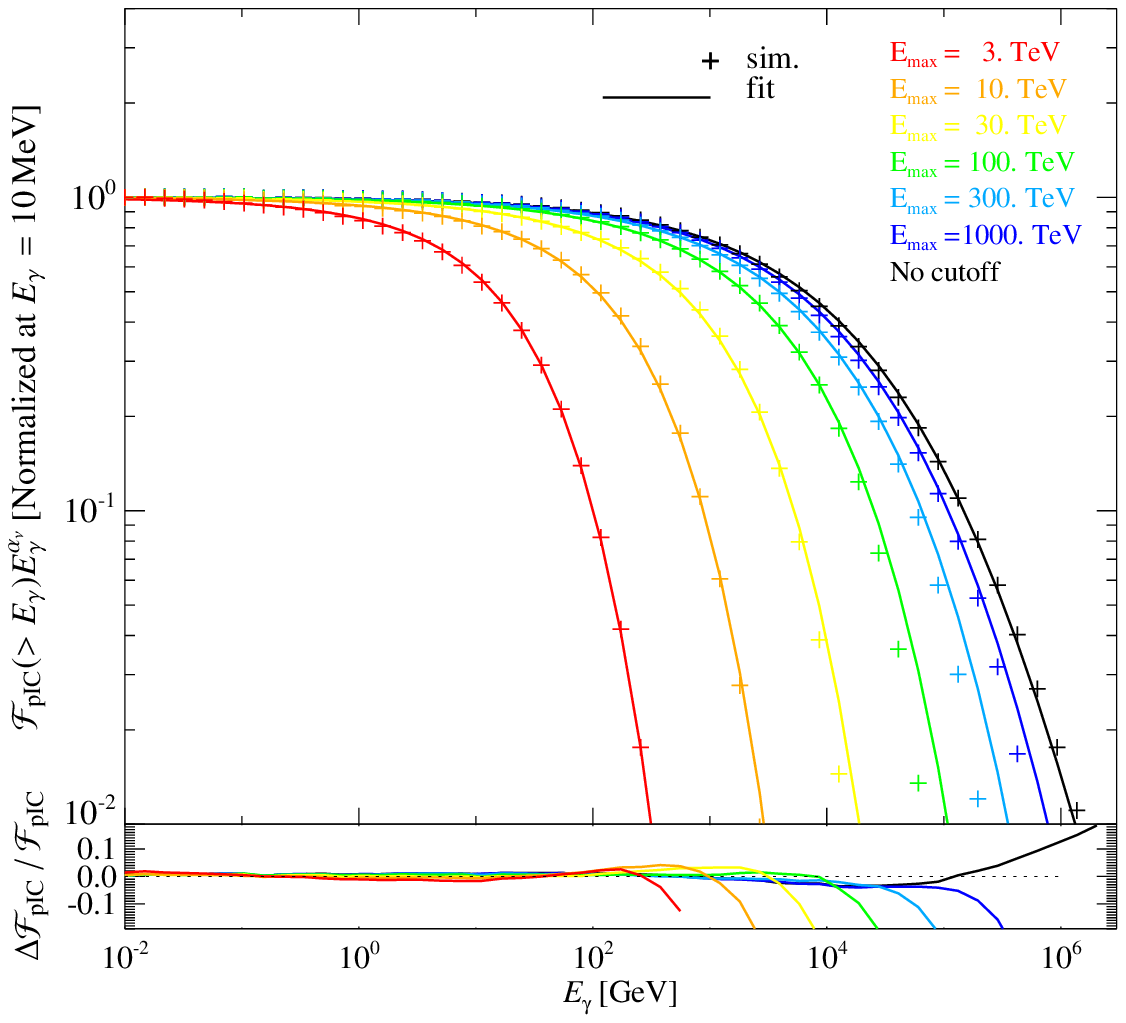}
  \caption{The primary inverse Compton $\gamma$-ray number flux weighted by
    photon energy for different electron cutoff energies. \emph{Main panel:} The
    pIC emission are shown for a fixed electron spectral index
    ($\alpha_\e=3.1$) and varying electron cutoff energies ($E_\rmn{max}$) in
    raising order; 3~TeV (red), 10~TeV (orange), 30~TeV (yellow), 100~TeV
    (green), 300~TeV (light blue), 1000~TeV (blue), and no cutoff (black). The
    numerically calculated data is represented by crosses and the fits are shown
    with solid lines. \emph{Bottom panel:} The difference between the data and
    the fits. The flux from the fits agree within a few percent with the data in
    the dominating flux regime.}
    \label{fig:pIC_fit}
\end{minipage}
\end{figure}

In Fig.~\ref{fig:pIC_fit} we test our analytic formula for pIC emission, given by
equations~(\ref{eq:IC_f_norm}), (\ref{eq:IC_l_norm}) and
(\ref{eq:pIC_fit_appendix}-\ref{eq:fit_trans_KN}), to the numerically calculated
pIC for different electron cutoff energies. We find that the fits agree within a
few percent with the numerically calculated IC emission in the dominating flux
regime where the flux is larger than 10 percent of the maximum pIC flux (that have
been normalized with $E_\gamma^{\alpha_\nu})$.

\bsp

\label{lastpage}

\end{document}